\documentclass[journal]{IEEEtran}
\usepackage[T1]{fontenc}
\usepackage{textcomp}
\usepackage[latin9]{inputenc}
\usepackage{color}
\usepackage{array}
\usepackage{booktabs}
\usepackage{multirow}
\usepackage{varwidth}
\usepackage{amsmath}
\usepackage{amssymb}
\usepackage{graphicx}
\usepackage{setspace}
\usepackage[bookmarks=true,bookmarksnumbered=true,bookmarksopen=true,bookmarksopenlevel=1,
 breaklinks=false,pdfborder={0 0 0},pdfborderstyle={},backref=false,colorlinks=false]
 {hyperref}
\hypersetup{pdftitle={A Dual-Band Reconfigurable Shared-Aperture Antenna Array With Independent Sub-6-GHz and Centimeter-Wave Beam Control},
 pdfauthor={Your Name},
 pdfpagelayout=OneColumn, pdfnewwindow=true, pdfstartview=XYZ, plainpages=false}

\makeatletter

\providecommand{\tabularnewline}{\\}
\newenvironment{cellvarwidth}[1][t]
    {\begin{varwidth}[#1]{\linewidth}}
    {\@finalstrut\@arstrutbox\end{varwidth}}

\let\oldforeign@language\foreign@language
\DeclareRobustCommand{\foreign@language}[1]{%
  \lowercase{\oldforeign@language{#1}}}
\newenvironment{lyxlist}[1]
	{\begin{list}{}
		{\settowidth{\labelwidth}{#1}
		 \setlength{\leftmargin}{\labelwidth}
		 \addtolength{\leftmargin}{\labelsep}
		 }}
	{\end{list}}

\usepackage[caption=false,font=footnotesize]{subfig}
\usepackage{cite}

\usepackage{varwidth}

\newcommand{\yesmark}{\textcolor{green}{$\checkmark$}}
\newcommand{\nomark}{\textcolor{red}{$\times$}}

\makeatother

\begin{document}
\title{A Dual-Band Reconfigurable Shared-Aperture Antenna Array With Independent
Sub-6-GHz and Centimeter-Wave Beam Control}
\author{Zhaoyang~Ming,~\IEEEmembership{Student Member,~IEEE,} Junhui~Rao,~\IEEEmembership{Member,~IEEE,} Jichen~Zhang,~\IEEEmembership{Student Member,~IEEE,}
\\
 Yijun~Chen,~\IEEEmembership{Student Member,~IEEE,} Alikhan~Umirbayev,~\IEEEmembership{Student Member,~IEEE,}\\
Chi Yuk~Chiu,~\IEEEmembership{Senior Member,~IEEE,} and Ross~Murch,~\IEEEmembership{Fellow,~IEEE}\thanks{This work was supported by Hong Kong Research Grants Council for the
Area of Excellence Grant, AoE/E-601/22-R.}\thanks{Zhaoyang Ming, Junhui Rao, Jichen Zhang, Yijun Chen, Alikhan Umirbayev and Chi Yuk Chiu are with
the Department of Electronic and Computer Engineering, University
of Hong Kong Science and Technology, Hong Kong. (e-mail: \protect\href{mailto:zming@connect.ust.hk}{zming@connect.ust.hk},
\protect\href{mailto:jraoaa@connect.ust.hk}{jraoaa@connect.ust.hk}).}\thanks{R. Murch is with the Department of Electronic and Computer Engineering
and Institute for Advanced Study (IAS) at the Hong Kong University
of Science and Technology, Hong Kong. (e-mail: \protect\href{http://eermurch@ust.hk}{eermurch@ust.hk}).}}
\markboth{}{Zhaoyang Ming \MakeLowercase{\emph{et al.}}: bababababa}
\maketitle
\begin{abstract}
A planar dual-band reconfigurable shared-aperture antenna array is proposed
for compact next-generation wireless front ends that require both sub-6-GHz
and centimeter-wave (cm-wave) coverage. The array integrates
a 2\texttimes 2 sub-6-GHz microstrip dipole array and a 4\texttimes 4
cm-wave stacked patch array within the same aperture, while providing
independent beam control in the two bands without conventional T/R modules
or beamforming networks. Slot-coupled feeding is employed to separate
the radiating aperture from the reconfigurable RF feeding networks and
DC bias circuits. PIN-diode-loaded split feeding rings first provide independent
1-bit phase reconfigurability for both bands. A compact reconfigurable
$90^{\circ}$ phase shifter is then introduced as an additional phase-control
stage, resulting in 2-bit phase control for sub-6 GHz elements and cm-wave subarrays. To reduce cross-band coupling in the compact shared aperture,
a double-layer electromagnetic band-gap (EBG) structure is used to suppress
cm-wave surface waves and higher-order sub-6-GHz modes excited by the
cm-wave elements. A prototype is fabricated and measured. In the sub-6-GHz
band, 11 reconfigurable radiation patterns are obtained, including two
difference patterns and nine directional beams, with a peak broadside
gain of 10.5 dBi. In the cm-wave band, two-dimensional beam scanning
up to $\pm40^{\circ}$ is demonstrated with a peak gain of 14.6 dBi
in both the E-plane and H-plane. These results show that the proposed
architecture can combine dual-band shared-aperture integration and independent
reconfigurable beam control in a compact antenna platform.
\end{abstract}


\section{Introduction}

\begin{figure}[t]
	\begin{centering}
		\textcolor{black}{\includegraphics[width=1\columnwidth]{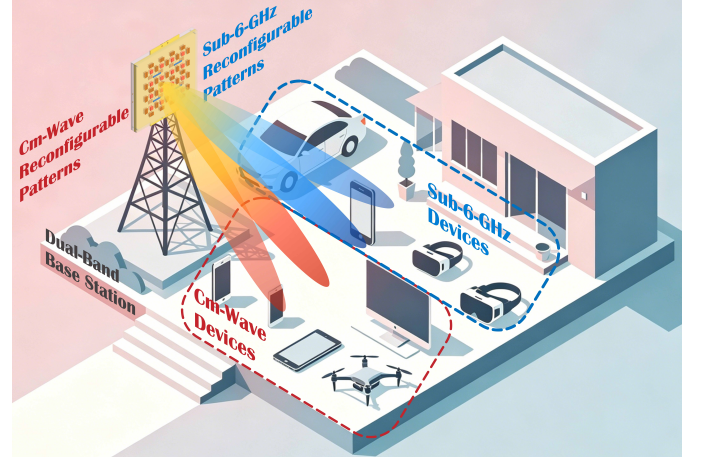}}
		\par\end{centering}
	\centering{}\textcolor{black}{\caption{\label{Application scenario} Application scenario of the proposed dual-band
			reconfigurable shared-aperture antenna array for sub-6-GHz and cm-wave
			base-station communications. Independent beams can be formed for to support sub-6 GHz and cm-wave devices.}
	}
\end{figure}

\begin{singlespace}
\IEEEPARstart{N}{ext-generation} wireless systems require antenna front
ends that can support multiple frequency bands, dynamic beam control,
and compact integration \cite{guo2025antenna}. In cellular systems,
the sub-6-GHz spectrum in Frequency Range 1 (FR1, 0.41-7.125 GHz) provides wide-area coverage, while the centimeter-wave (cm-wave) spectrum, often associated with the emerging Frequency Range 3 (FR3, approximately 7-24 GHz),
can provide high capacity with more manageable propagation loss than
the millimeter-wave bands associated with Frequency Range 2 (FR2, 24.25-52.6 GHz)
\cite{3gpp_ts38.521-1_v19.3.0,semaan20236g,cui20256g}. Therefore,
integrating sub-6-GHz and cm-wave functions within a shared aperture is
an attractive solution for space-limited base stations, as illustrated
in Fig. \ref{Application scenario}. In addition to dual-band operation,
the antenna front end should also support independent beam control in
both bands, so that the coverage and capacity links can be adjusted
separately.
\end{singlespace}

Shared-aperture antennas provide a compact way to integrate multiple
radiating functions within the same physical aperture. They have been
realized using interlaced layouts
\cite{mao2017dual,mao2017low,he2021low}, stacked arrangements
\cite{zou2023dual,li2018metasurface,chen2019novel,mao2017shared,cao2021compact},
nested configurations
\cite{naishadham2012shared,loui2003dual,deng2023singly,ding2021broadside,rodriguez2020mm},
and structure-reuse techniques
\cite{xiang2017flexible,zhang2020dual,ma2023microwave,zhang2018dual,ding2019ku}.
These approaches can reduce the aperture size of multi-band antenna
systems, but they also introduce strong cross-band coupling and make it
difficult to preserve bandwidth, scanning performance, aperture
efficiency, and low-profile integration at the same time. To address
these issues, previous studies have used enclosed structures, shielding
vias, defected ground structures, filtering feeders, frequency selective
surfaces, and metamaterial-based structures
\cite{ding2019ku,liu2019dual,hao2022k,zhang2023k,ding2022dual,bai2020metamaterial,li2024broadband}.

However, many shared-aperture phased arrays still rely on conventional
beam-steering hardware, such as distributed transmit/receive (T/R)
modules or complex beamforming networks, to realize dynamic beam
scanning. These architectures are effective, but they increase hardware
cost, power consumption, and integration complexity. This becomes a
clear limitation for compact and cost-sensitive base-station systems,
where both aperture sharing and low-complexity beam control are needed.

Reconfigurable phased arrays provide a promising way to reduce this
hardware burden. By embedding PIN diodes, varactors, or other tunable
devices into the radiating elements or feeding structures, these arrays
can switch discrete phase states without using a dedicated T/R module
behind each element
\cite{hu2018compact,yang20161,wu2021circularly,wu20232,wang2019design,wang2020novel,diaby20192,xiao2021design,mmwave-wang2022low,wang2022low,pan2021low,liu2019circularly,yin2021low,fang2023design,xi20232,li2023two,yin2023modular,wang2024low}.
Planar reconfigurable phased arrays \cite{hu2018compact,mmwave-wang2022low,wang2022low,pan2021low,liu2019circularly,yin2021low,fang2023design,xi20232,li2023two,yin2023modular,wang2024low} are especially attractive because
they avoid the additional spatial feeding structures used in 3-D reflectarray
or transmitarray architectures \cite{yang20161,wu2021circularly,wu20232,wang2019design,wang2020novel,diaby20192,xiao2021design},
making them suitable for low-profile integration.
Recent designs have demonstrated 1-bit
\cite{hu2018compact,mmwave-wang2022low,pan2021low}, 2-bit
\cite{liu2019circularly,yin2021low,fang2023design,xi20232,li2023two,yin2023modular,wang2024low},
and multi-bit \cite{wang2022low} phase control. In particular, 2-bit
arrays can achieve a 2-D scanning range up to $\pm60^{\circ}$
\cite{li2023two,yin2023modular}. However, most reported planar
reconfigurable phased arrays operate in a single frequency band, which
limits their use in multi-band systems that require independent beam
control across separated spectra.

Therefore, the key challenge is not only to integrate two antenna arrays
within the same aperture, but also to provide independent beam control
in both bands without relying on conventional T/R modules or complex
beamforming networks. Shared-aperture designs address the aperture
integration problem, while planar reconfigurable phased arrays address
the hardware-complexity problem. However, these two advantages have not
been fully combined in a dual-band shared-aperture array. This is
especially challenging when the two bands are widely separated and must
be reconfigured independently within a compact planar aperture.

To address this challenge, this paper proposes a dual-band
reconfigurable shared-aperture antenna array that supports independent
beam control in the sub-6-GHz and cm-wave bands. Fig. \ref{Application scenario}
illustrates the intended deployment of the proposed array in a space-limited
base station that jointly uses FR1 and emerging FR3 spectra. Specifically,
this array integrates a 2$\times$2 sub-6-GHz microstrip dipole array
and a 4$\times$4 cm-wave stacked patch array within the same aperture. Each band uses
switch-controlled 1-bit phase reconfiguration, and a compact
reconfigurable $90^{\circ}$ phase shifter is further introduced to improve the
available phase states. As a result, the proposed array realizes 2-bit
phase control, while keeping the two bands independently
reconfigurable. The main contributions are summarized as follows:

\textit{1) Dual-Band Reconfigurable Shared-Aperture Architecture:}
A dual-band shared-aperture array is developed by interleaving a
2$\times$2 sub-6-GHz microstrip dipole array with a 4$\times$4 cm-wave
stacked patch array. To the best of the authors' knowledge, this is one
of the first reported dual-band reconfigurable shared-aperture arrays
that provides independent beam control in both the sub-6-GHz and cm-wave
bands without using conventional T/R modules or beamforming networks.

\textit{2) Compact Phase-Reconfigurable Feeding Design:}
A miniaturized reconfigurable $90^{\circ}$ phase shifter is developed for
both frequency bands. By combining this phase shifter with the 1-bit
switching mechanism of the antenna elements, 2-bit phase control of the sub-6 GHz elements and cm-wave subarrays is achieved.

\textit{3) Double-Layer EBG for Cross-Band Coupling Suppression:}
A double-layer electromagnetic band-gap (EBG) structure is introduced
inside the shared aperture. It suppresses cm-wave surface waves and
reduces higher-order sub-6-GHz modes excited by the cm-wave elements,
which helps maintain the radiation performance of the closely integrated
dual-band array.

\textit{4) Beamforming Optimization and Experimental Verification:}
An optimization method based on full-wave element radiation patterns is
used to select the fixed phase delays and reconfigurable phase states
for beam steering. A prototype is fabricated and measured. It achieves
11 representative sub-6-GHz patterns with a peak broadside gain of
10.5 dBi, and 2-D cm-wave beam scanning up to $\pm40^{\circ}$ with a
peak gain of 14.6 dBi.

The remainder of this paper is organized as follows. Section II
describes the overall configuration of the proposed antenna array.
Section III presents the shared-aperture and phase-reconfigurable
design. Section IV introduces the beamforming design and optimization
method. Section V reports the simulated and measured results of the
fabricated prototype. Section VI compares the proposed design with
related work. Finally, Section VII concludes the paper.

\begin{figure}[t]
\begin{centering}
\textcolor{black}{\includegraphics[width=1\columnwidth]{\string"Figures/Overall_configuration_v3-01\string".jpg}}
\par\end{centering}
\begin{raggedright}
\hspace*{0.48\columnwidth} (a)
\par\end{raggedright}
\begin{centering}
\textcolor{black}{\includegraphics[width=1\columnwidth]{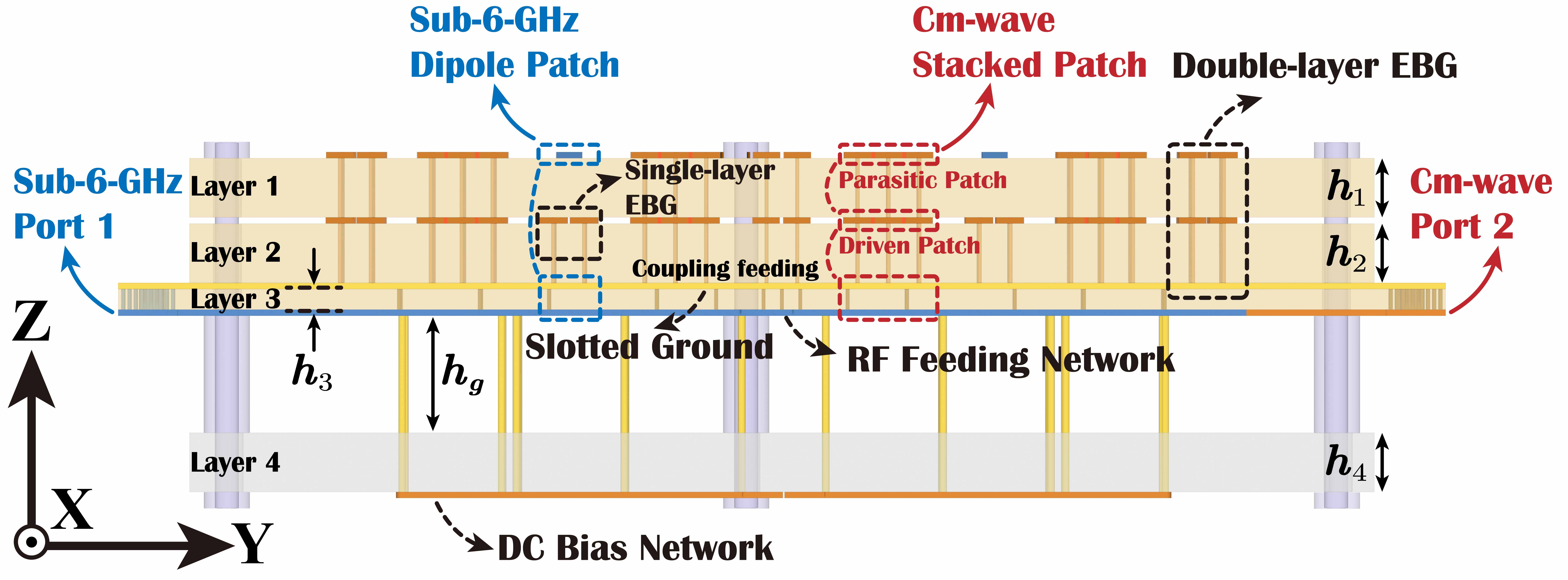}}
\par\end{centering}
\begin{raggedright}
\hspace*{0.48\columnwidth} (b)
\par\end{raggedright}
\centering{}\textcolor{black}{\caption{\label{Overall structure} Overall configuration of the proposed dual-band reconfigurable
shared-aperture antenna array. (a) Three-dimensional view. (b) Multi-layer
cross-sectional view. The substrate thicknesses and air gap are $h_{1}=1.524$,
$h_{2}=1.524$, $h_{3}=0.508$, $h_{4}=1.524$, and $h_{g}=3$ (unit:
mm). Further dimension details are provided in Fig. \ref{Multi layers} and \ref{RF feeding layer}.}
}
\end{figure}

\section{Overall Configuration}

\begin{figure}[tp]
\begin{centering}
\textsf{\includegraphics[width=0.49\columnwidth]{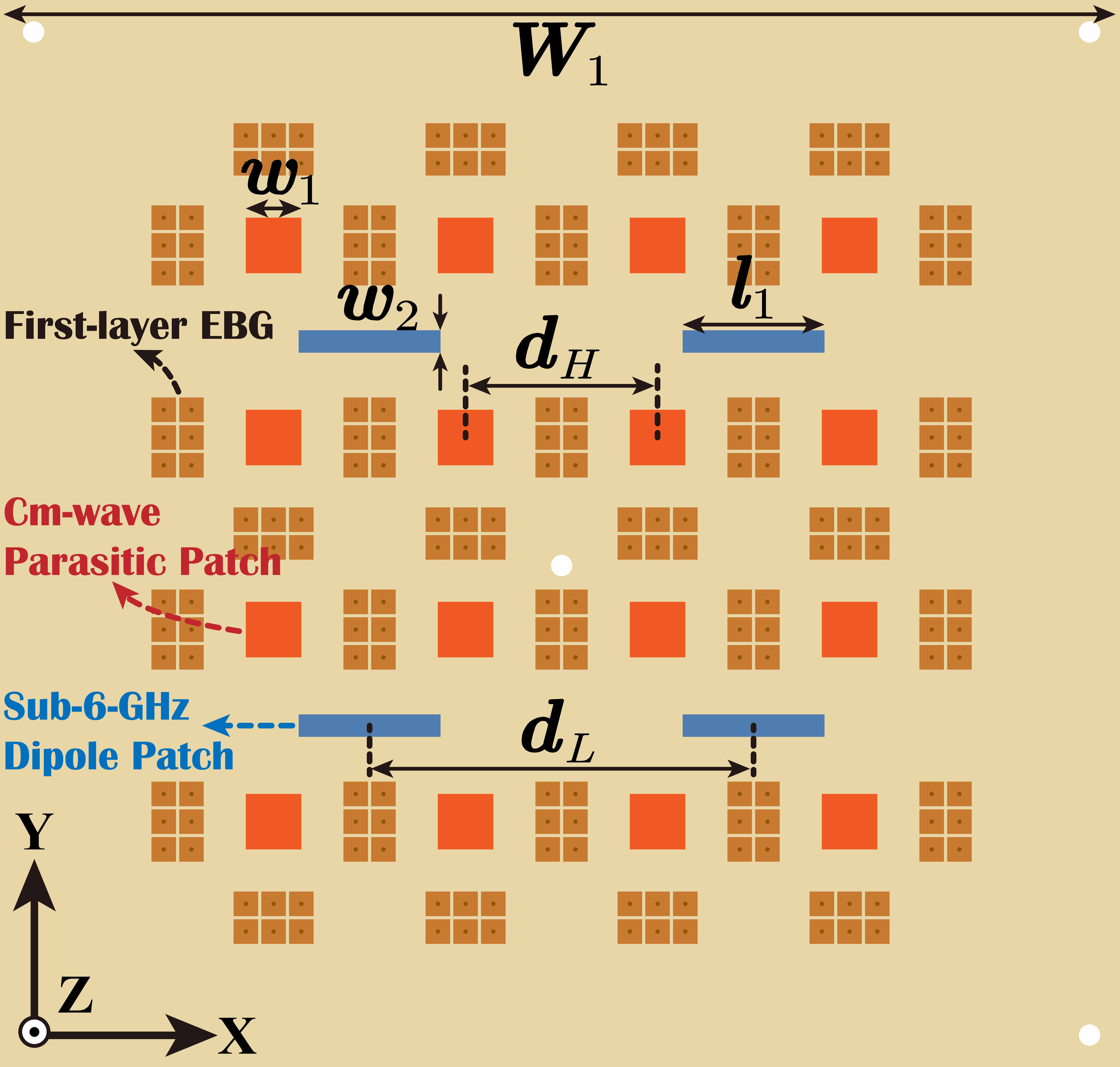}\hspace*{0.01\columnwidth}\includegraphics[width=0.49\columnwidth]{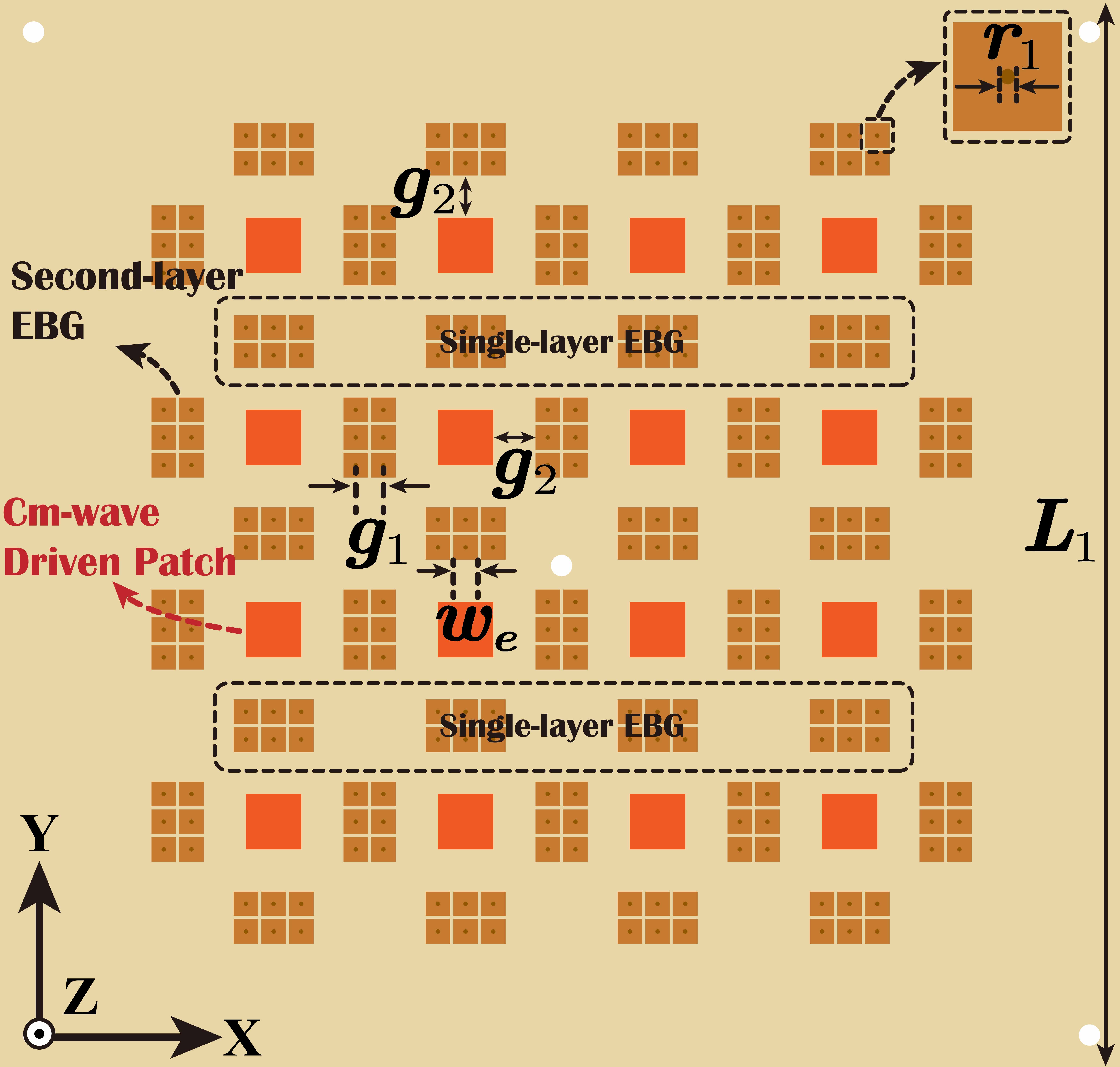}}
\par\end{centering}
\begin{raggedright}
\hspace*{0.22\columnwidth} (a)\hspace*{0.45\columnwidth} (b)
\par\end{raggedright}
\begin{centering}
\textsf{\includegraphics[width=0.49\columnwidth]{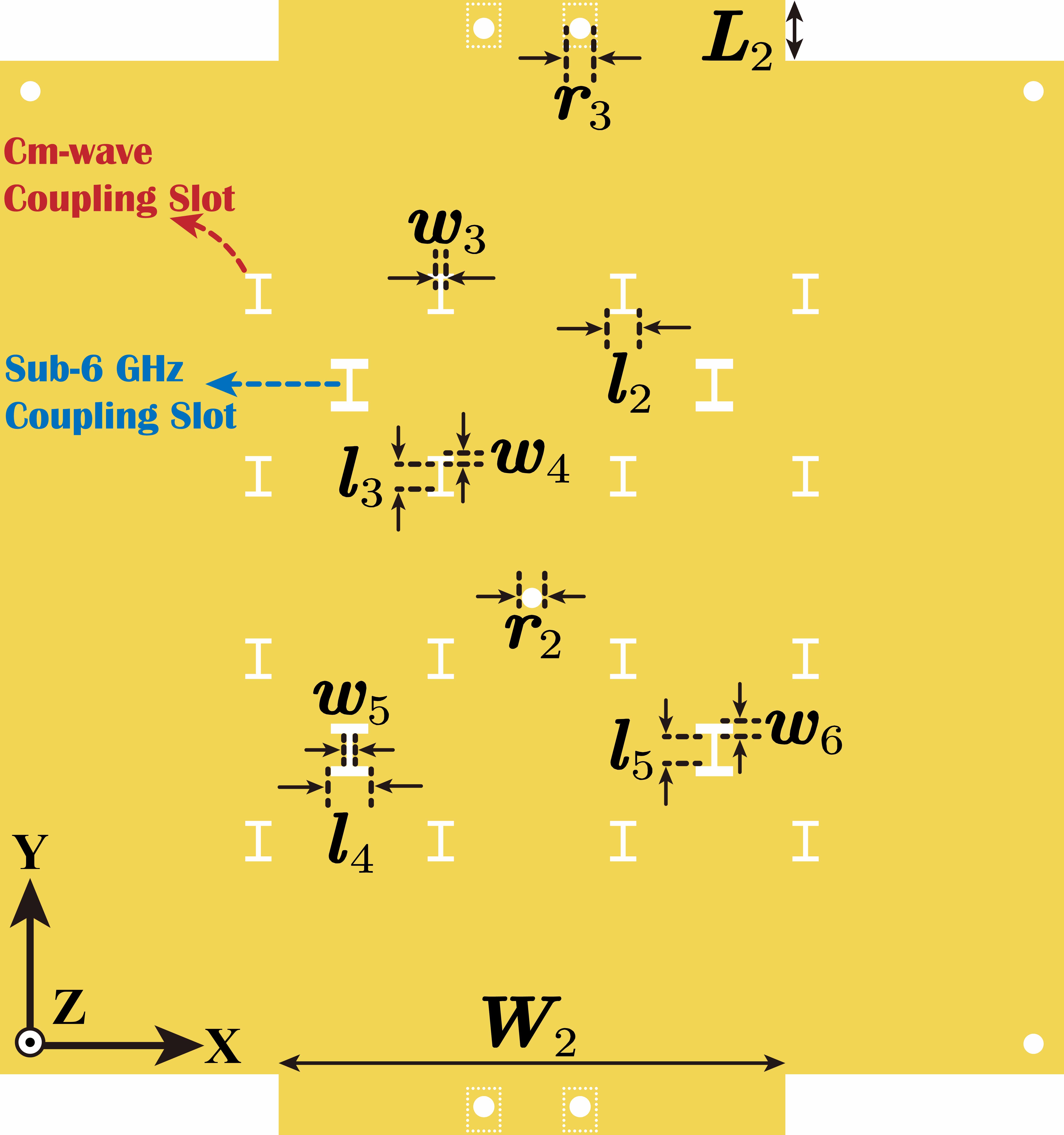}\hspace*{0.01\columnwidth}\includegraphics[width=0.49\columnwidth]{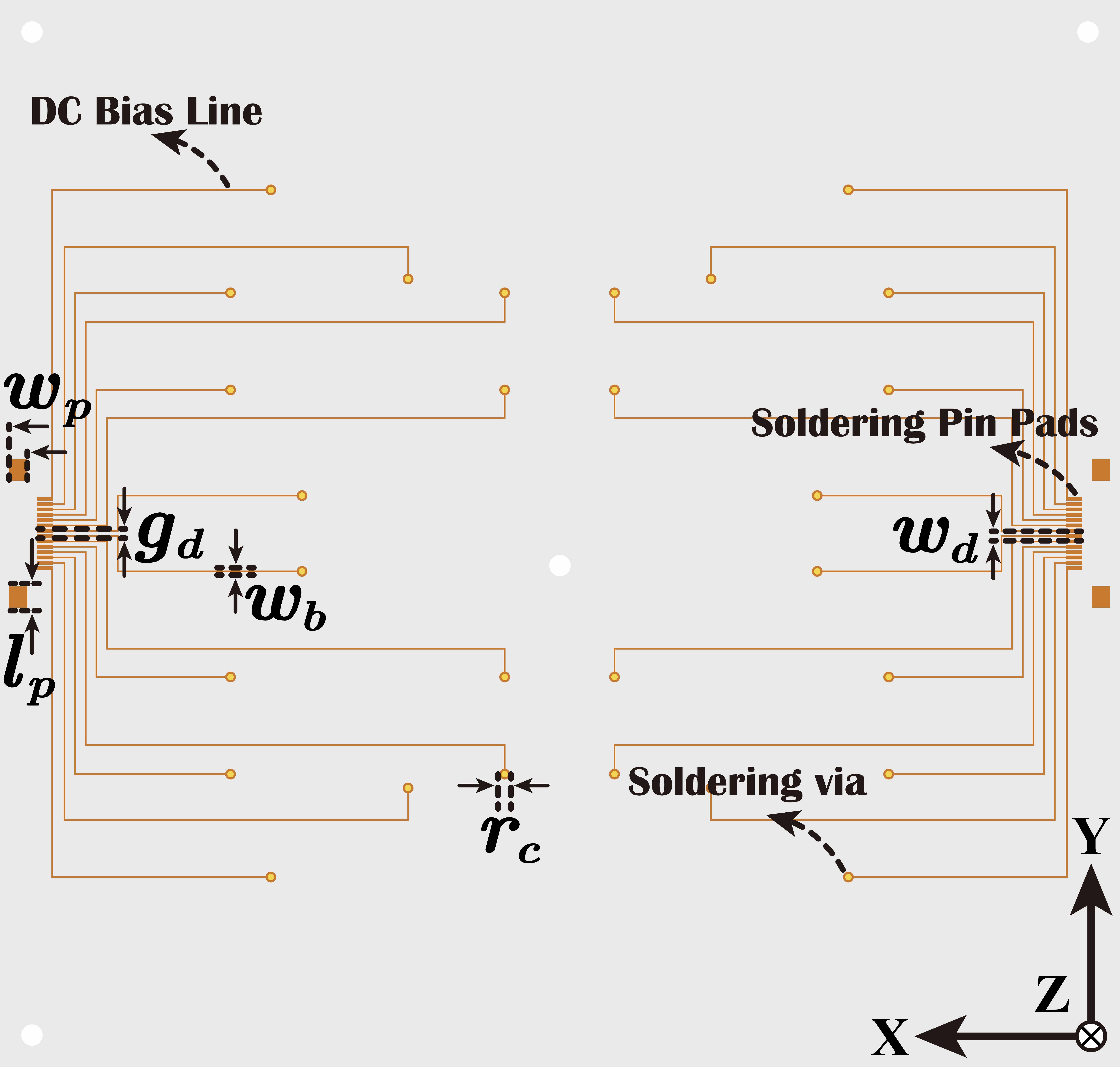}}
\par\end{centering}
\begin{raggedright}
\hspace*{0.22\columnwidth} (c)\hspace*{0.45\columnwidth} (d)
\par\end{raggedright}
\caption{Layer-by-layer layout of the proposed dual-band reconfigurable shared-aperture
antenna array. (a) Top view of substrate layer 1. (b) Top view of
substrate layer 2. (c) Slotted ground plane. (d) DC bias network on
substrate layer 4. The geometrical parameters are $W_{1}=105$, $W_{2}=50$,
$L_{1}=108$, $L_{2}=6$, $d_{L}=36$, $d_{H}=18$, $w_{1}=5.2$,
$w_{2}=2.1$, $w_{3}=0.5$, $w_{4}=0.5$, $w_{5}=0.6$, $w_{6}=1$,
$w_{e}=2.3$, $w_{b}=0.15$, $w_{d}=0.35$, $w_{p}=1.7$,
$l_{1}=13.3$, $l_{2}=2.6$, $l_{3}=3$, $l_{4}=3.7$, $l_{5}=3.2$,
$l_{p}=2$, $r_{1}=0.3$, $r_{2}=2$, $r_{3}=2.1$, $r_{c}=0.6$,
$g_{1}=2.6$, $g_{2}=3.95$, and $g_{d}=0.5$ (unit: mm).}
\label{Multi layers}
\end{figure}

\begin{figure*}[tp]
\begin{centering}
\textsf{\includegraphics[width=0.99\columnwidth]{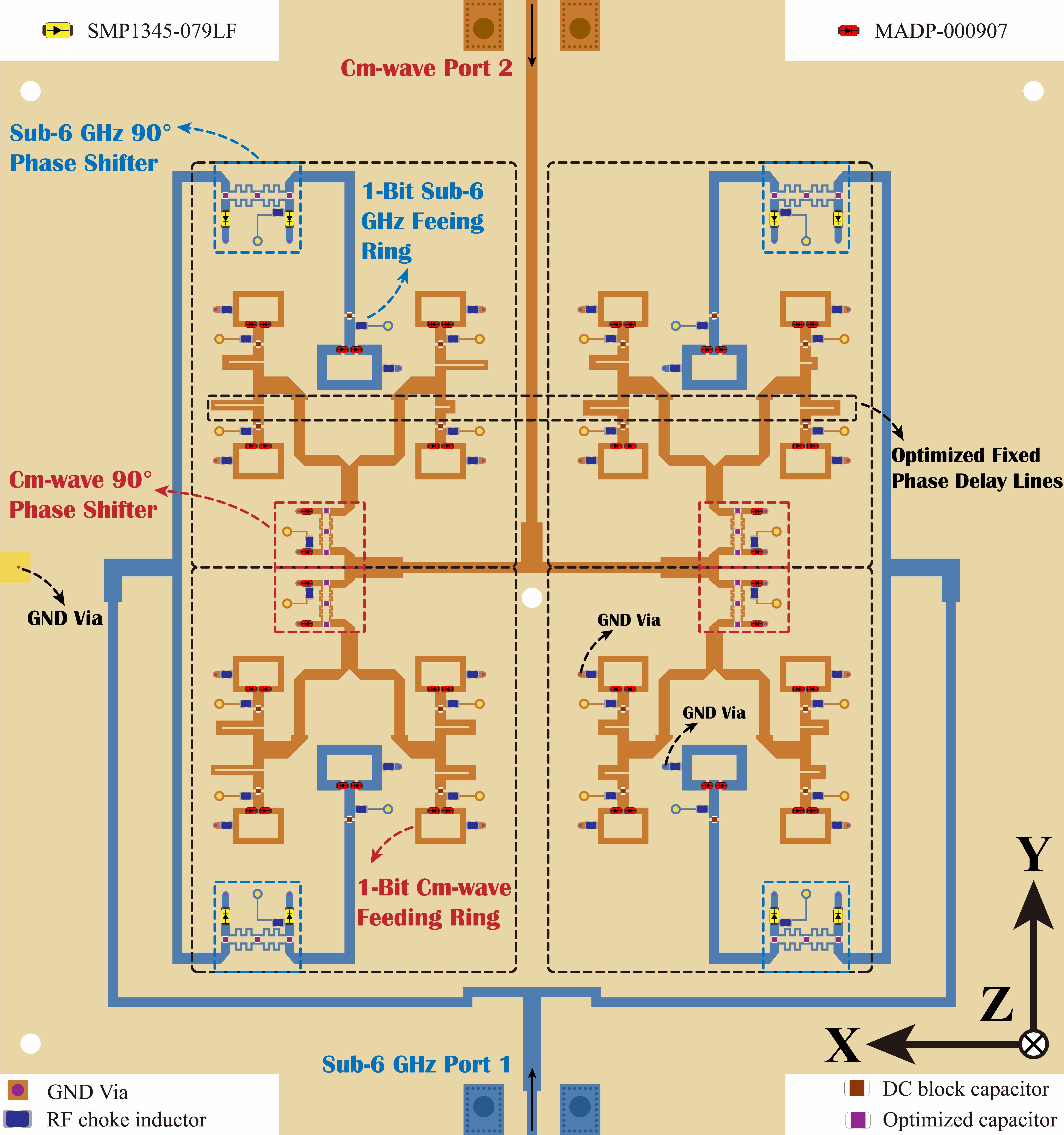}\hspace*{0.02\columnwidth}\includegraphics[width=0.99\columnwidth]{\string"Figures/Subarray_1\string".jpg}}
\par\end{centering}
\begin{raggedright}
\hspace*{0.485\columnwidth} (a)\hspace*{0.98\columnwidth} (b)
\par\end{raggedright}
\caption{RF feeding network and subarray-level phase-control layout on substrate
layer 3. (a) Overall RF feeding network. (b) Zoomed-in view of the
reconfigurable feeding network for one shared-aperture subarray. The
geometrical parameters are $w_{1}=1.1$, $w_{2}=1$, $w_{3}=1$,
$w_{4}=1.1$, $w_{5}=0.6$, $l_{1}=17.975$, $l_{2}=16$, $l_{3}=2.5$,
$l_{4}=4.4$, $l_{5}=2.4$, $l_{6}=3.8$, $l_{7}=4.85$,
$w_{d_{1}}=1.1$, $w_{d_{2}}=2.1$, $w_{d_{3}}=1$, $w_{d_{4}}=1.7$,
$w_{d_{5}}=1.1$, $l_{d_{1}}=3.9$, $l_{d_{2}}=4.8$, $l_{d_{3}}=3.6$,
$l_{d_{4}}=5.45$, $l_{d_{5}}=2.65$, $w_{f_{1}}=1.75$, $w_{f_{2}}=30$,
$l_{f_{1}}=0.95$, $l_{f_{2}}=40.95$, $l_{f_{3}}=39.95$,
$l_{f_{4}}=4.3$, $l_{f_{5}}=6.75$, and $l_{f_{6}}=38.225$ (unit: mm).}
\label{RF feeding layer}
\end{figure*}

The overall configuration of the proposed antenna array is shown in
Fig. \ref{Overall structure}, with detailed orthographic views given in
Fig. \ref{Multi layers} and Fig. \ref{RF feeding layer}. The array
uses a dual-band shared-aperture architecture in which two
phase-reconfigurable arrays are interleaved within the same physical
aperture. The first is a 2$\times$2 sub-6-GHz microstrip dipole
array operating at 5.2 GHz with an element spacing of 36 mm
($0.624\lambda_{L}$), where $\lambda_{L}$ is the free-space wavelength
at 5.2 GHz. The second is a 4$\times$4 cm-wave stacked patch array
operating at 10.4 GHz with an element spacing of 18 mm
($0.624\lambda_{H}$), where $\lambda_{H}$ is the free-space wavelength
at 10.4 GHz. A double-layer EBG structure is also included to suppress cross-band coupling. The total size of the antenna array is
108$\times$105$\times$8.19 $\mathrm{mm}^{3}$
(1.872$\times$1.82$\times$0.142 $\lambda_{L}^{3}$).

The antenna is implemented using a four-layer structure. Three Rogers
4003C layers ($\varepsilon_{r}=3.55$, $\tan\delta=0.0027$) form the
radiating elements and the RF feeding networks for the two bands. A
fourth FR-4 layer ($\varepsilon_{r}=4.3$, $\tan\delta=0.025$) is used
to support the DC bias network. Short copper posts connect the DC bias
pads in the reconfigurable RF feeding networks to the DC bias layer.
This arrangement separates the radiating aperture, RF feeding networks,
and digital bias network while keeping the overall profile compact.

The sub-6-GHz array is fed through port 1, and the cm-wave array is fed
through port 2. The phase states of the radiating elements and subarrays
are controlled by the DC bias network, allowing the two bands to be
reconfigured independently for the target beamforming states. The
shared-aperture topology, phase-reconfigurable feeding design, and
beamforming optimization are detailed in the following sections.


\section{Shared-Aperture and Phase Reconfigurable Design}

This section presents the shared-aperture and phase-reconfigurable design
of the proposed array. The discussion first introduces the interleaved
sub-6-GHz and cm-wave antenna elements and their 1-bit switching mechanism.
It then describes the miniaturized reconfigurable $90^{\circ}$ phase
shifter, followed by the double-layer EBG structure used to suppress
cross-band coupling. Finally, the sub-6-GHz element and 
cm-wave $2 \times 2$ subarray are summarized as the basic reconfigurable building
blocks of the complete array.

\subsection{1-Bit Sub-6 GHz and Centimeter-Wave Band Antenna Elements}

Fig. \ref{Multi layers} and Fig. \ref{RF feeding layer} show the detailed
geometry and layer arrangement of the proposed shared-aperture array.
The first design step is to integrate two independently reconfigurable
radiating structures within the same aperture. As shown in Fig. \ref{Multi layers}(a)-(b),
a 2$\times$2 sub-6-GHz microstrip dipole array is interlaced with
a 4$\times$4 cm-wave stacked square patch array. The two upper Rogers
4003C layers, each with a thickness of 1.524 mm, support the main
radiators and the embedded double-layer EBG structures. The sub-6-GHz
dipoles are placed on the first dielectric layer, which is also used
by the parasitic patches of the cm-wave elements, whereas the cm-wave
driven patches are placed on the second dielectric layer. This arrangement
keeps the shared aperture compact and reduces direct coupling between
the sub-6-GHz dipoles and the cm-wave driven patches.

To reduce parasitic radiation from the RF feeding networks and DC bias
lines, slot-coupled feeding is used for both bands. The shared H-shaped
slotted ground plane, located on the bottom of the second Rogers 4003C
layer as shown in Fig. \ref{Multi layers}(c), serves as the coupling
interface between the radiators and the feeding networks. With this
ground plane, the radiating elements are separated from the DC bias
network in Fig. \ref{Multi layers}(d) and the RF feeding networks in
Fig. \ref{RF feeding layer}(a). This separation prevents the reconfigurable
feeding circuits from being directly stacked with the radiators and
helps preserve the radiation performance. The third Rogers 4003C layer
in Fig. \ref{RF feeding layer} is made thinner, with a thickness of 0.508 mm, to ensure sufficient coupling between the slotted ground plane and the lower feeding networks.

Based on the shared-aperture topology and slot-coupled feeding scheme,
each radiating element is further provided with 1-bit phase reconfigurability.
As shown in Fig. \ref{RF feeding layer}(b), one pair of series PIN diodes
is embedded in the symmetrical split feeding ring of each sub-6-GHz
or cm-wave element. By turning on one diode and turning off the other,
the feeding path through the split ring is reversed, producing two phase
states separated by $180^{\circ}$. PIN diode MADP-000907 from
MACOM is used in the RF switches for both bands, together with band-specific
RF chokes and DC-blocking capacitors. For each element, the split feeding
ring is connected to the reference ground through a via biased at 0 V,
while the control voltage between the two series diodes is switched
between -1.6 V and 1.6 V. In the prototype, this voltage is supplied
by one I/O pin of a commercial FPGA with a 3.3-/0-V output and a ground
reference biased at -1.6 V. A total of 20 I/O pins are required to
realize the 1-bit phase reconfiguration: 4 pins for the 2$\times$2
sub-6-GHz array and 16 pins for the 4$\times$4 cm-wave array.

\subsection{\texorpdfstring{Miniaturized Reconfigurable $90^{\circ}$ Phase Shifter}{Miniaturized Reconfigurable 90-degree Phase Shifter}}

\begin{figure}[t]
\begin{centering}
\textsf{\includegraphics[width=0.49\columnwidth]{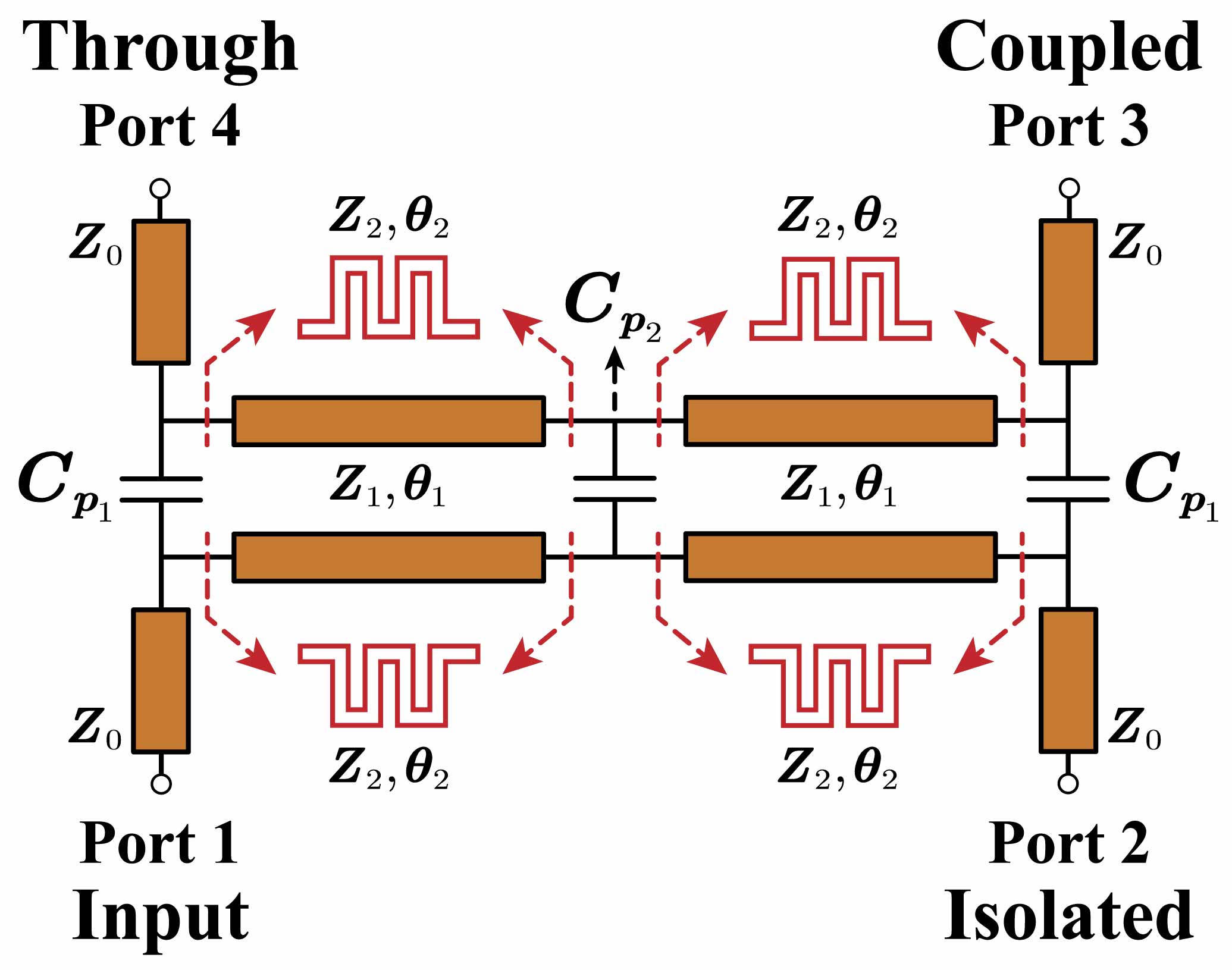}\hspace*{0.02\columnwidth}\includegraphics[width=0.49\columnwidth]{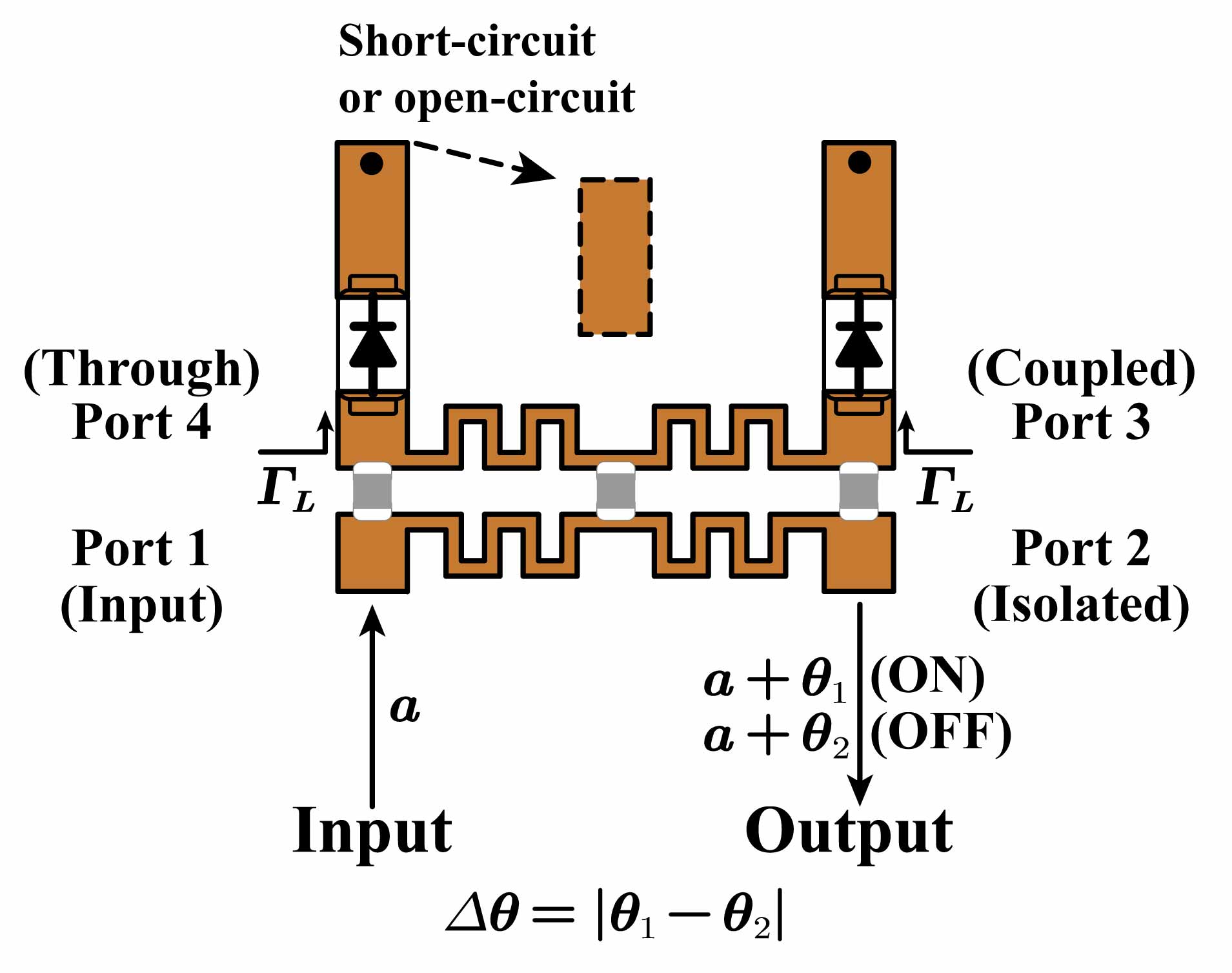}}
\par\end{centering}
\begin{raggedright}
\hspace*{0.21\columnwidth} (a)\hspace*{0.45\columnwidth} (b)
\par\end{raggedright}
\caption{Design principle of the miniaturized reconfigurable $90^{\circ}$
phase shifter. (a) Compact planar quadrature coupler used as the phase-shifter
core. (b) Reflection-type phase-shifting mechanism with switchable open-
and short-circuited loads.}
\label{Design principle of PS}
\end{figure}

The 1-bit phase states described above provide only a $180^{\circ}$
phase difference, which limits the beam scanning range and sidelobe
suppression of both bands. To increase the phase resolution, a miniaturized
reconfigurable $90^{\circ}$ reflection-type phase shifter is introduced
and integrated with the 1-bit shared-aperture array, as marked by
the blue and red dashed lines in Fig. \ref{RF feeding layer}. The
proposed phase shifter is based on the compact planar quadrature coupler
in \cite{lin2018compact}, as shown in Fig. \ref{Design principle of PS}(a).
In this coupler, the vertical transmission-line sections of a conventional
branch-line coupler are replaced by capacitors $C_{P_{1}}$
and $C_{P_{2}}$ based on even-odd-mode analysis \cite{pozar2011microwave}.
Meandered coupled lines are also used to further reduce the circuit
size. As a result, the compact coupler can serve as the core component
that makes up the proposed reflection-type phase shifter.

As shown in Fig. \ref{Design principle of PS}(b), the through and
coupled ports of the miniaturized coupler are terminated by a pair
of switchable microstrip loads. Each load can be configured as a short-
or open-circuit line with an optimized electrical length. By changing
the states of the RF switches, the reflected signals from the two
ports acquire different phases, and the two-port network provides
the required phase shift between the ON and OFF states. Because the
quadrature coupler provides a $90^{\circ}$ phase difference between
the through and coupled paths, the reflected waves are canceled at
the input port and combined at the isolated port. This reflection-type
operation helps maintain good impedance matching and low insertion
loss while enabling compact phase reconfiguration.

\begin{figure}[t]
\begin{centering}
\textsf{\includegraphics[width=0.47\columnwidth]{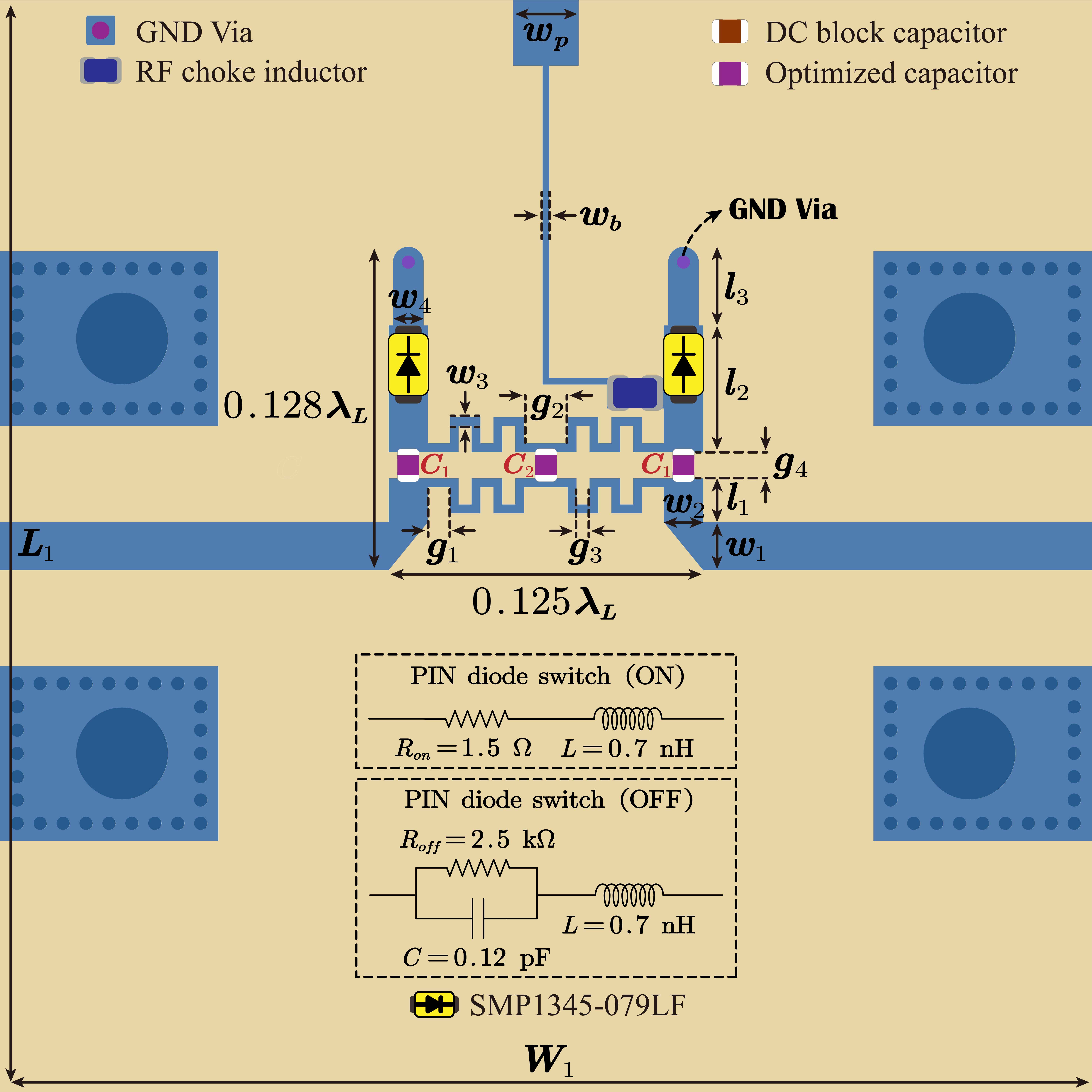}\hspace*{0.01\columnwidth}\includegraphics[width=0.52\columnwidth]{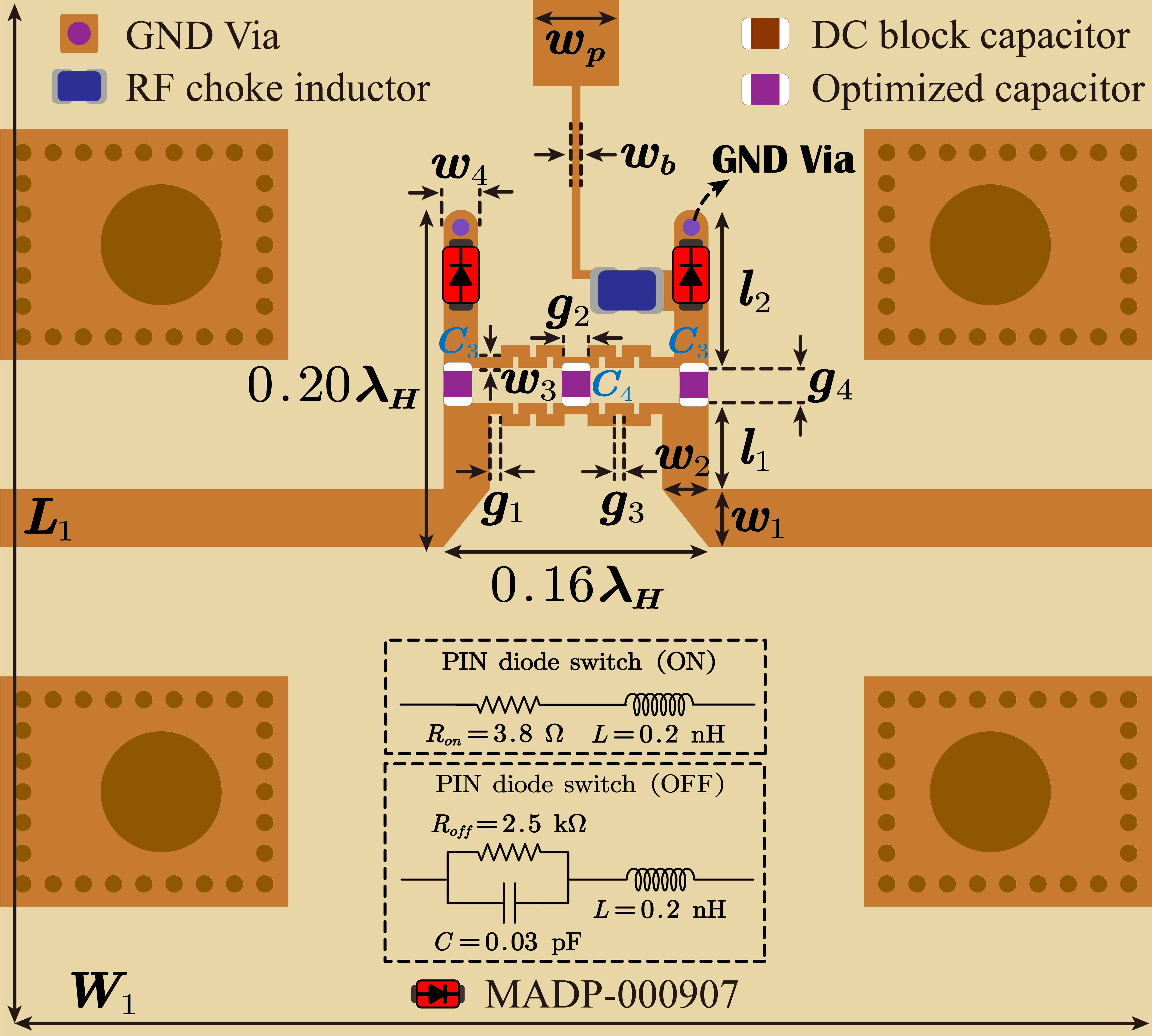}}
\par\end{centering}
\begin{raggedright}
\hspace*{0.2\columnwidth} (a)\hspace*{0.46\columnwidth} (b)
\par\end{raggedright}
\centering{}\textcolor{black}{\caption{\label{PS schematics} Schematics of the proposed reconfigurable
$90^{\circ}$ phase shifters for both bands. (a) Sub-6-GHz phase shifter
with optimized capacitors of $C_{1}=0.6$ pF and $C_{2}=0.3$ pF. The
geometrical parameters are $W_{1}=25$, $L_{1}=25$, $w_{1}=1.1$,
$w_{2}=0.9$, $w_{3}=0.2$, $w_{4}=0.9$, $w_{b}=0.15$, $w_{p}=1.5$,
$l_{1}=1$, $l_{2}=2.9$, $l_{3}=1.8$, $g_{1}=0.5$, $g_{2}=1$,
$g_{3}=0.3$, and $g_{4}=0.6$ (unit: mm). (b) Cm-wave phase shifter
with optimized capacitors of $C_{3}=0.3$ pF and $C_{4}=0.2$ pF. The
geometrical parameters are $W_{1}=20$, $L_{1}=18$, $w_{1}=1$,
$w_{2}=0.8$, $w_{3}=0.2$, $w_{4}=0.6$, $w_{b}=0.15$, $w_{p}=1.5$,
$l_{1}=1.5$, $l_{2}=2.75$, $g_{1}=0.2$, $g_{2}=0.4$, $g_{3}=0.1$,
and $g_{4}=0.6$ (unit: mm).}
}
\end{figure}

Two miniaturized reconfigurable phase shifters are then designed for
the sub-6-GHz and cm-wave bands, as shown in Fig. \ref{PS schematics}(a)
and (b), respectively. The two designs use the same reflection-type
topology, but their dimensions, lumped capacitors, and switching devices
are optimized separately for the two frequency bands. For the sub-6-GHz
phase shifter, the capacitors $C_{1}$ and $C_{2}$ are used to replace
the vertical transmission-line sections of the compact coupler. For
the cm-wave phase shifter, the corresponding capacitors are $C_{3}$
and $C_{4}$. The SMP1345-079LF diode from Skyworks is selected for
the sub-6-GHz design because of its low ON-state resistance and good
OFF-state isolation. The MADP-000907 diode from MACOM is used for
the cm-wave design because it provides better OFF-state isolation at
the higher operating frequency while maintaining acceptable ON-state
resistance. Band-specific RF chokes and DC-blocking capacitors are also
added to support the bias and RF paths of the two phase shifters.

\begin{figure}[t]
\begin{centering}
\textsf{\includegraphics[width=0.49\columnwidth]{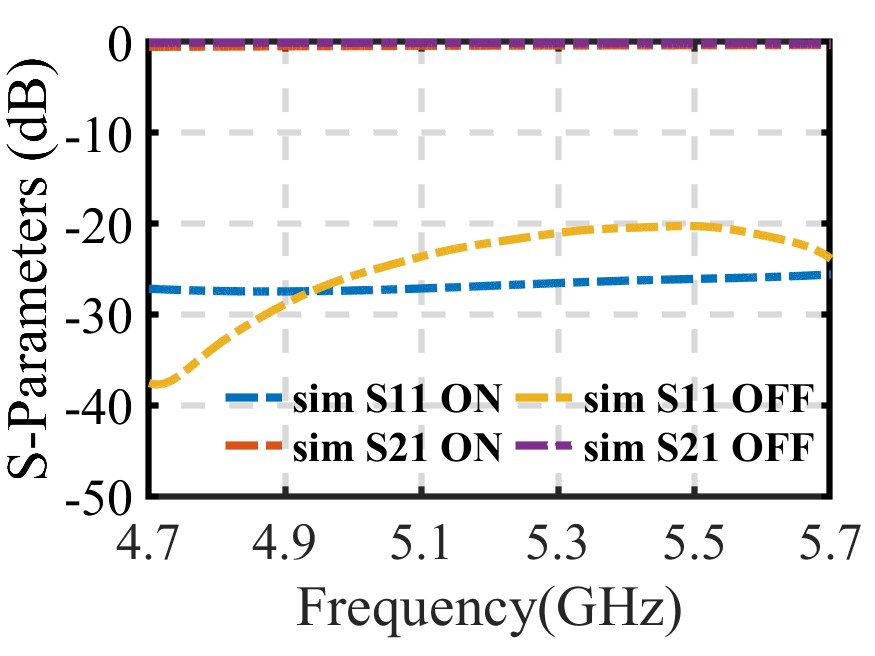}\hspace*{0.02\columnwidth}\includegraphics[width=0.49\columnwidth]{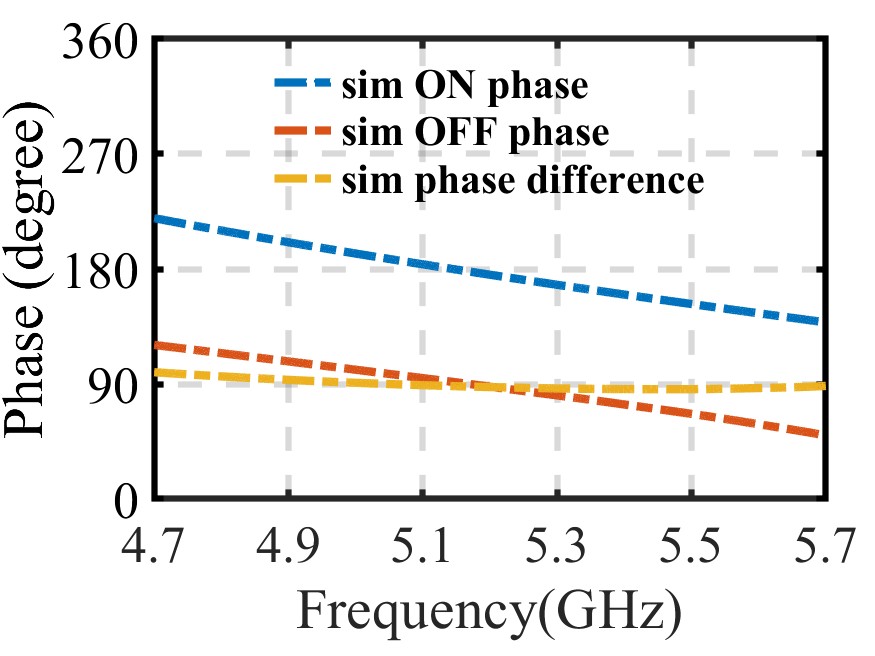}}
\par\end{centering}
\begin{raggedright}
\hspace*{0.24\columnwidth} (a)\hspace*{0.45\columnwidth} (b)
\par\end{raggedright}
\begin{centering}
\textsf{\includegraphics[width=0.49\columnwidth]{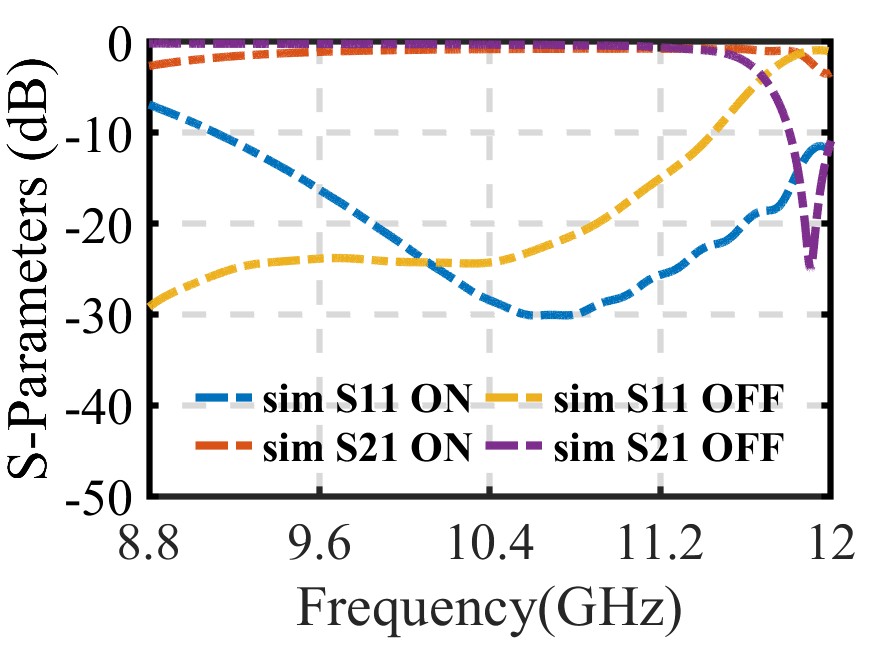}\hspace*{0.02\columnwidth}\includegraphics[width=0.49\columnwidth]{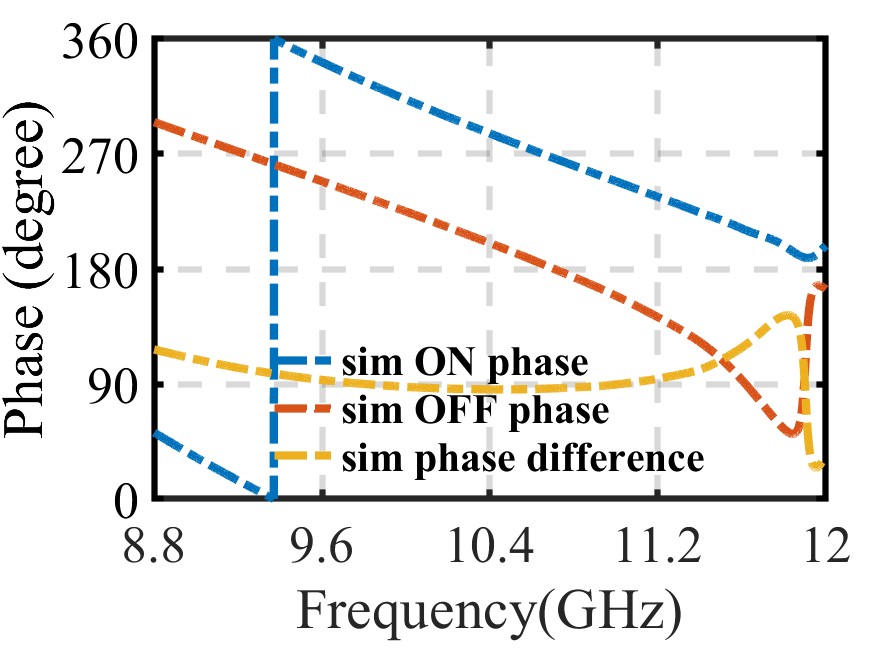}}
\par\end{centering}
\begin{raggedright}
\hspace*{0.24\columnwidth} (c)\hspace*{0.45\columnwidth} (d)
\par\end{raggedright}
\caption{Simulated performance of the proposed reconfigurable $90^{\circ}$
phase shifters. (a) S-parameters of the sub-6-GHz phase shifter in
the ON and OFF states. (b) Phase responses and phase difference between
the two states of the sub-6-GHz phase shifter. (c) S-parameters of
the cm-wave phase shifter in the ON and OFF states. (d) Phase responses
and phase difference between the two states of the cm-wave phase shifter.}
\label{Sim Performance of PS}
\end{figure}

The simulated performance of the two phase shifters is shown in Fig.
\ref{Sim Performance of PS}. Both phase shifters are placed on the
bottom side of the third Rogers 4003C layer, whose thickness is 0.508
mm. For the sub-6-GHz design, the phase shifter occupies only 7.2$\times$7.4
$\mathrm{mm}^{2}$, corresponding to 0.125$\times$0.128 $\lambda_{L}^{2}$.
The ON and OFF states exhibit good impedance matching and an insertion
loss below 0.5 dB, while the phase difference remains close to $90^{\circ}$
from 4.9 to 5.5 GHz. For the cm-wave design, the phase shifter has
a compact size of 5.85$\times$4.6 $\mathrm{mm}^{2}$, corresponding
to 0.16$\times$0.20 $\lambda_{H}^{2}$. It also exhibits good impedance
matching, an insertion loss below 0.82 dB, and a stable $90^{\circ}$
phase difference from 9.8 to 11 GHz. These results verify that the
proposed phase shifters provide compact and low-loss $90^{\circ}$
phase reconfiguration for both operating bands.

The proposed $90^{\circ}$ phase shifters are used as an additional
phase-control stage in the shared-aperture array. As shown in Fig.
\ref{RF feeding layer}, the sub-6-GHz phase shifter is integrated
with each 1-bit sub-6-GHz element, while the cm-wave phase shifter
is shared by each 2$\times$2 cm-wave subarray because of the limited
aperture space. For the cm-wave band we refer to this as quasi-2-bit phase control since 1-bit control is available for all elements but the second bit can only control its 2$\times$2 cm-wave subarrays. This additional phase-control stage requires 8 I/O
pins in total, including 4 pins for the sub-6-GHz elements and 4 pins
for the cm-wave subarrays. The resulting 2-bit sub-6-GHz element and
quasi-2-bit cm-wave subarray are discussed in the following subsection.

\subsection{Double-Layer EBG}

\begin{figure}[t]
\begin{centering}
\textcolor{black}{\includegraphics[width=1\columnwidth]{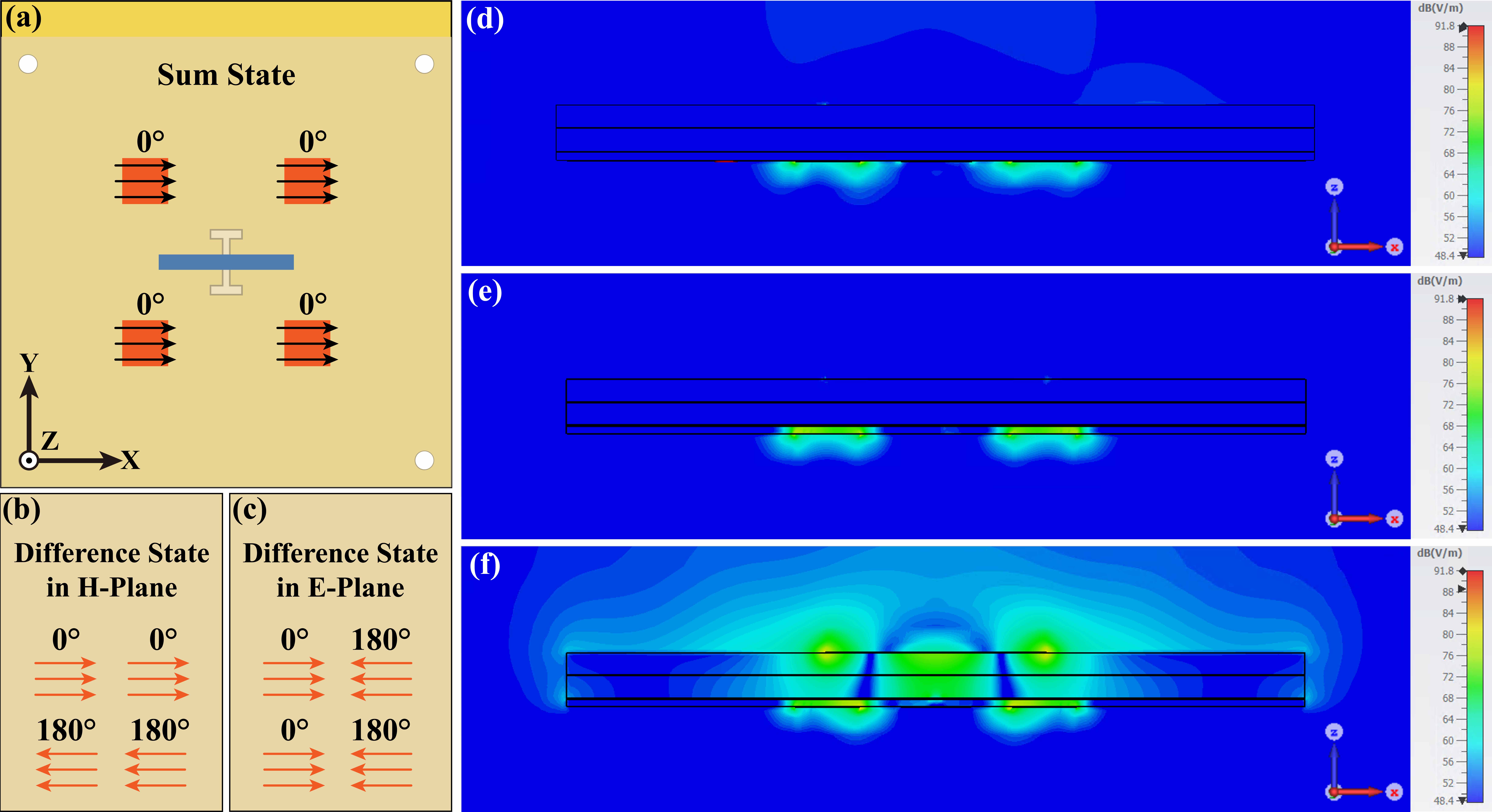}}
\par\end{centering}
\caption{Three representative phase states of the cm-wave subarray without
EBG structures and their electric-field distributions. (a) Sum state.
(b) H-plane difference state in the YOZ plane. (c) E-plane difference
state in the XOZ plane. (d)-(f) Corresponding central cross-sectional
electric-field distributions for the sum, H-plane difference, and E-plane
difference states, respectively.}
\label{coupling problem}
\end{figure}

The proposed array has an approximate 2:1 frequency ratio between
the cm-wave band at 10.4 GHz and the sub-6-GHz band at 5.2 GHz. Because
of this frequency relationship and the compact shared-aperture layout,
the cm-wave fields can excite higher-order modes on the sub-6-GHz
dipole and surface waves along the substrate. To identify the most
critical coupling condition, the 2$\times$2 cm-wave subarray is configured
into three representative states: the sum state, the H-plane difference
state, and the E-plane difference state, as shown in Fig. \ref{coupling problem}(a)-(c).
The corresponding electric-field distributions along the central x-axis
cross-section are shown in Fig. \ref{coupling problem}(d)-(f). Among
these cases, the E-plane difference state produces the strongest field
concentration around the central sub-6-GHz dipole and excites pronounced
surface-wave propagation along the substrate. As a result, the cm-wave
radiation performance is degraded, with pattern distortion, reduced
gain, and increased cross-polarization.

\begin{figure}[t]
\begin{centering}
\textcolor{black}{\includegraphics[width=1\columnwidth]{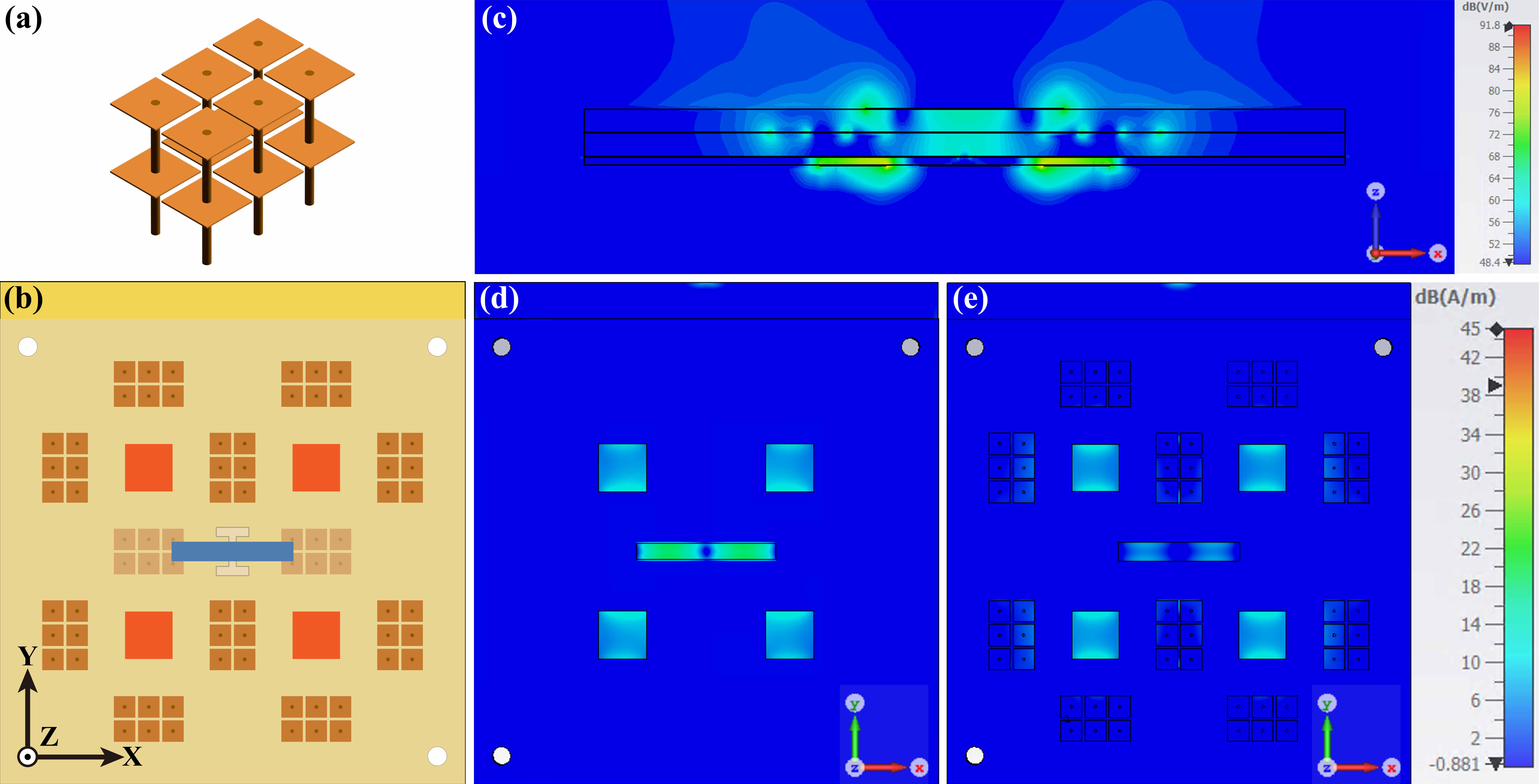}}
\par\end{centering}
\caption{Double-layer EBG structure and its suppression effect on the E-plane
difference state. (a) Double-layer EBG unit structure. (b) Cm-wave subarray
integrated with the double-layer EBG structures. (c) Central cross-sectional
electric-field distribution of the E-plane difference state with EBG. (d)
and (e) Surface-current distributions of the E-plane difference state without
and with EBG, respectively.}
\label{fields and currents with EBG}
\end{figure}

To suppress the cm-wave surface waves and reduce the higher-order modes
excited on the sub-6-GHz radiator, a double-layer EBG structure is
introduced, as shown in Fig. \ref{fields and currents with EBG}(a).
The structure is implemented within each cm-wave subarray, as shown
in Fig. \ref{fields and currents with EBG}(b), and is also included
in the complete shared-aperture array in Fig. \ref{Multi layers}(a)-(b).
The double-layer configuration is selected because each cm-wave element
uses a stacked-patch structure, while the sub-6-GHz dipole is placed
on the same layer as the cm-wave parasitic patch. Therefore, EBG mushrooms
are arranged around both the driven-patch layer and the parasitic-patch
layer of the cm-wave elements. This arrangement forms an effective high-impedance boundary within the limited shared-aperture configuration.
The EBG dimensions, spacing, and via diameters are optimized together
with the antenna geometry. Near the central sub-6-GHz dipole, only
the lower EBG layer is retained to avoid physical overlap with the
upper dipole, as shown in Fig. \ref{Multi layers}(a)-(b) and Fig. \ref{fields and currents with EBG}(b).

Fig. \ref{fields and currents with EBG}(c) shows the electric-field
distribution of the E-plane difference state after adding the double-layer
EBG. Compared with the case without EBG in Fig. \ref{coupling problem}(f),
the field concentration around the central sub-6-GHz dipole is clearly
reduced. The surface-current distributions in Fig. \ref{fields and currents with EBG}(d)
and (e) further show that the EBG suppresses current spreading along
the driven and parasitic patch layers. This indicates that the double-layer
EBG weakens the near-field interaction between the cm-wave radiators
and the centrally located sub-6-GHz dipole. Therefore, the higher-order
mode excitation and surface-wave propagation observed in the original
shared-aperture subarray are effectively mitigated.

\begin{figure}[t]
	\begin{centering}
		\textsf{\includegraphics[width=0.49\columnwidth]{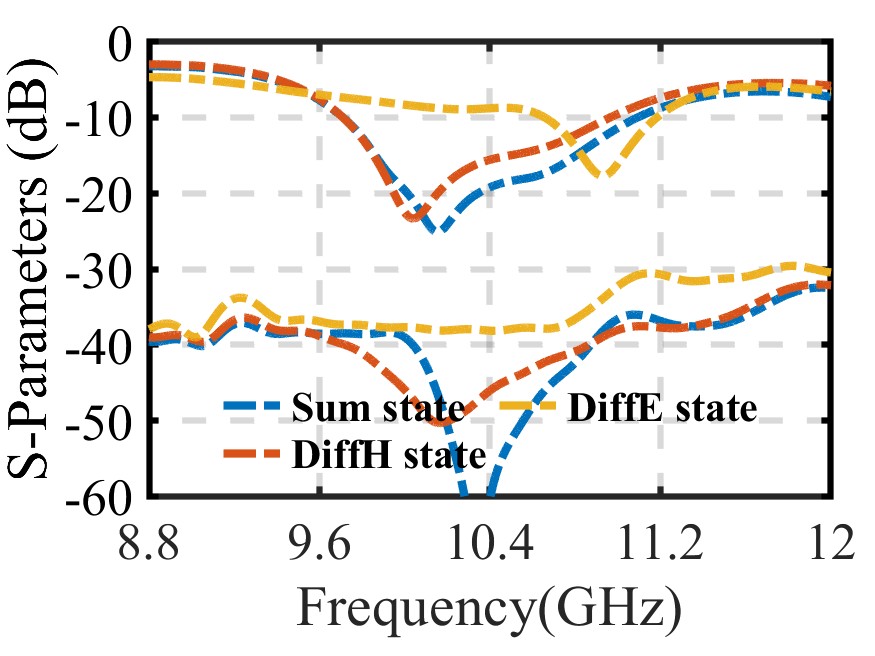}\hspace*{0.02\columnwidth}\includegraphics[width=0.49\columnwidth]{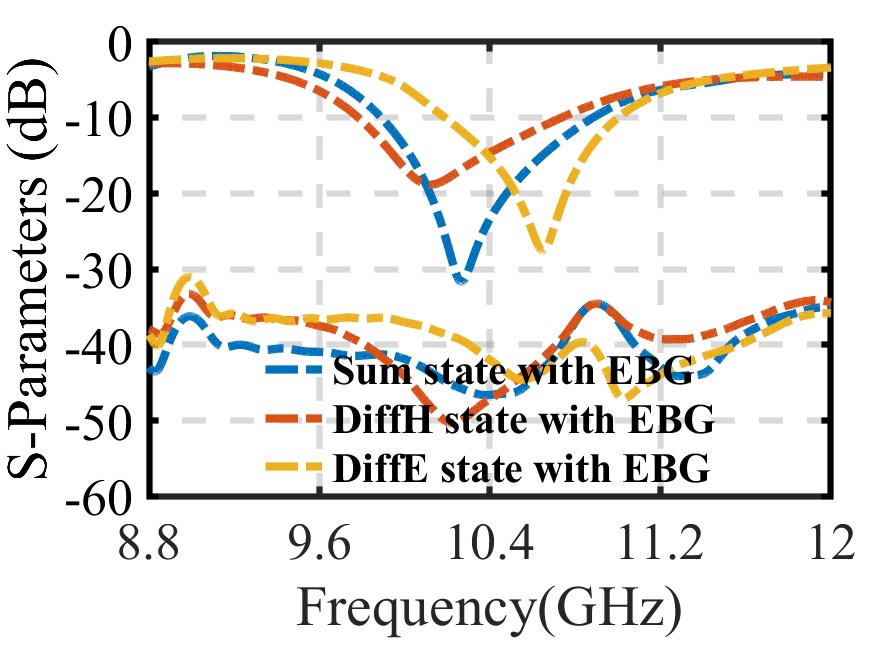}}
		\par\end{centering}
	\begin{raggedright}
		\hspace*{0.24\columnwidth} (a)\hspace*{0.45\columnwidth} (b)
		\par\end{raggedright}
	\begin{centering}
		\textsf{\includegraphics[width=0.33\columnwidth]{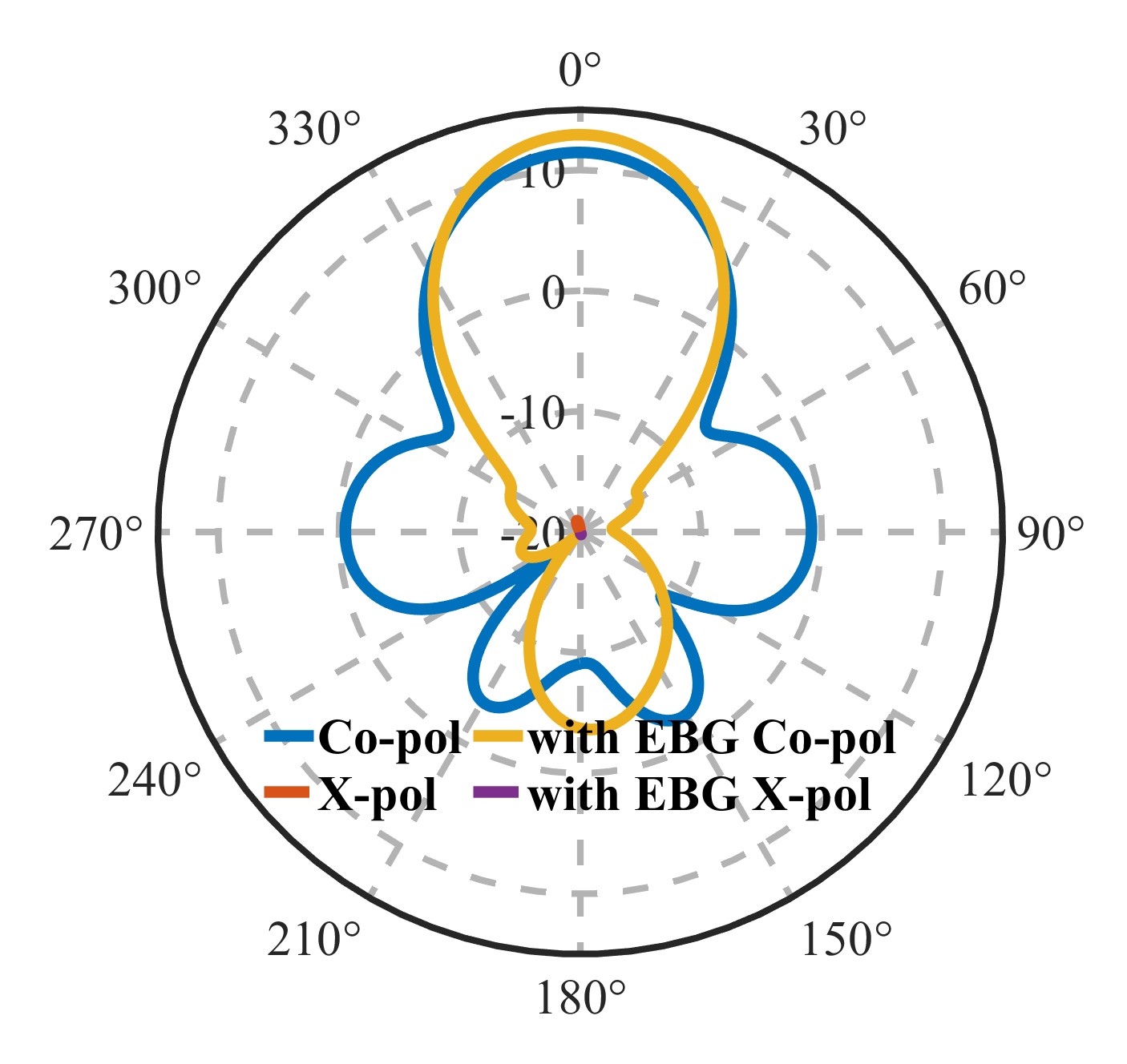}\includegraphics[width=0.33\columnwidth]{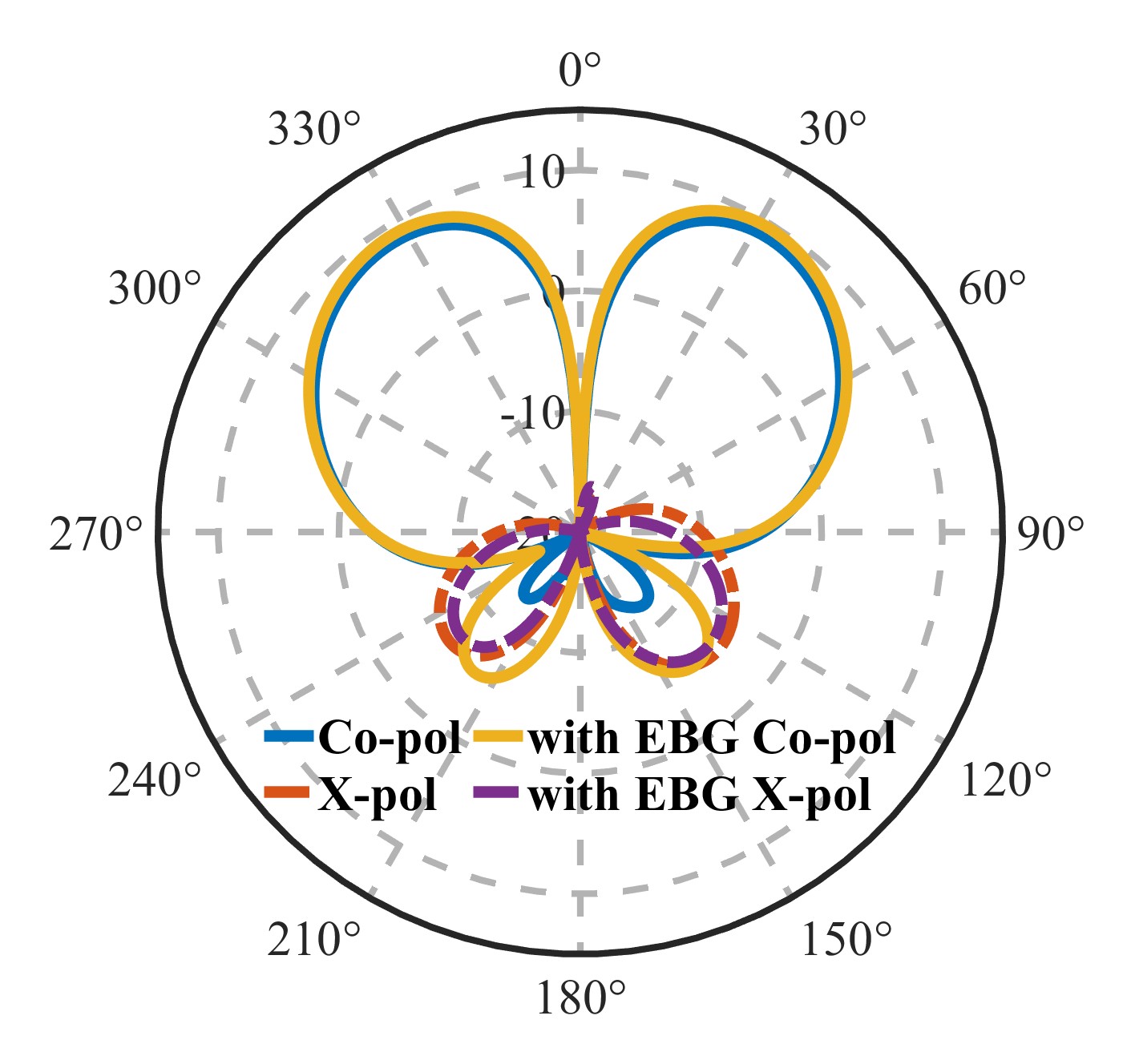}\includegraphics[width=0.33\columnwidth]{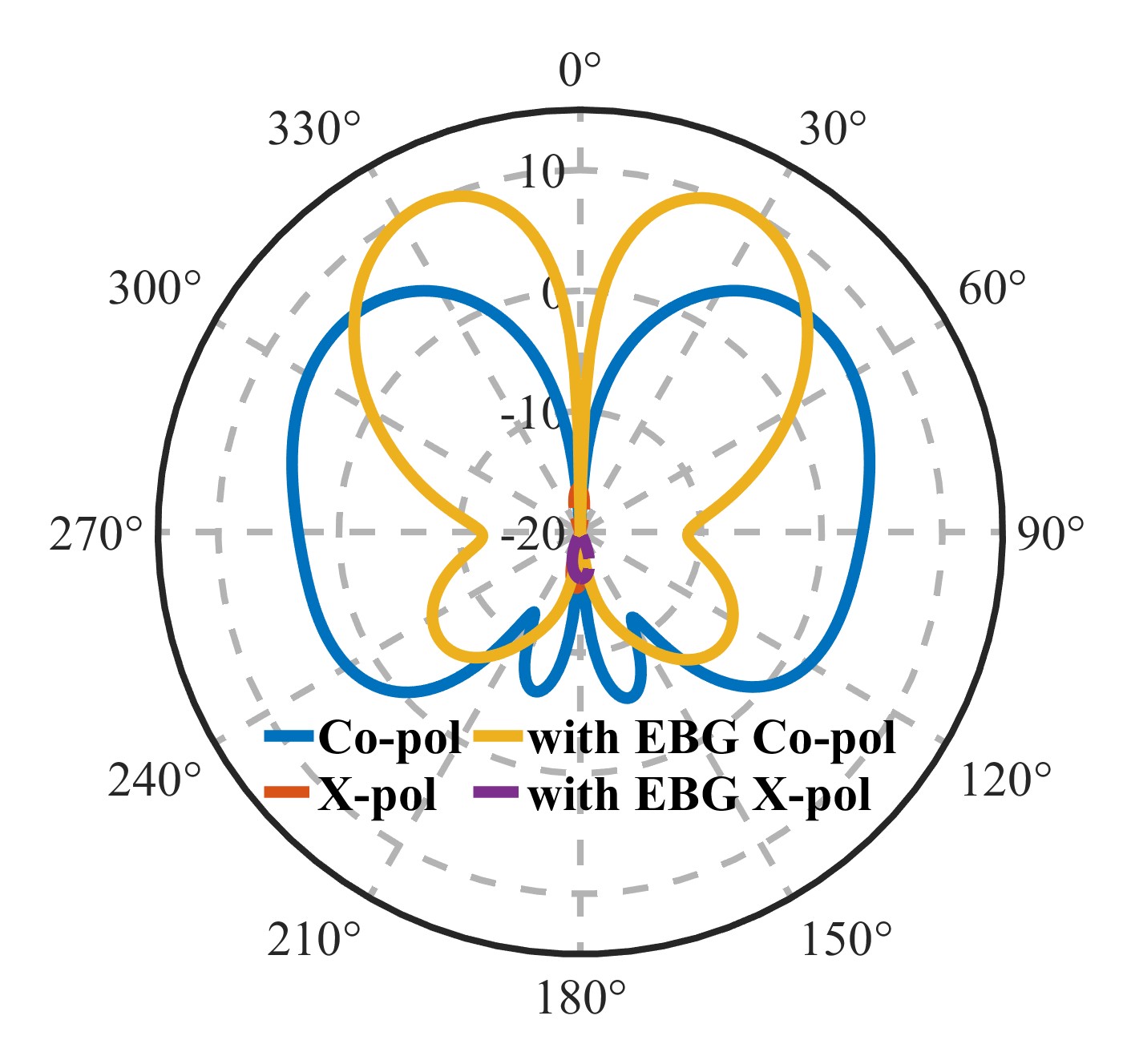}}
		\par\end{centering}
	\begin{raggedright}
		\hspace*{0.14\columnwidth} (c)\hspace*{0.27\columnwidth} (d)\hspace*{0.27\columnwidth}
		(e)
		\par\end{raggedright}
	\caption{Simulated cm-wave subarray performance with and without the proposed
		EBG structures. (a) S-parameters of the three representative states without
		EBG. (b) S-parameters of the three representative states with EBG. (c)-(e)
		Radiation-pattern comparison at 10.4 GHz for the sum state, H-plane difference
		state, and E-plane difference state, respectively (unit: dBi).}
	\label{EBG improvement}
\end{figure}

The improvement is also reflected in the simulated S-parameters and
radiation patterns in Fig. \ref{EBG improvement}. Without the EBG,
the E-plane difference state shows poor impedance matching around 10.4
GHz, as shown in Fig. \ref{EBG improvement}(a), even though the cm-wave
elements and the subarray feeding network have been optimized. After
the double-layer EBG is introduced, all three representative states
are well matched around 10.4 GHz, as shown in Fig. \ref{EBG improvement}(b).
The radiation patterns in Fig. \ref{EBG improvement}(c)-(e) further
show improved pattern shape, higher gain, and lower cross-polarization
for the cm-wave subarray. These results confirm that the double-layer
EBG effectively suppresses the cross-band coupling and surface-wave
effects that limit the cm-wave radiation performance.

\begin{figure}[t]
	\begin{centering}
		\textsf{\includegraphics[width=0.49\columnwidth]{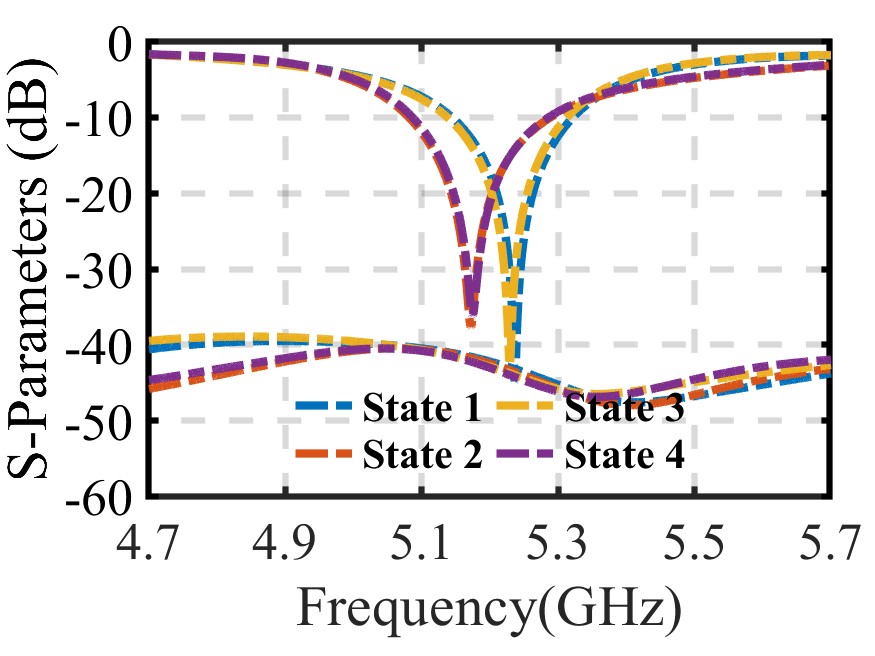}\hspace*{0.02\columnwidth}\includegraphics[width=0.49\columnwidth]{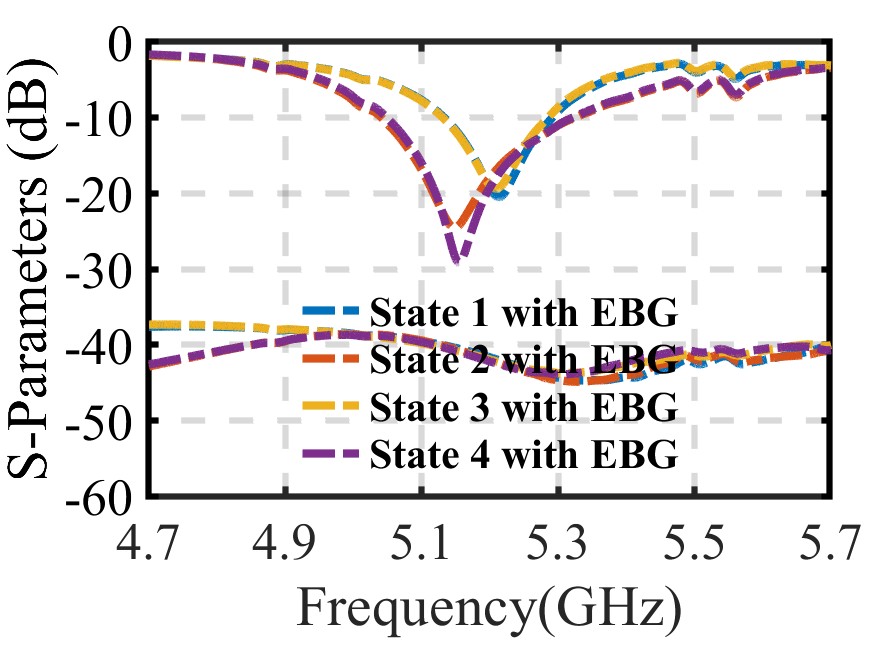}}
		\par\end{centering}
	\begin{raggedright}
		\hspace*{0.22\columnwidth} (a)\hspace*{0.45\columnwidth} (b)
		\par\end{raggedright}
	\begin{centering}
		\textsf{\includegraphics[width=0.49\columnwidth]{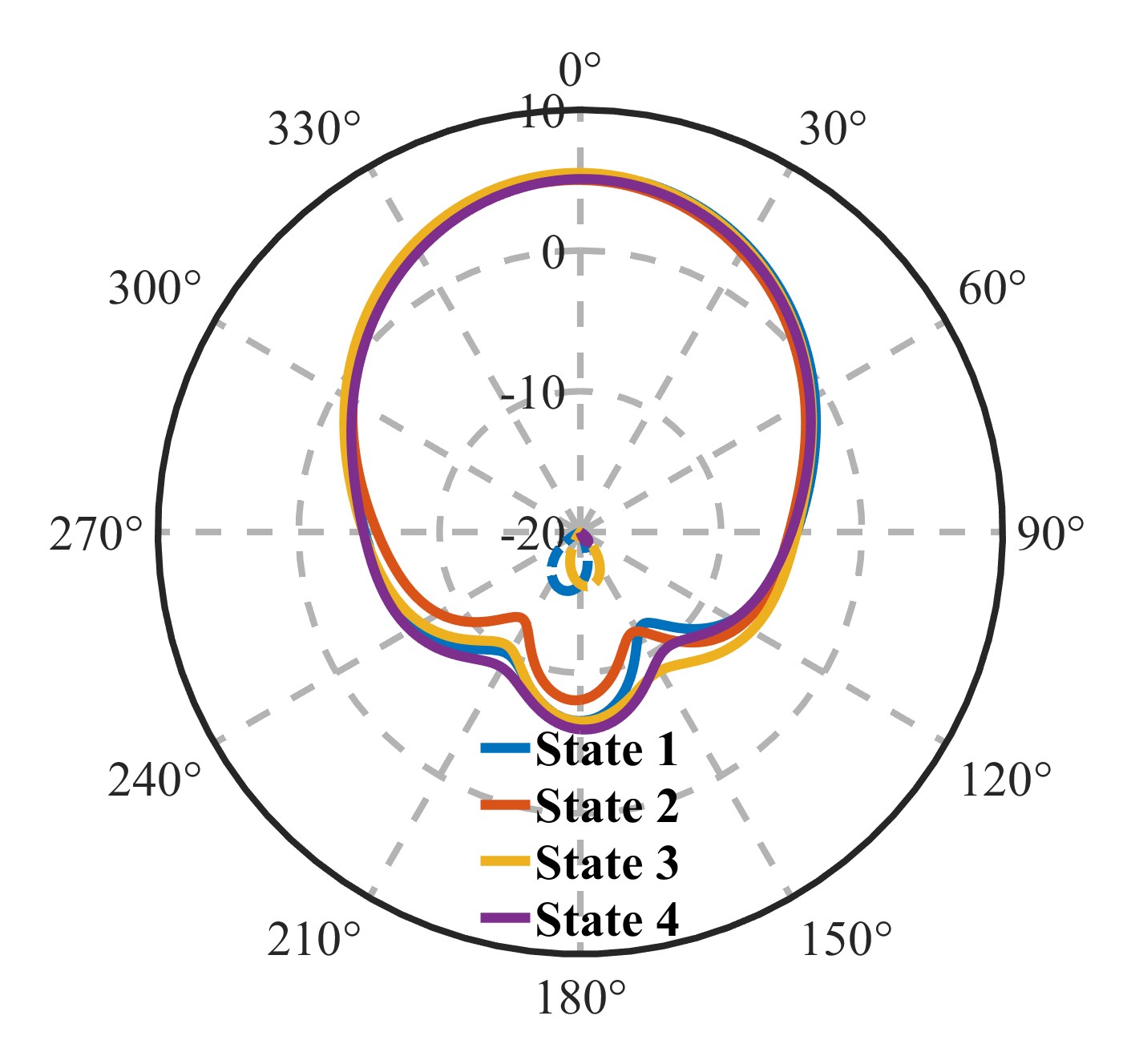}\hspace*{0.02\columnwidth}\includegraphics[width=0.49\columnwidth]{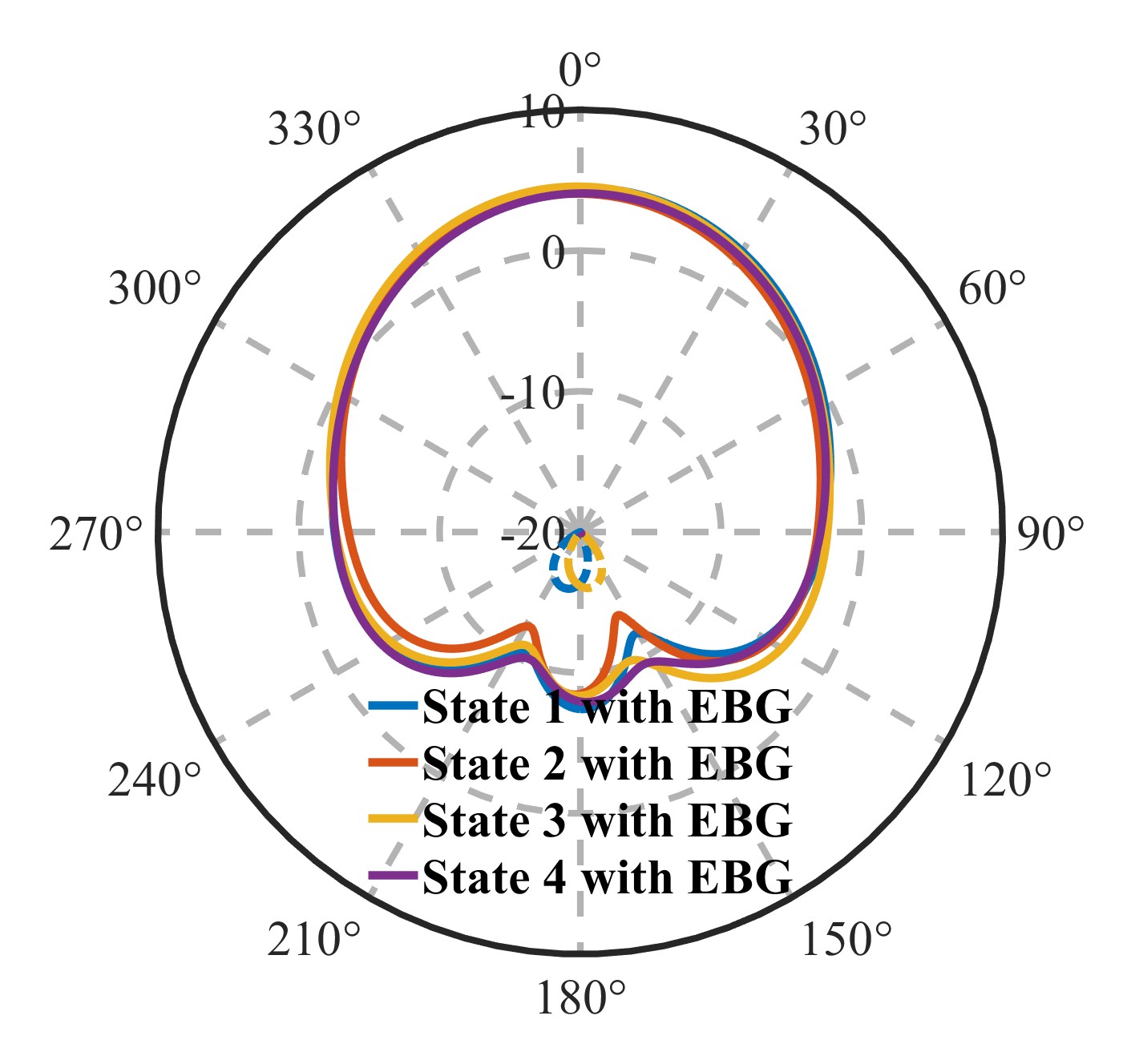}}
		\par\end{centering}
	\begin{raggedright}
		\hspace*{0.22\columnwidth} (c)\hspace*{0.45\columnwidth} (d)
		\par\end{raggedright}
	\begin{centering}
		\textsf{\includegraphics[width=0.49\columnwidth]{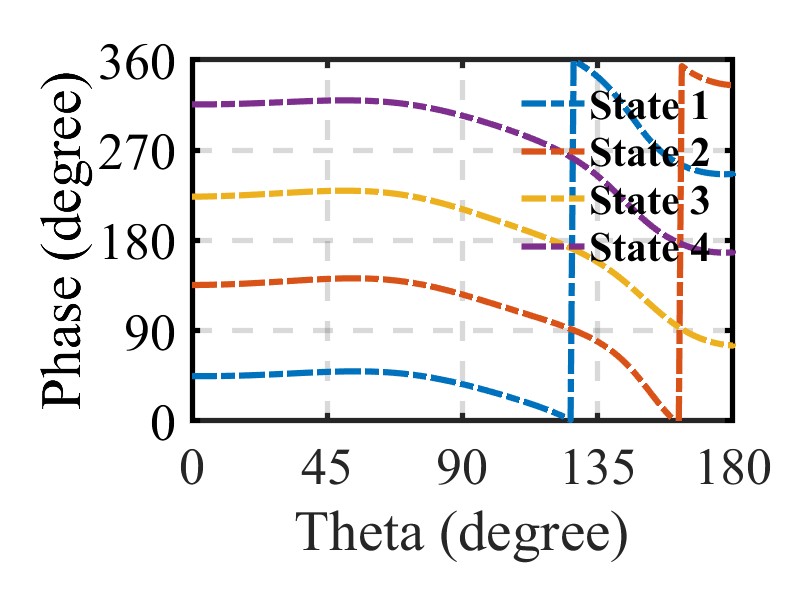}\hspace*{0.02\columnwidth}\includegraphics[width=0.49\columnwidth]{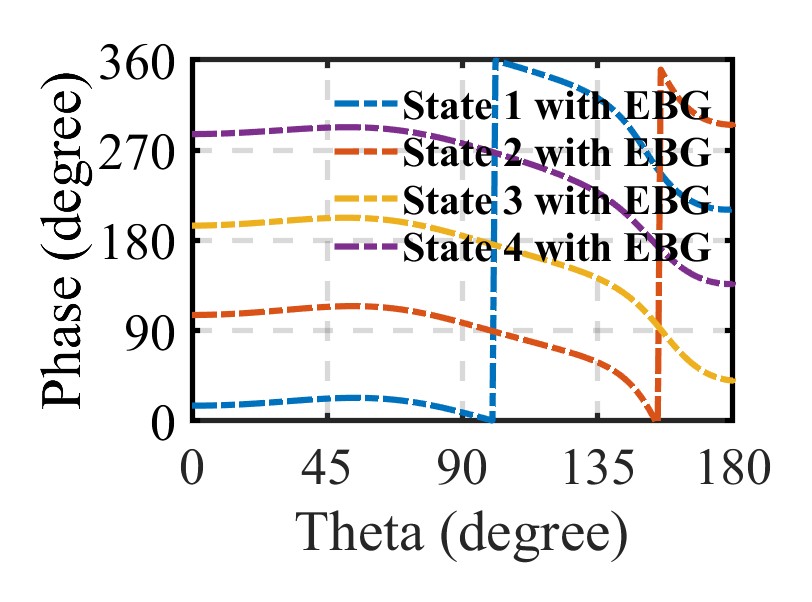}}
		\par\end{centering}
	\begin{raggedright}
		\hspace*{0.22\columnwidth} (e)\hspace*{0.45\columnwidth} (f)
		\par\end{raggedright}
	\caption{Simulated influence of the proposed EBG structures on the 2-bit
		sub-6-GHz element within the shared-aperture subarray. (a) and (b) S-parameters
		of the four reconfigurable states without and with EBG, respectively. (c)
		and (d) Radiation patterns at 5.2 GHz without and with EBG, respectively (unit: dBi). (e) and (f) Phase responses of the four states without and with EBG,
		respectively.}
	\label{EBG effects for low band}
\end{figure}

The double-layer EBG also affects the sub-6-GHz dipole because the
EBG mushrooms are placed close to the dipole and its coupling slot.
As shown by the comparison between Fig. \ref{coupling problem}(a)
and Fig. \ref{fields and currents with EBG}(b), the integrated EBG
helps reduce the footprint of the sub-6-GHz dipole and the associated coupling
slot. At the same time, the nearby EBG mushrooms act as parasitic elements
and therefore influence the impedance bandwidth and radiation response
of the sub-6-GHz element. This trade-off is evaluated in Fig. \ref{EBG effects for low band}.
Although the EBG slightly modifies the sub-6-GHz performance, the
2-bit sub-6-GHz element still maintains acceptable impedance matching,
stable realized gain, and an approximately $90^{\circ}$ phase step
among its four states.

\subsection{2-Bit Sub-6-GHz Antenna and Quasi-2-Bit Centimeter-Wave Band Subarray}

Based on the shared-aperture layout, the 1-bit switching elements,
the miniaturized $90^{\circ}$ phase shifters, and the double-layer
EBG structure described above, the proposed array forms two phase-reconfigurable
building blocks within the same aperture. For the sub-6-GHz band,
each 1-bit dipole element is cascaded with a $90^{\circ}$ phase shifter,
resulting in a 2-bit reconfigurable antenna element. For the cm-wave
band, the $90^{\circ}$ phase shifter is shared by each 2$\times$2
subarray because of the limited aperture space, leading to a quasi-2-bit
reconfigurable cm-wave subarray. The double-layer EBG further supports
this integration by suppressing the cross-band coupling and maintaining
the radiation performance of the cm-wave subarray, as discussed in
Fig. \ref{EBG improvement}. The resulting sub-6-GHz element performance
is evaluated in Fig. \ref{EBG effects for low band}.

\begin{figure}[t]
	\begin{centering}
		\textsf{\includegraphics[width=0.5\columnwidth]{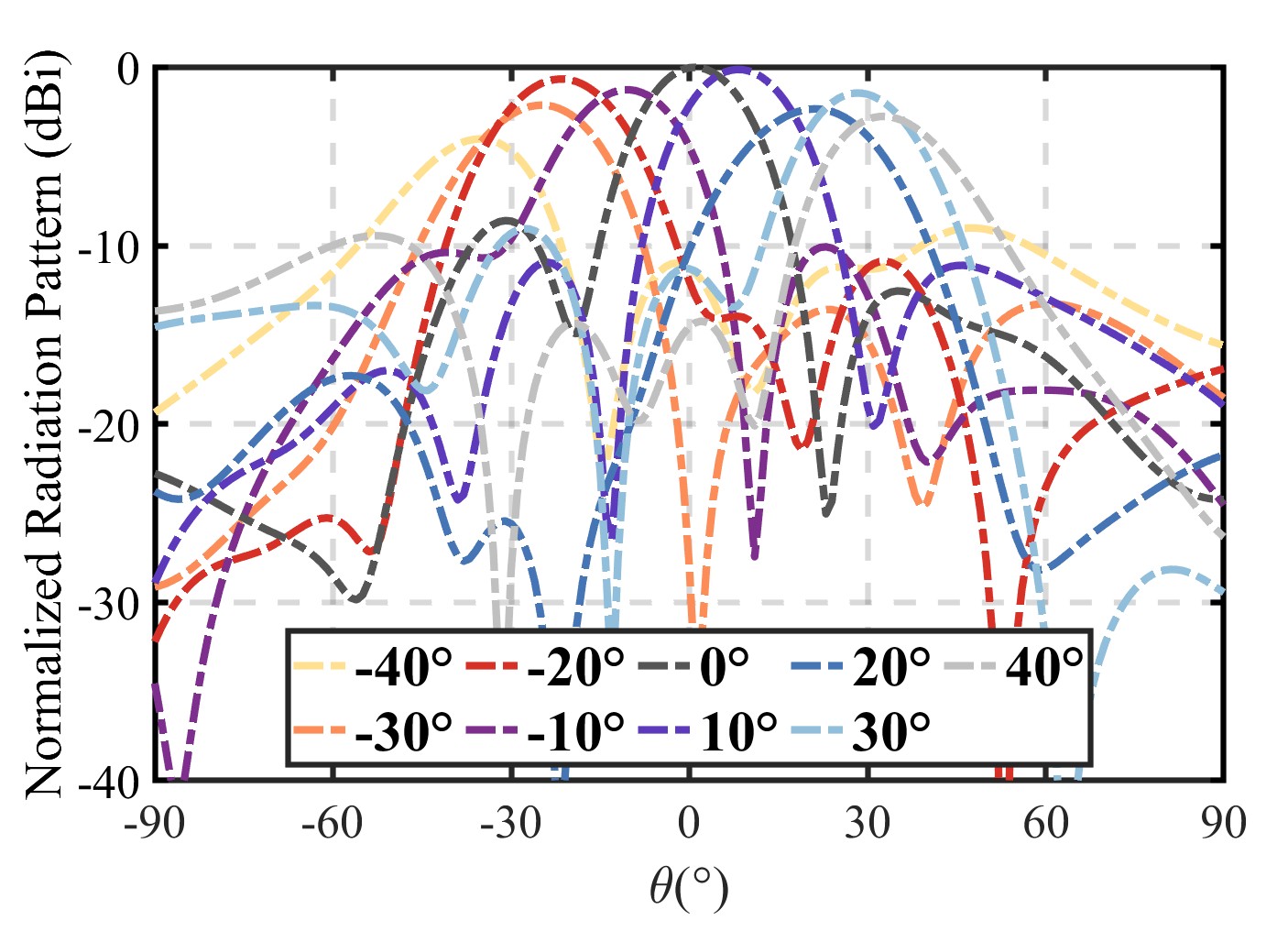}\includegraphics[width=0.5\columnwidth]{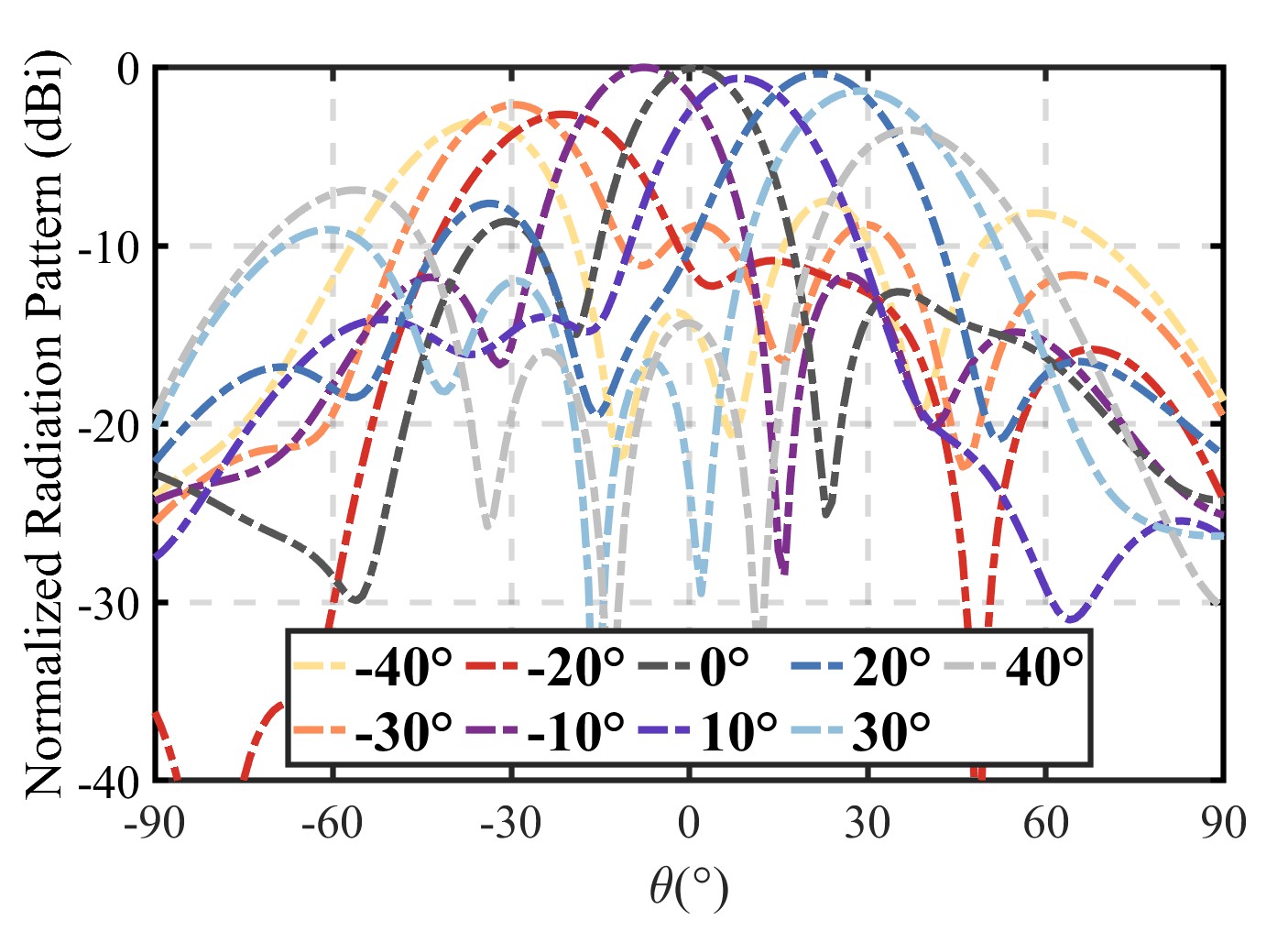}}
		\par\end{centering}
	\begin{raggedright}
		\hspace*{0.24\columnwidth} (a)\hspace*{0.44\columnwidth} (b)
		\par\end{raggedright}
	\caption{Optimized cm-wave beam scanning performance of the 4$\times$4
		quasi-2-bit array at 10.4 GHz. (a) E-plane scanning. (b) H-plane scanning.}
	\label{optimized scanning performance}
\end{figure}

\begin{figure}[t]
	\begin{centering}
		\textsf{\includegraphics[width=1\columnwidth]{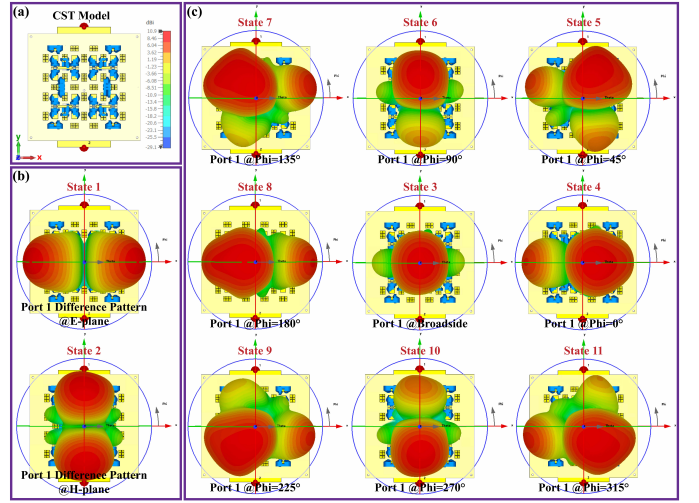}}
		\par\end{centering}
	\caption{Simulated reconfigurable radiation patterns of the 2-bit sub-6-GHz
		array. (a) Full-wave simulation model with lumped switching elements.
		(b) Two difference-beam states. (c) Nine representative directional-beam
		states.}
	\label{sim sub-6 beam steering}
\end{figure}

The resulting 2-bit sub-6-GHz array and quasi-2-bit cm-wave array provide
the reconfigurable phase states used in the following beamforming design.

\section{Beamforming Design and Optimization}

Although the proposed structure improves the phase resolution in both
bands, phase quantization error remains an important limitation for
reconfigurable antenna arrays. It can lead to gain degradation, increased
sidelobes, and limited scanning range \cite{smith1983comparison,mailloux1984array}.
Several methods have been reported to reduce quantization effects,
including additional fixed phase-delay lines for each element \cite{smith1983comparison,li20221,li2023two,mmwave-wang2022low}
and the rotation of elements or subarrays \cite{yin2023modular}. In
this work, the shared-aperture layout and the reconfigurable feeding
networks are already arranged within a compact physical aperture. Therefore,
this section introduces a beamforming model and an optimization method
for selecting the reconfigurable phase states and the fixed phase delays
of the cm-wave array. For
the cm-wave array, the phase of the $n$th 1-bit antenna element is
denoted as $\alpha_{n}\in\{0,\pi\}$. The phase introduced by the
$90^{\circ}$ phase shifter shared by each 2$\times$2 subarray is denoted
as $\beta_{n}\in\{0,\pi/2\}$, where $\beta_{n}$ is constrained to be
identical for the four elements within the same subarray. The additional
fixed phase delay introduced for the $n$th cm-wave element is denoted
as $\gamma_{n}$, where $n=1,\ldots,16$.

A conventional array-factor-based model can be written as
\begin{equation}
\begin{array}{c}
\mathbf{E}_{H}\left(\Omega\right)=\sum_{n=1}^{N=16}\mathbf{F}_{n}\left(\Omega\right)
\exp\left[jk_{0}\left(x_{n}u+y_{n}v\right)\right. \\
\left.+j\left(\alpha_{n}+\beta_{n}+\gamma_{n}\right)\right]
\end{array},
\label{array factor}
\end{equation}
where $\Omega=(\theta,\varphi)$, $\theta\in[0,180^{\circ}]$,
$\varphi\in[0,360^{\circ}]$, $k_{0}$ is the wavenumber, $x_{n}$ and
$y_{n}$ denote the phase center of the $n$th element,
$u=\sin\theta\cos\varphi$ and $v=\sin\theta\sin\varphi$, and $\mathbf{F}_{n}\left(\Omega\right)$ is the element pattern approximated by the same cosine-q pattern for each element.

However, this model relies on an approximated element pattern and does
not fully capture the element displacement, mutual coupling, and shared-aperture
perturbation effects. Therefore, the cm-wave radiation pattern is calculated
using the embedded electric field of each element:
\begin{equation}
\mathbf{E}_{H}\left(\Omega\right)=\sum_{n=1}^{N=16}\mathbf{E}_{n}\left(\Omega\right)
\exp\left[j\left(\alpha_{n}+\beta_{n}+\gamma_{n}\right)\right],
\label{accurate pattern summation}
\end{equation}
where $\mathbf{E}_{n}(\Omega)$ is the electric field radiated by the
$n$th cm-wave antenna port when it is excited by a unit current source
and all the other ports are matched. The embedded field $\mathbf{E}_{n}(\Omega)$
is obtained from full-wave simulation and already includes the phase-center
information and coupling effects of each element. This model provides
a more accurate basis for optimizing the beamforming states of the proposed
shared-aperture array.

Based on \eqref{accurate pattern summation}, a standard genetic algorithm (GA) is used to optimize the reconfigurable phase states and the fixed phase
delays. For each target beam direction $\Omega_{m}=(\theta_{m},\varphi_{m})$,
the reconfigurable phase states $\alpha_{n}^{\Omega_{m}}$ and
$\beta_{n}^{\Omega_{m}}$ are selected. The target scanning angles are
defined by $\theta_{m}\in\{\pm40^{\circ},\pm30^{\circ},\pm20^{\circ},\pm10^{\circ},0^{\circ}\}$.
The cases with $\varphi_{m}=0^{\circ}$ and $\varphi_{m}=90^{\circ}$
correspond to E-plane and H-plane scanning, respectively. The fixed
phase delay $\gamma_{n}$ is discretized by $Q$ binary bits with a phase
step $P$:
\begin{equation}
\gamma_{n}\in\{0,P,2P,\ldots,(2^{Q}-1)P\}.
\end{equation}
In this design, $Q=5$ and $P=5^{\circ}$, giving
$\gamma_{n}\in\{0^{\circ},5^{\circ},10^{\circ},\ldots,155^{\circ}\}$.
The binary variables corresponding to $\alpha_{n}^{\Omega_{m}}$ and
$\beta_{n}^{\Omega_{m}}$ are denoted as
$a_{n}^{\Omega_{m}},b_{n}^{\Omega_{m}}\in\{0,1\}$. The variable
$b_{n}^{\Omega_{m}}$ is constrained to be identical within each
2$\times$2 subarray because the $90^{\circ}$ phase shifter is shared
by the subarray. The state vector $\mathbf{x}$ collects all the optimized
variables, including $\alpha_{n}^{\Omega_{m}}$, $\beta_{n}^{\Omega_{m}}$,
and $\gamma_{n}$.

\begin{figure}[t]
	\begin{centering}
		\textsf{\includegraphics[width=1\columnwidth]{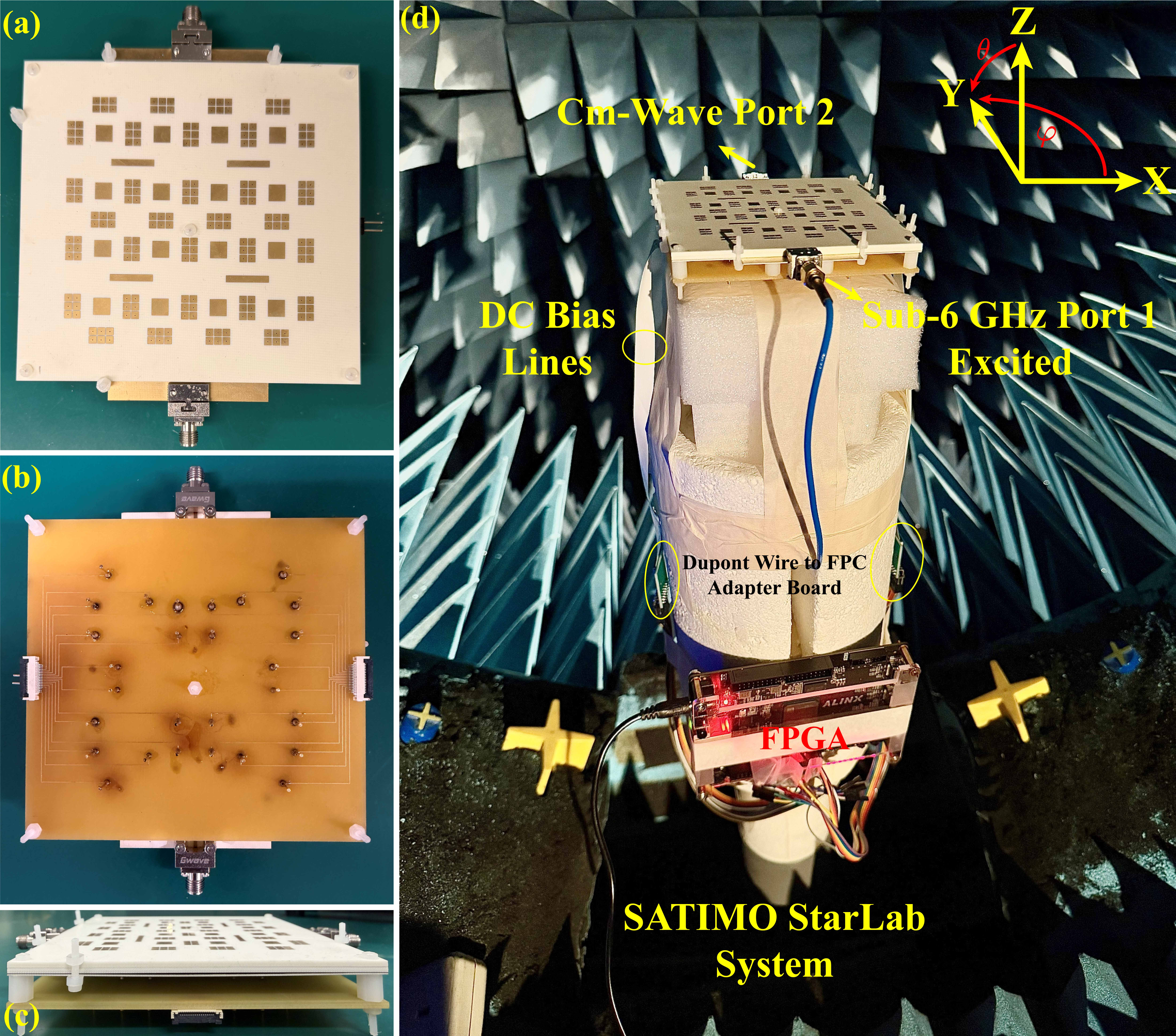}}
		\par\end{centering}
	\caption{Fabricated prototype of the proposed dual-band shared-aperture antenna
		array and measurement setup. (a) Top view. (b) Bottom view. (c) Side
		view. (d) Anechoic-chamber measurement setup.}
	\label{Fabricated antenna array and Mea setup}
\end{figure}

\begin{figure}[t]
	\begin{centering}
		\textsf{\includegraphics[width=1\columnwidth]{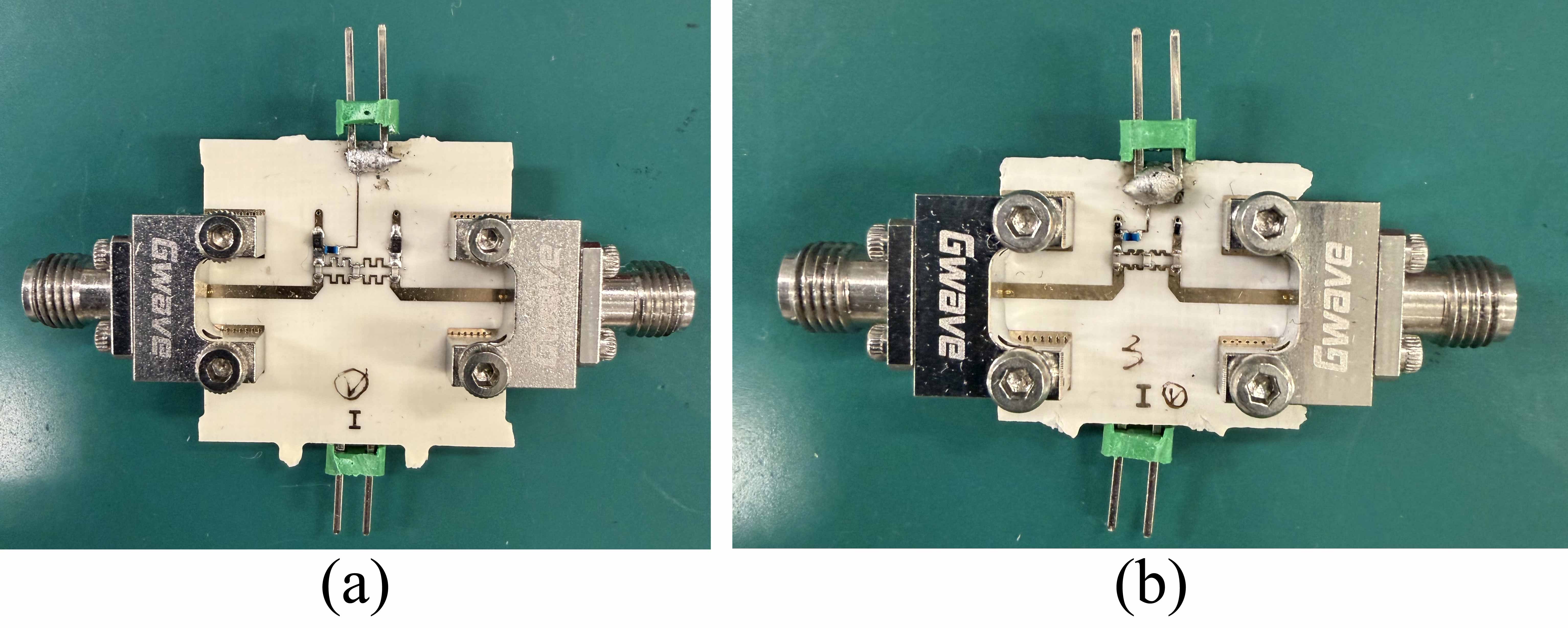}}
		\par\end{centering}
	\caption{Fabricated miniaturized reconfigurable $90^{\circ}$ phase shifters.
		(a) Sub-6-GHz phase shifter. (b) Cm-wave phase shifter.}
	\label{Fabricated PS}
\end{figure}

The optimization problem is formulated as
\begin{equation}
\begin{array}{cc}
\underset{\mathbf{x}}{\min} &
\sum_{m=1}^{M}
\left(
\max\left(\mathbf{E}_{H}^{m}(\Omega_{\mathrm{sidelobe}}^{m},\mathbf{x})\right)
-\left|E_{H}^{m}(\Omega_{m},\mathbf{x})\right|
\right)\\
\mathrm{s.t.} &
\mathbf{x}\in\{0,1\}^{(N+4)M+QN}
\end{array},
\label{beamforming optimization}
\end{equation}
where $\Omega_{\mathrm{sidelobe}}^{m}$ is the sidelobe region defined as
\begin{equation}
\left(u-u_{m}\right)^{2}+\left(v-v_{m}\right)^{2}>r_{\mathrm{mainlobe}}^{2}=0.4^{2}.
\label{sidelobe region}
\end{equation}
This objective function maximizes the field level at the target beam
direction while suppressing the maximum sidelobe level outside the
main-lobe region.

After optimization, the fixed phase delays $\gamma_{n}$ are implemented
by the phase-delay lines shown in Fig. \ref{RF feeding layer}. The
optimized cm-wave scanning performance is shown in Fig. \ref{optimized scanning performance}.
Nine beam states are obtained for E-plane scanning, and another nine
beam states are obtained for H-plane scanning. With the optimized fixed
phase delays and reconfigurable phase states, the 4$\times$4 quasi-2-bit
cm-wave array achieves beam scanning up to $\pm40^{\circ}$ with a
sidelobe level below -7 dB.

\begin{figure}[t]
	\begin{centering}
		\textsf{\includegraphics[width=0.49\columnwidth]{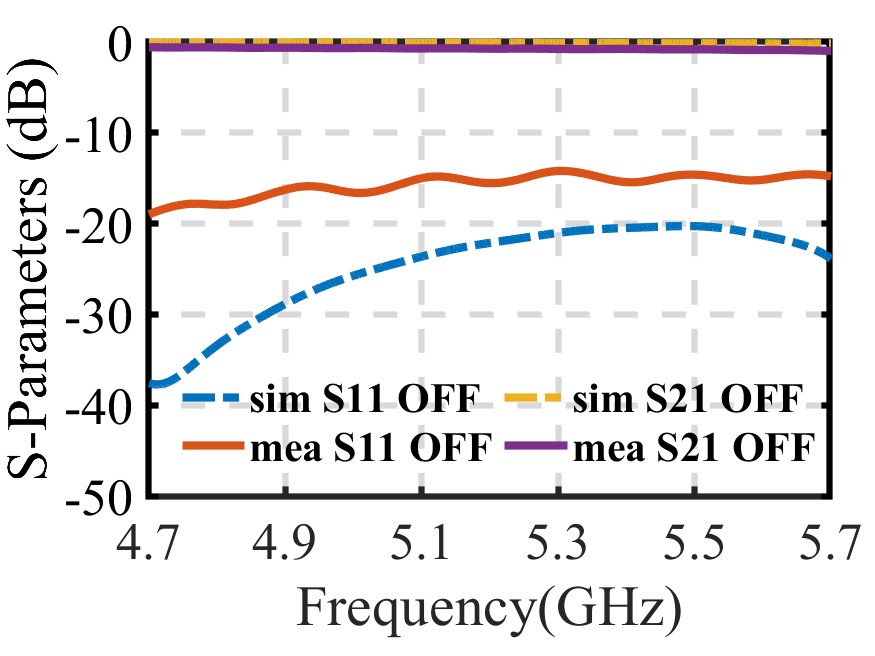}\hspace*{0.02\columnwidth}\includegraphics[width=0.49\columnwidth]{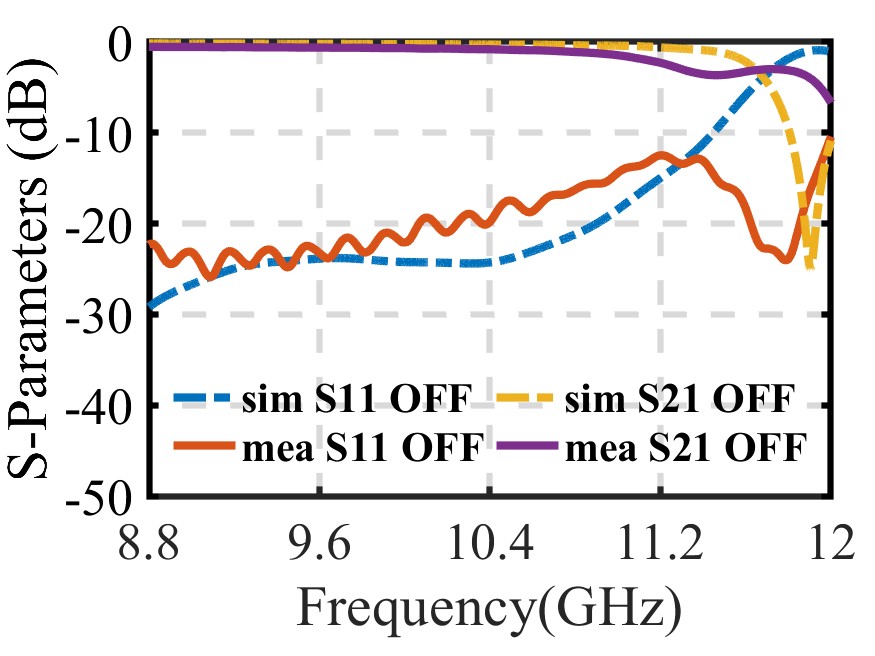}}
		\par\end{centering}
	\begin{raggedright}
		\hspace*{0.24\columnwidth} (a)\hspace*{0.45\columnwidth} (d)
		\par\end{raggedright}
	\begin{centering}
		\textsf{\includegraphics[width=0.49\columnwidth]{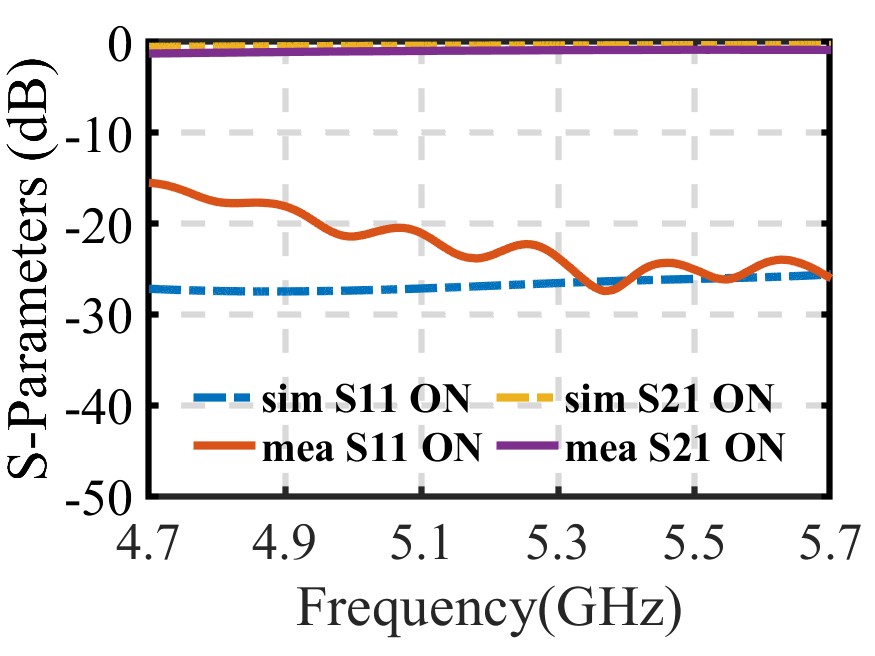}\hspace*{0.02\columnwidth}\includegraphics[width=0.49\columnwidth]{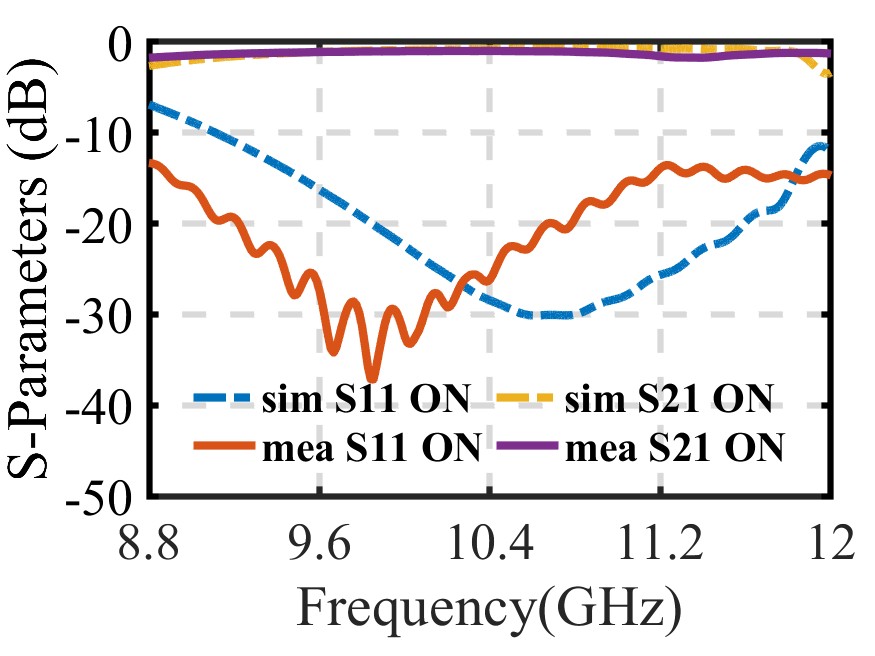}}
		\par\end{centering}
	\begin{raggedright}
		\hspace*{0.24\columnwidth} (b)\hspace*{0.45\columnwidth} (e)
		\par\end{raggedright}
	\begin{centering}
		\textsf{\includegraphics[width=0.49\columnwidth]{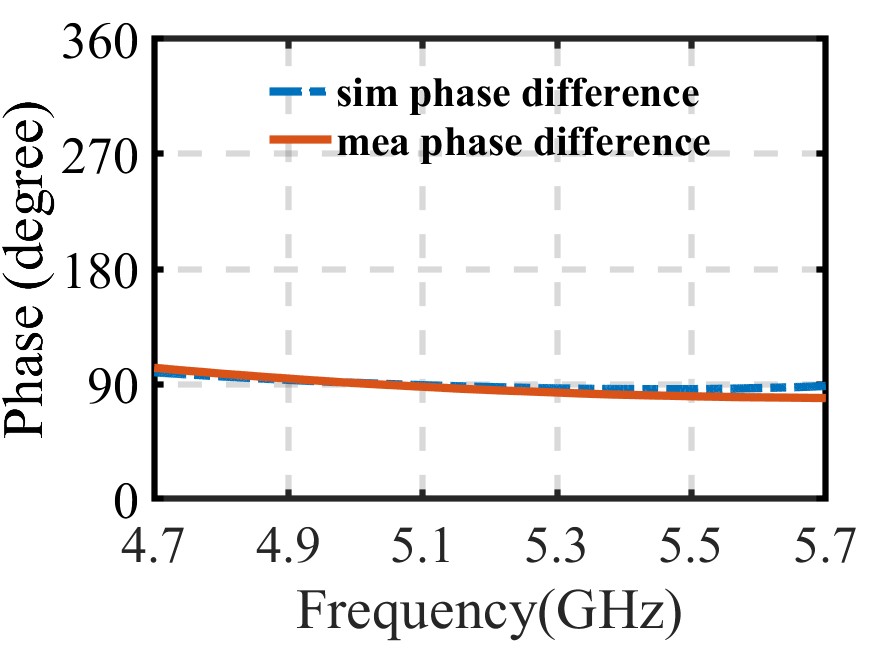}\hspace*{0.02\columnwidth}\includegraphics[width=0.49\columnwidth]{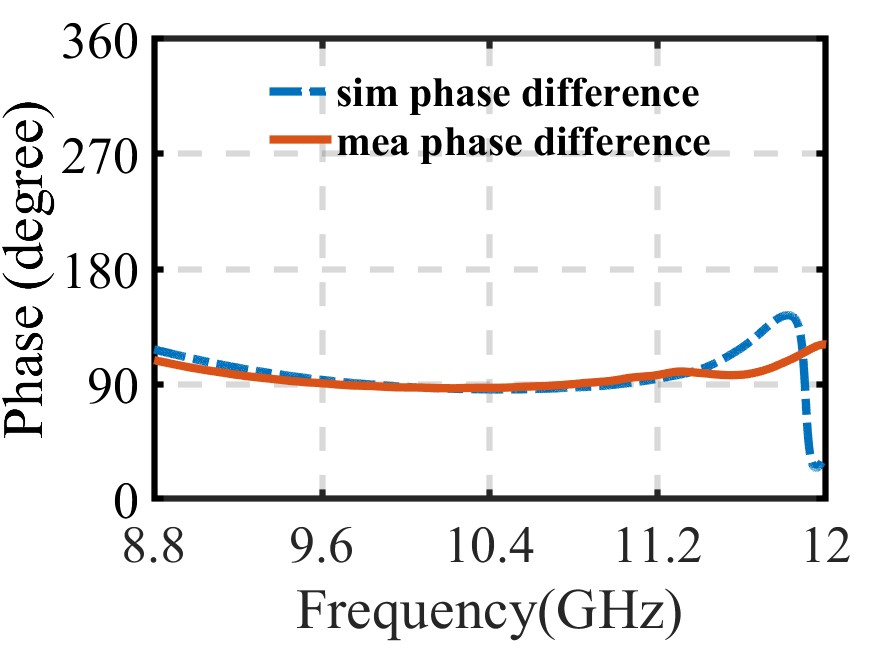}}
		\par\end{centering}
	\begin{raggedright}
		\hspace*{0.24\columnwidth} (c)\hspace*{0.45\columnwidth} (f)
		\par\end{raggedright}
	\caption{Simulated and measured performance of the fabricated reconfigurable
		$90^{\circ}$ phase shifters. (a) and (b) S-parameters of the sub-6-GHz
		phase shifter in the OFF and ON states, respectively. (c) Simulated
		and measured phase difference between the two states of the sub-6-GHz
		phase shifter. (d) and (e) S-parameters of the cm-wave phase shifter
		in the OFF and ON states, respectively. (f) Simulated and measured phase
		difference between the two states of the cm-wave phase shifter.}
	\label{Mea Performance of PS}
\end{figure}

For the sub-6-GHz array, the beamforming capability is mainly limited
by the small 2$\times$2 aperture. Therefore, instead of applying the
same fixed-delay optimization, representative reconfigurable patterns
are evaluated directly. As shown in Fig. \ref{sim sub-6 beam steering},
the 2-bit sub-6-GHz array provides two difference patterns and nine
directional beam states, giving 11 representative reconfigurable radiation
patterns in total.

\section{Measurement Results}

A prototype of the proposed dual-band reconfigurable shared-aperture
antenna array was fabricated and measured to validate the phase shifters,
the sub-6-GHz operation, and the cm-wave beam scanning performance.
The fabricated array and the measurement setup are shown in Fig. \ref{Fabricated antenna array and Mea setup}.
All the RF switches were controlled by an Artix-7 FPGA, and the radiation
patterns were measured in an anechoic chamber.

\subsection{\texorpdfstring{Miniaturized Reconfigurable $90^{\circ}$ Phase Shifters}{Miniaturized Reconfigurable 90-degree Phase Shifters}}

The fabricated phase shifters for the sub-6-GHz and cm-wave bands are
shown in Fig. \ref{Fabricated PS}(a) and (b), respectively. They were
measured using a calibrated Rohde \& Schwarz ZVA40 four-port vector
network analyzer. For the sub-6-GHz phase shifter, the measured results
in Fig. \ref{Mea Performance of PS}(a)-(c) agree well with the simulated
$90^{\circ}$ phase difference, with an insertion loss about 0.6 dB
higher than in simulation. For the cm-wave phase shifter, the measured
results in Fig. \ref{Mea Performance of PS}(d)-(f) also show a stable
$90^{\circ}$ phase difference, with an insertion loss about 0.5 dB
higher than in simulation. The additional insertion loss is
mainly attributed to the accumulated parasitic effects of the lumped
capacitors and inductors, cable loss, and fabrication tolerances.

\subsection{Sub-6-GHz Performance}

\begin{figure}
	\begin{centering}
		\textsf{\includegraphics[width=0.5\columnwidth]{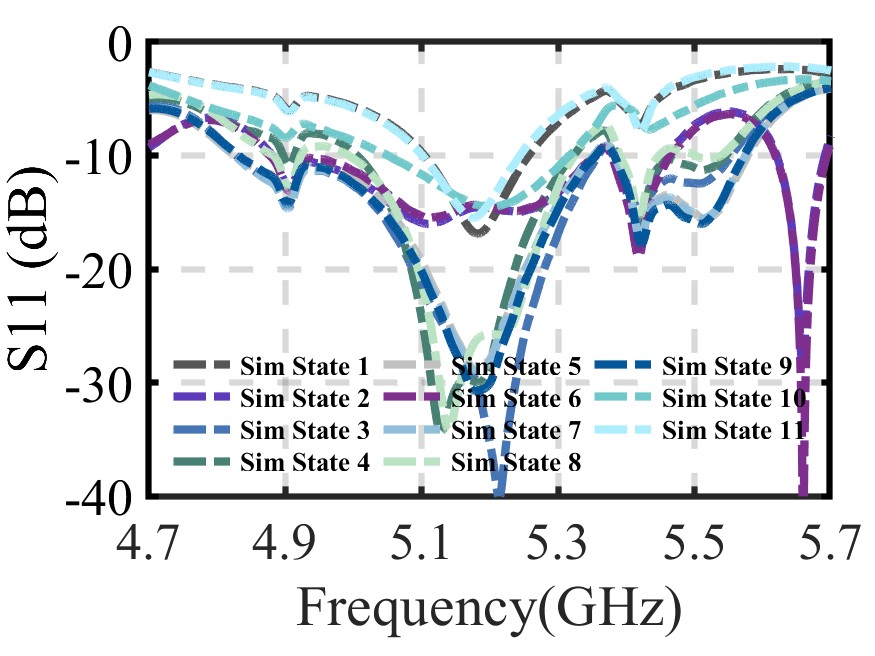}\includegraphics[width=0.5\columnwidth]{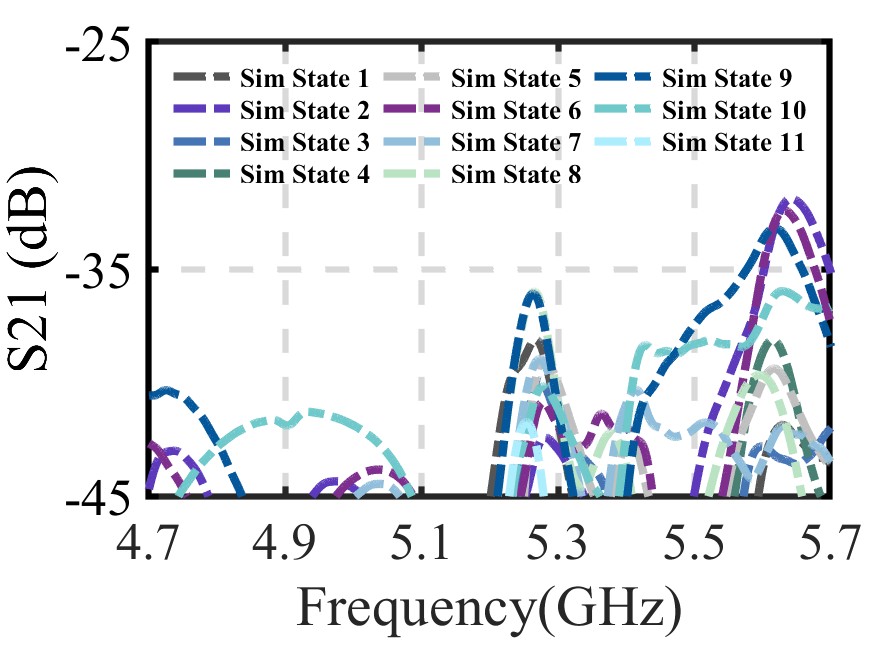}}
		\par\end{centering}
	\begin{raggedright}
		\hspace*{0.25\columnwidth} (a)\hspace*{0.45\columnwidth} (c)
		\par\end{raggedright}
	\begin{centering}
		\textsf{\includegraphics[width=0.5\columnwidth]{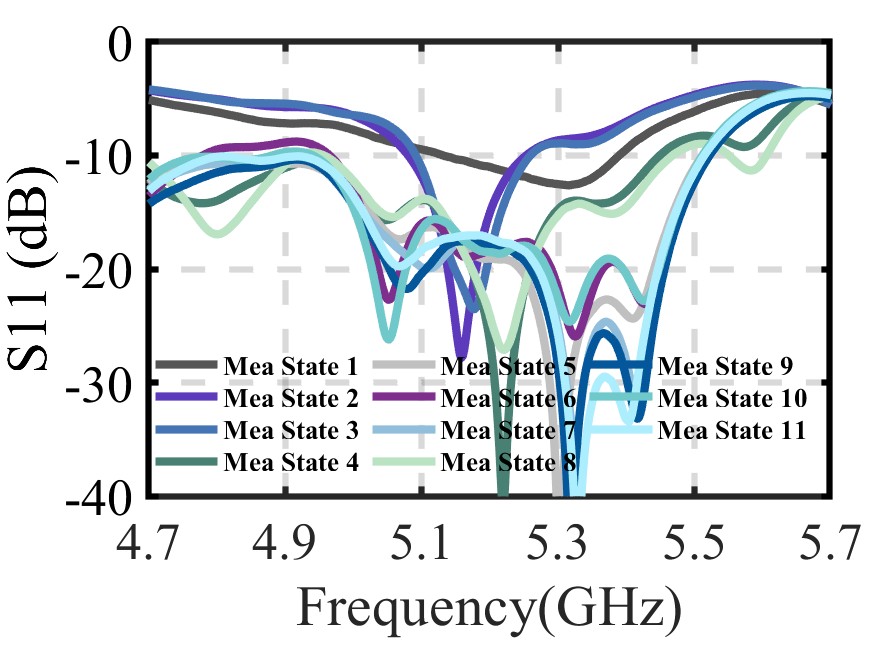}\includegraphics[width=0.5\columnwidth]{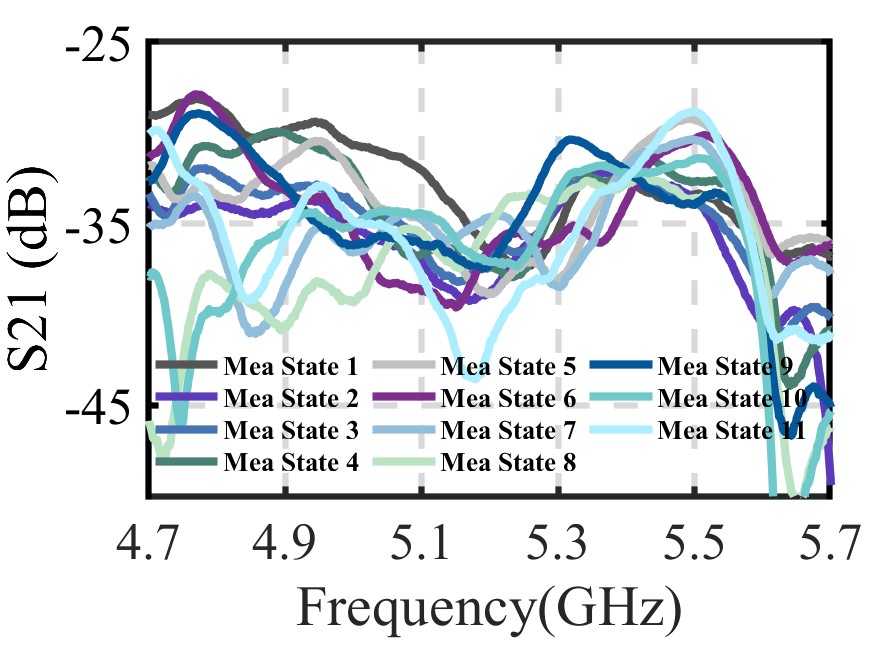}}
		\par\end{centering}
	\begin{raggedright}
		\hspace*{0.25\columnwidth} (b)\hspace*{0.45\columnwidth} (d)
		\par\end{raggedright}
	\caption{Simulated and measured S-parameters of the sub-6-GHz array for 11
		representative reconfigurable states. (a) Simulated $S_{11}$. (b) Measured
		$S_{11}$. (c) Simulated $S_{21}$. (d) Measured $S_{21}$.}
	\label{Sub-6 Mea S parameters}
\end{figure}

The sub-6-GHz performance was measured by feeding port 1 and sequentially
switching the FPGA control states. As shown in Fig. \ref{Sub-6 Mea S parameters}(a)
and (b), the simulated and measured $S_{11}$ responses of the 11 representative
states are well matched around 5.2 GHz. The simulated and measured
$S_{21}$ responses in Fig. \ref{Sub-6 Mea S parameters}(c) and (d)
show isolation higher than 30 dB between the two ports.

\begin{figure}[t]
	\begin{centering}
		\textsf{\includegraphics[width=1\columnwidth]{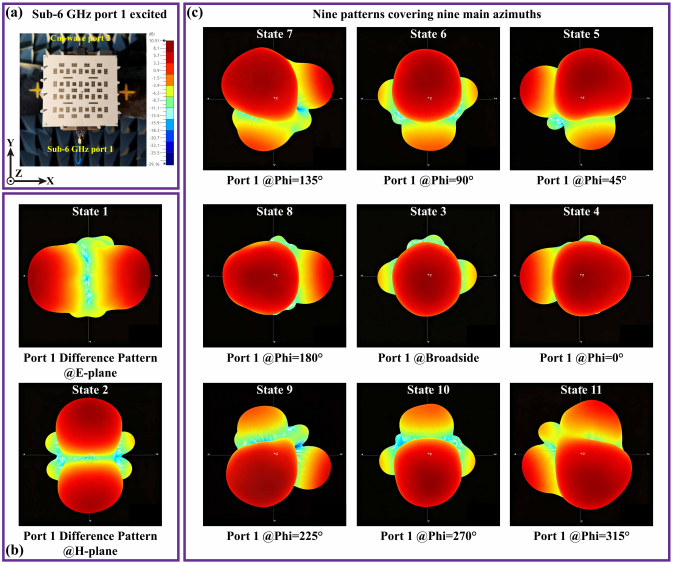}}
		\par\end{centering}
	\caption{Measured reconfigurable radiation patterns of the 2-bit sub-6-GHz
		array. (a) Sub-6-GHz measurement setup. (b) Two difference-beam states.
		(c) Nine representative directional-beam states.}
	\label{Mea sub-6 beam steering}
\end{figure}

\begin{figure}[t]
	\begin{centering}
		\textsf{\includegraphics[width=1\columnwidth]{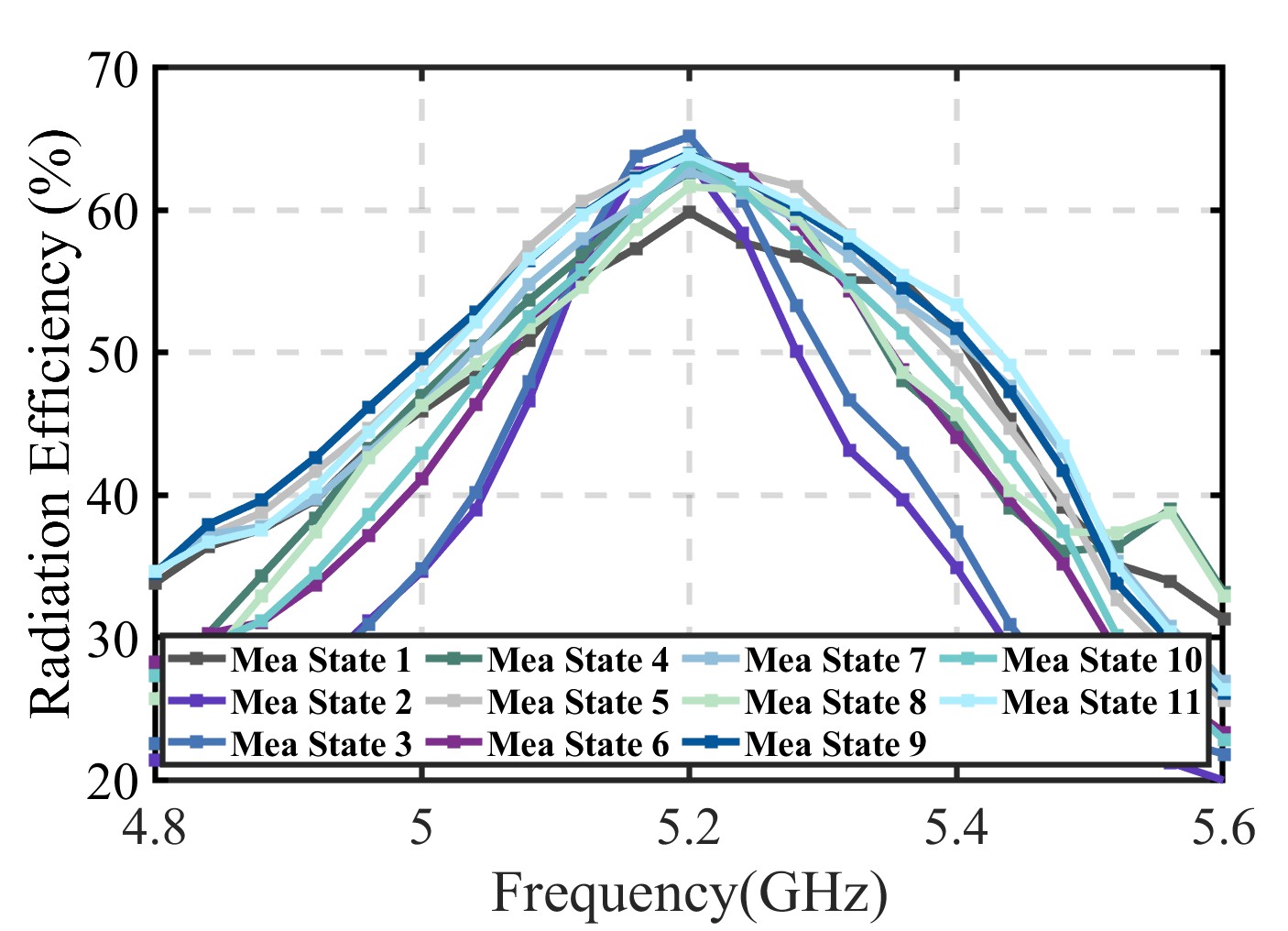}}
		\par\end{centering}
	\caption{Measured radiation efficiency of the 2-bit sub-6-GHz array for 11
		representative reconfigurable states.}
	\label{Mea sub-6 Efficiency}
\end{figure}

The measured radiation patterns of the 11 representative states are
shown in Fig. \ref{Mea sub-6 beam steering}. The realized gain ranges
from 7 to 10.5 dBi, which is consistent with the simulated reconfigurable
patterns in Fig. \ref{sim sub-6 beam steering}. The measured radiation
efficiencies in Fig. \ref{Mea sub-6 Efficiency} exceed 60\% for most
states. Compared with the simulated results, an efficiency degradation
of about 5\% to 10\% is observed, mainly due to the parasitic effects
of the embedded capacitors, inductors, and RF switches, as well as
cable loss. A larger sub-6-GHz array can be used in future designs
to further improve the beam steering capability.

\subsection{Centimeter-Wave Band Performance}

\begin{figure}
	\begin{centering}
		\textsf{\includegraphics[width=0.5\columnwidth]{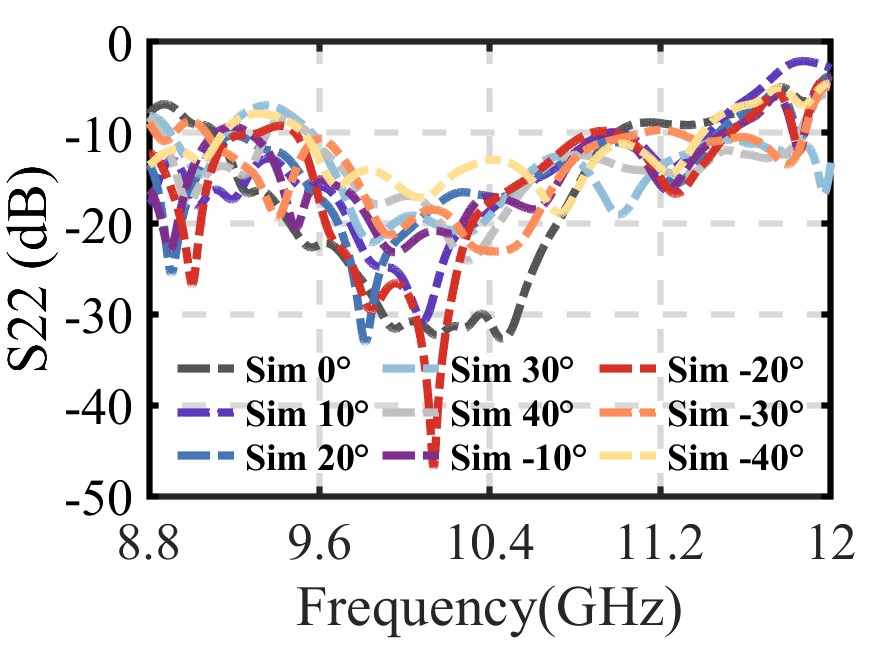}\includegraphics[width=0.5\columnwidth]{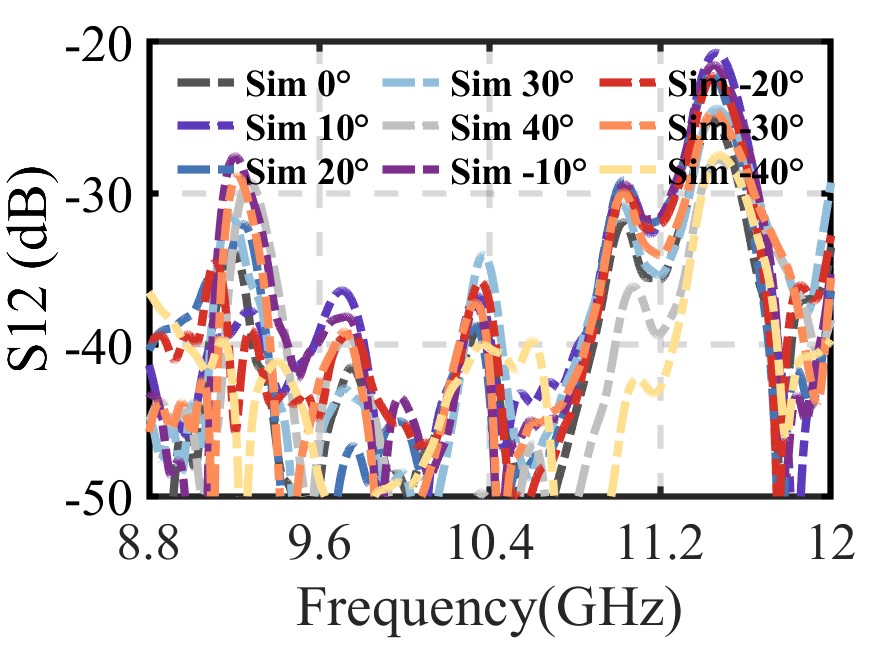}}
		\par\end{centering}
	\begin{raggedright}
		\hspace*{0.25\columnwidth} (a)\hspace*{0.45\columnwidth} (c)
		\par\end{raggedright}
	\begin{centering}
		\textsf{\includegraphics[width=0.5\columnwidth]{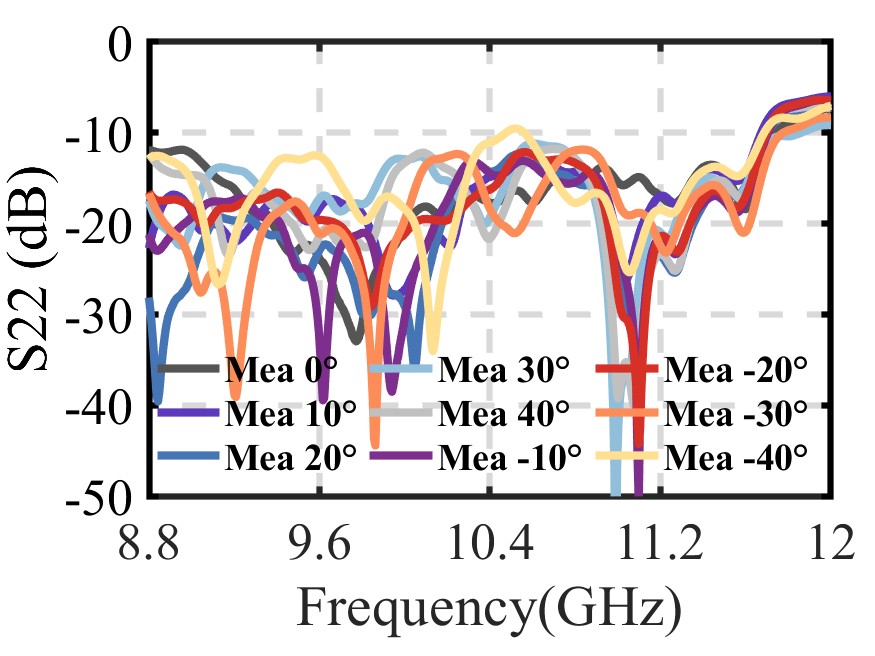}\includegraphics[width=0.5\columnwidth]{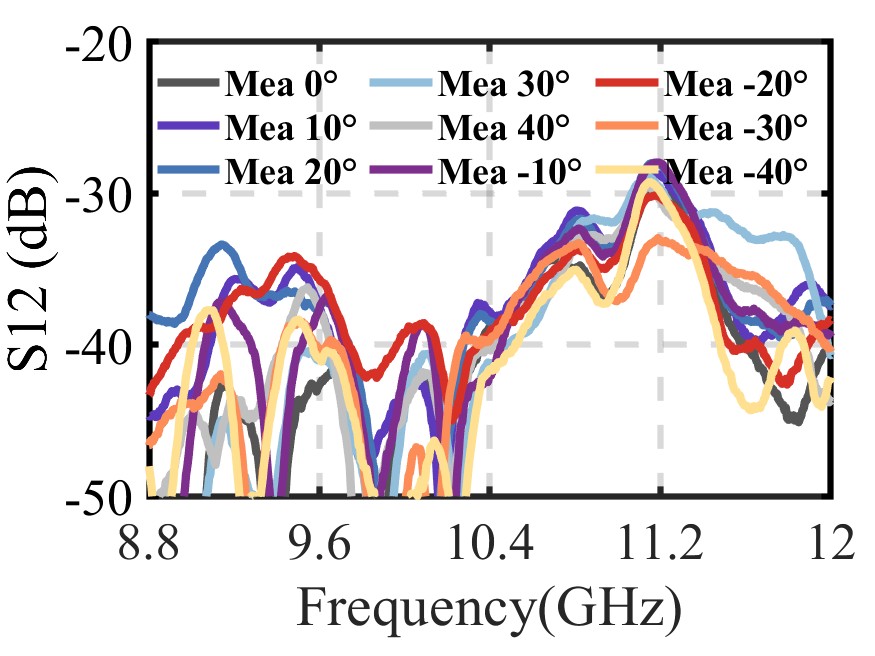}}
		\par\end{centering}
	\begin{raggedright}
		\hspace*{0.25\columnwidth} (b)\hspace*{0.45\columnwidth} (d)
		\par\end{raggedright}
	\caption{Simulated and measured S-parameters of the cm-wave array for E-plane
		scanning. (a) Simulated $S_{22}$. (b) Measured $S_{22}$. (c) Simulated
		$S_{12}$. (d) Measured $S_{12}$ for the nine E-plane scanning states.}
	\label{Mea cm Eplane S parameters}
\end{figure}

\begin{figure}
	\begin{centering}
		\textsf{\includegraphics[width=0.5\columnwidth]{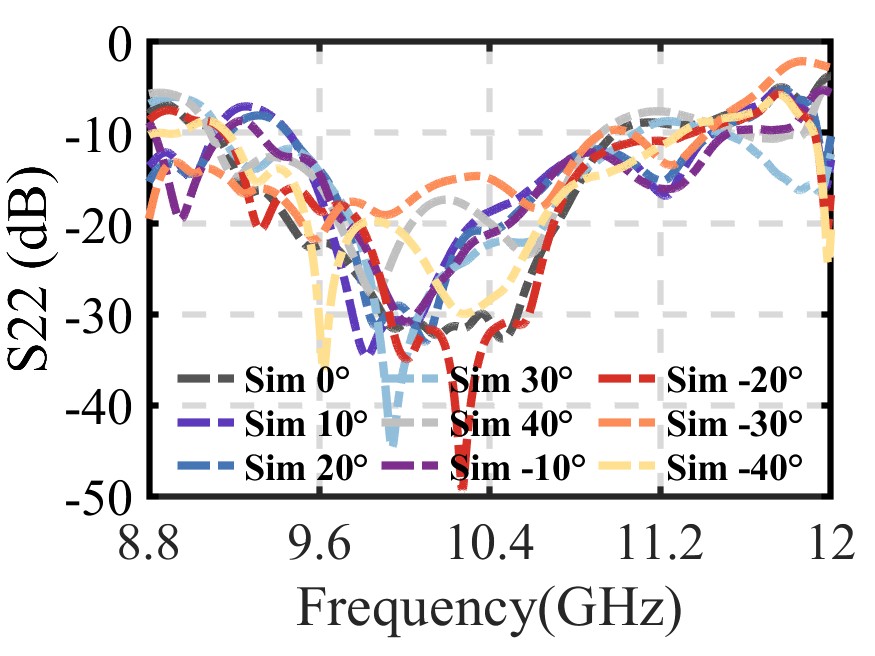}\includegraphics[width=0.5\columnwidth]{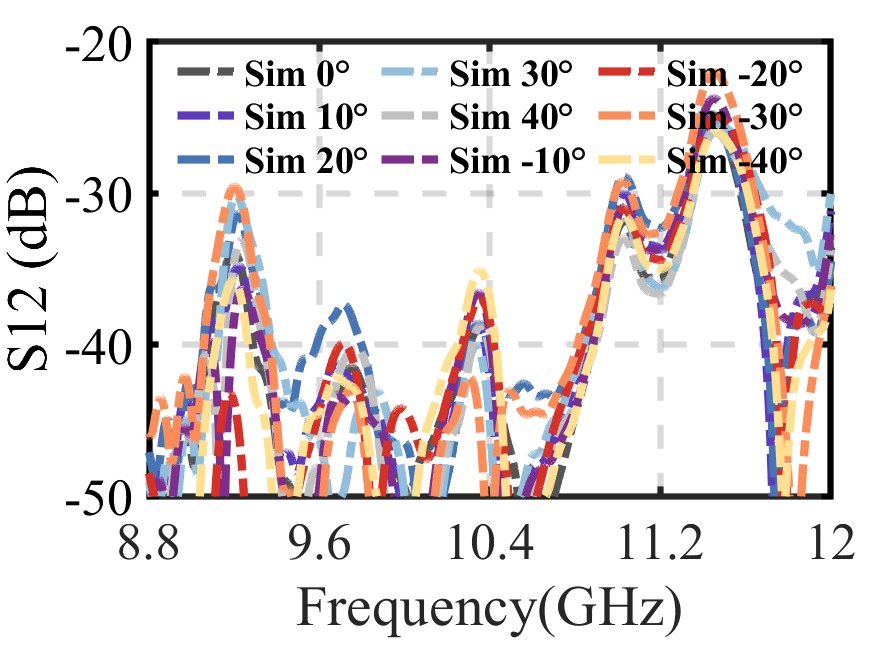}}
		\par\end{centering}
	\begin{raggedright}
		\hspace*{0.25\columnwidth} (a)\hspace*{0.45\columnwidth} (c)
		\par\end{raggedright}
	\begin{centering}
		\textsf{\includegraphics[width=0.5\columnwidth]{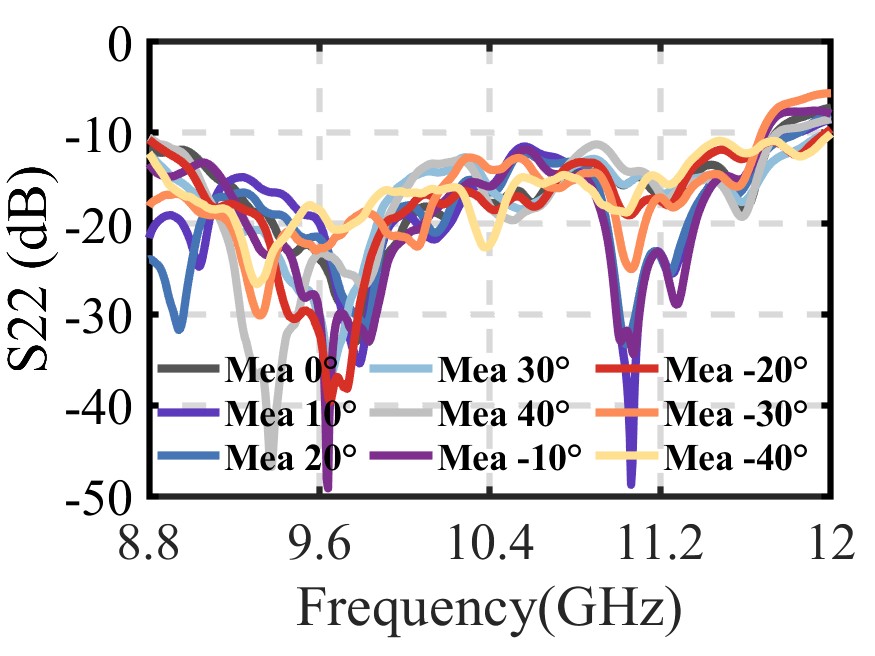}\includegraphics[width=0.5\columnwidth]{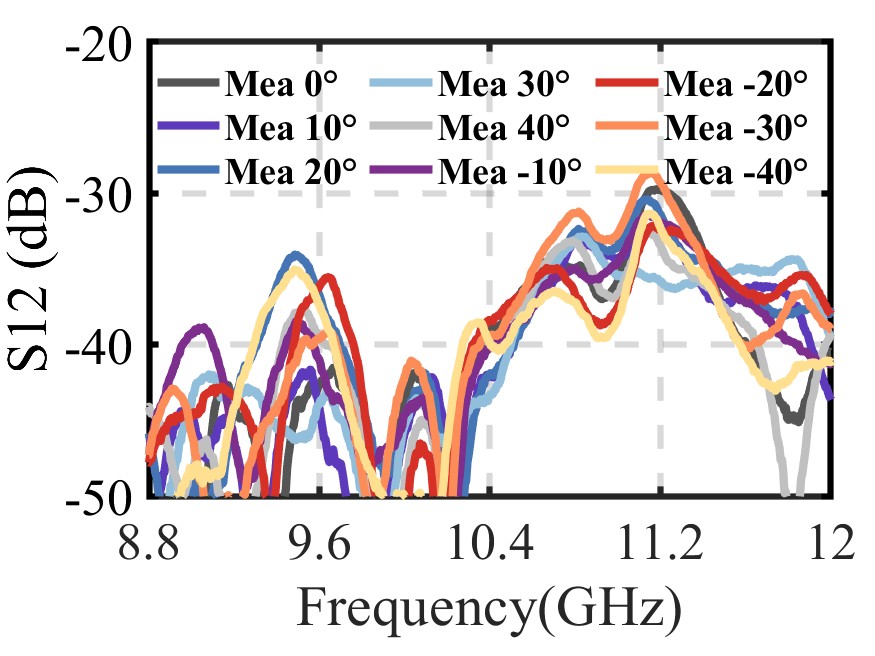}}
		\par\end{centering}
	\begin{raggedright}
		\hspace*{0.25\columnwidth} (b)\hspace*{0.45\columnwidth} (d)
		\par\end{raggedright}
	\caption{Simulated and measured S-parameters of the cm-wave array for H-plane
		scanning. (a) Simulated $S_{22}$. (b) Measured $S_{22}$. (c) Simulated
		$S_{12}$. (d) Measured $S_{12}$ for the nine H-plane scanning states.}
	\label{Mea cm Hplane S parameters}
\end{figure}

The cm-wave performance was measured by feeding port 2 and sequentially
switching the FPGA control states. For E-plane scanning, the simulated
and measured $S_{22}$ responses of all scanning states are shown in
Fig. \ref{Mea cm Eplane S parameters}(a) and (b), respectively, and
all states are well matched from 8.8 to 11.7 GHz. The corresponding
$S_{12}$ responses in Fig. \ref{Mea cm Eplane S parameters}(c) and
(d) show isolation higher than 30 dB. Similar impedance matching
and isolation performance are observed for H-plane scanning, as shown
in Fig. \ref{Mea cm Hplane S parameters}.

\begin{figure*}[t]
	\begin{centering}
		\textsf{\includegraphics[width=0.55\columnwidth]{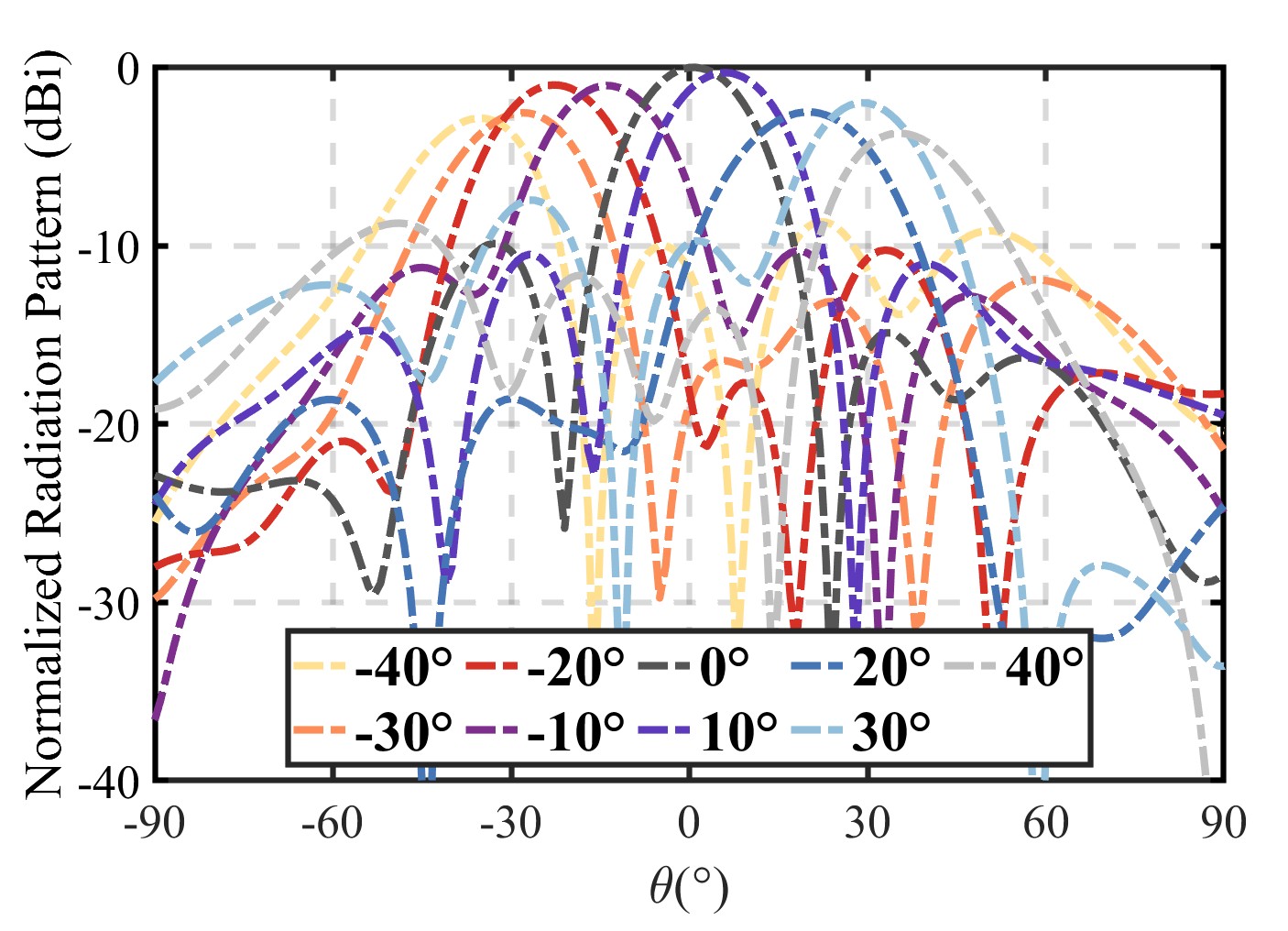}\includegraphics[width=0.55\columnwidth]{\string"Figures/10.4G_Eplane_sim_all_Board3\string".jpg}\includegraphics[width=0.55\columnwidth]{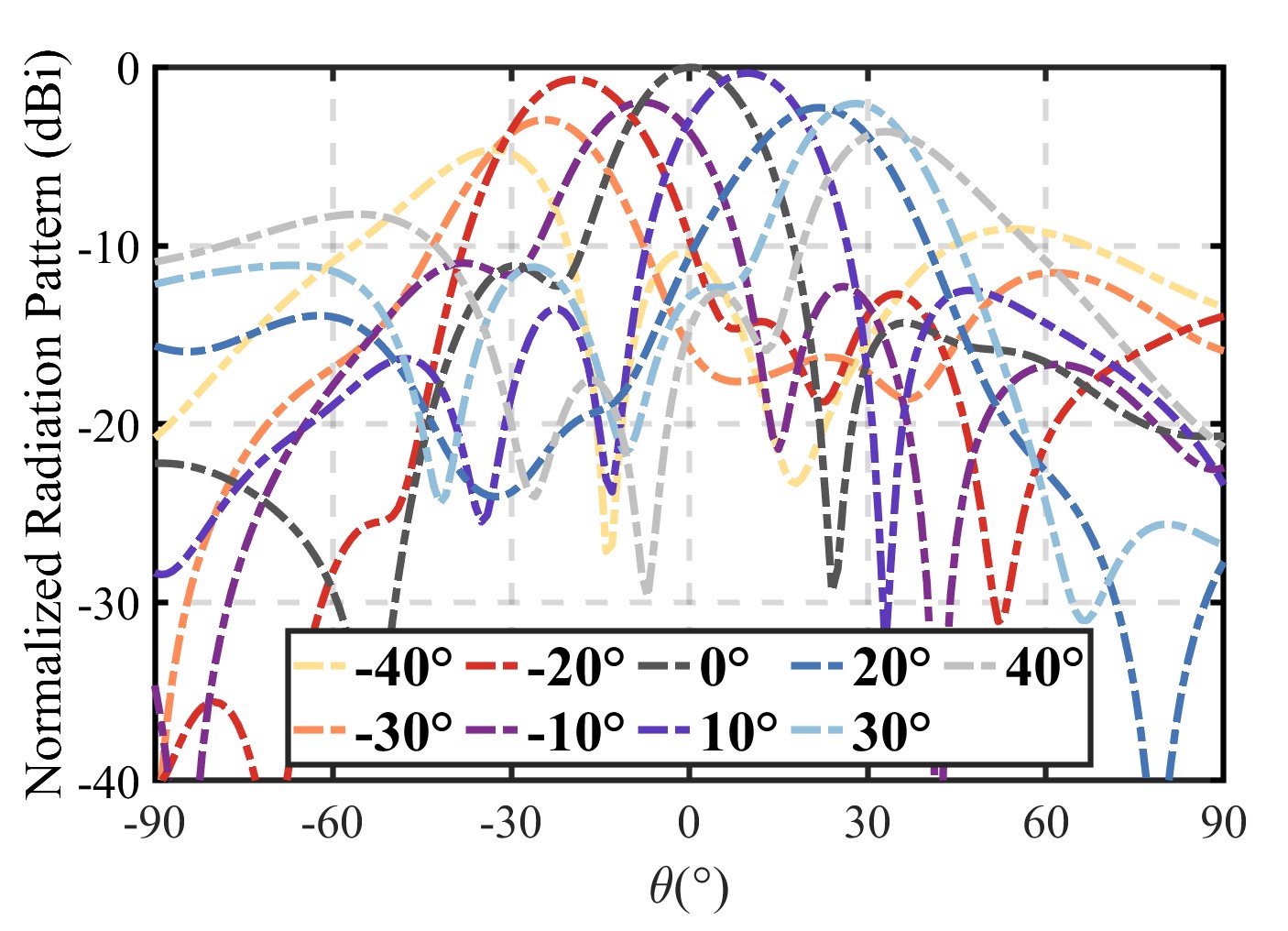}}
		\par\end{centering}
	\begin{raggedright}
		\hspace*{0.45\columnwidth} (a)\hspace*{0.5\columnwidth} (b)\hspace*{0.5\columnwidth}
		(c)
		\par\end{raggedright}
	\begin{centering}
		\textsf{\includegraphics[width=0.55\columnwidth]{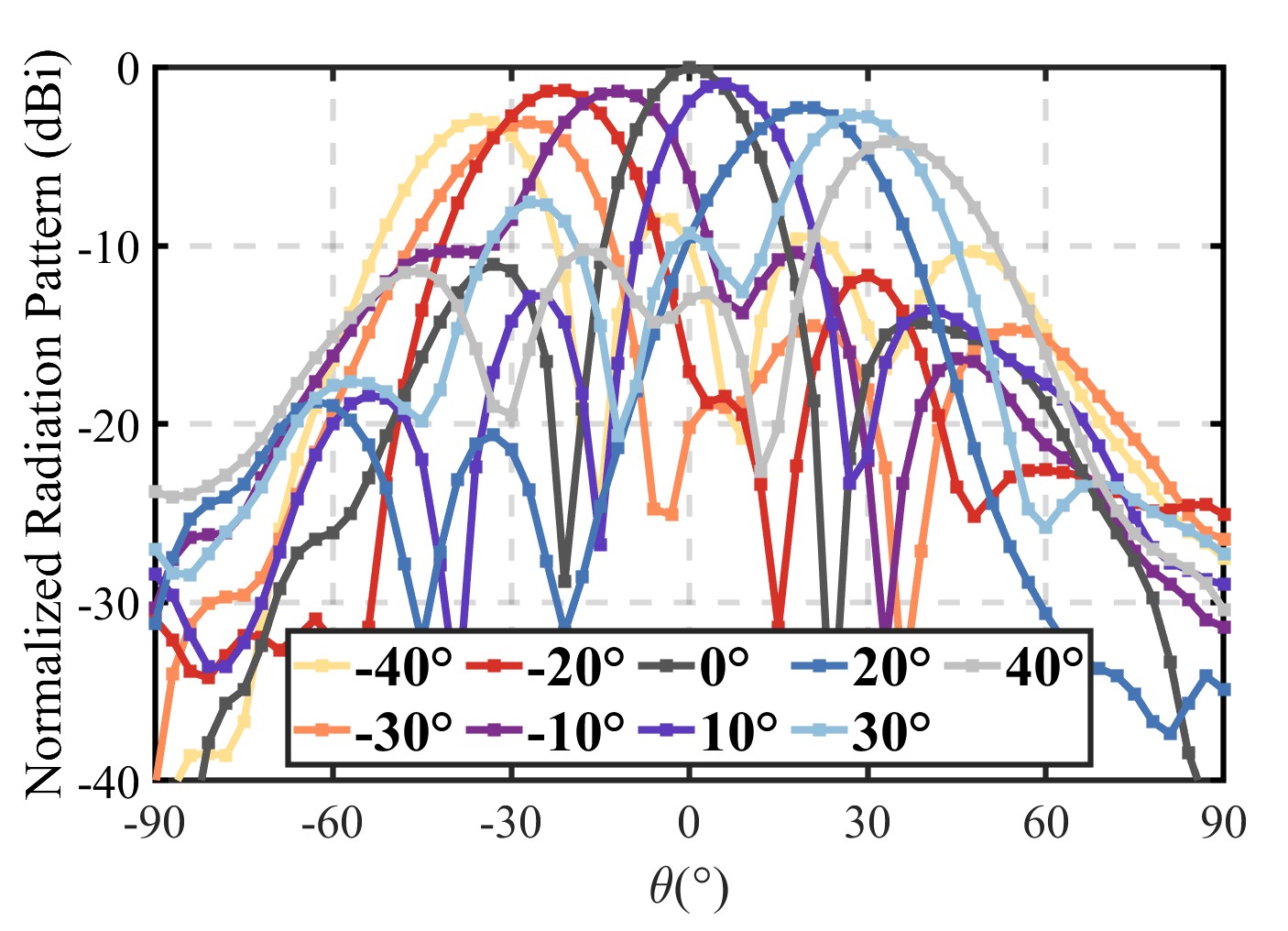}\includegraphics[width=0.55\columnwidth]{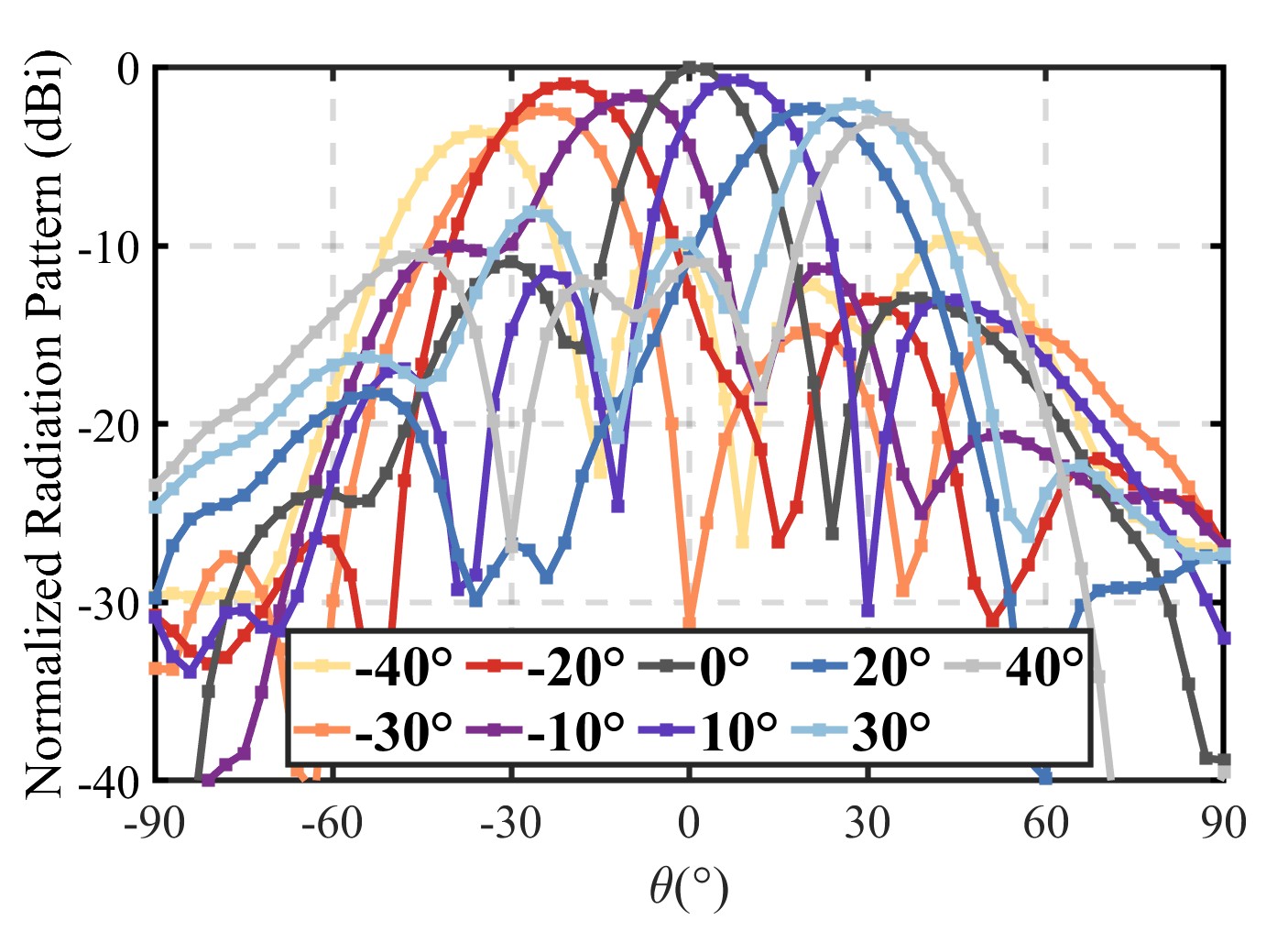}\includegraphics[width=0.55\columnwidth]{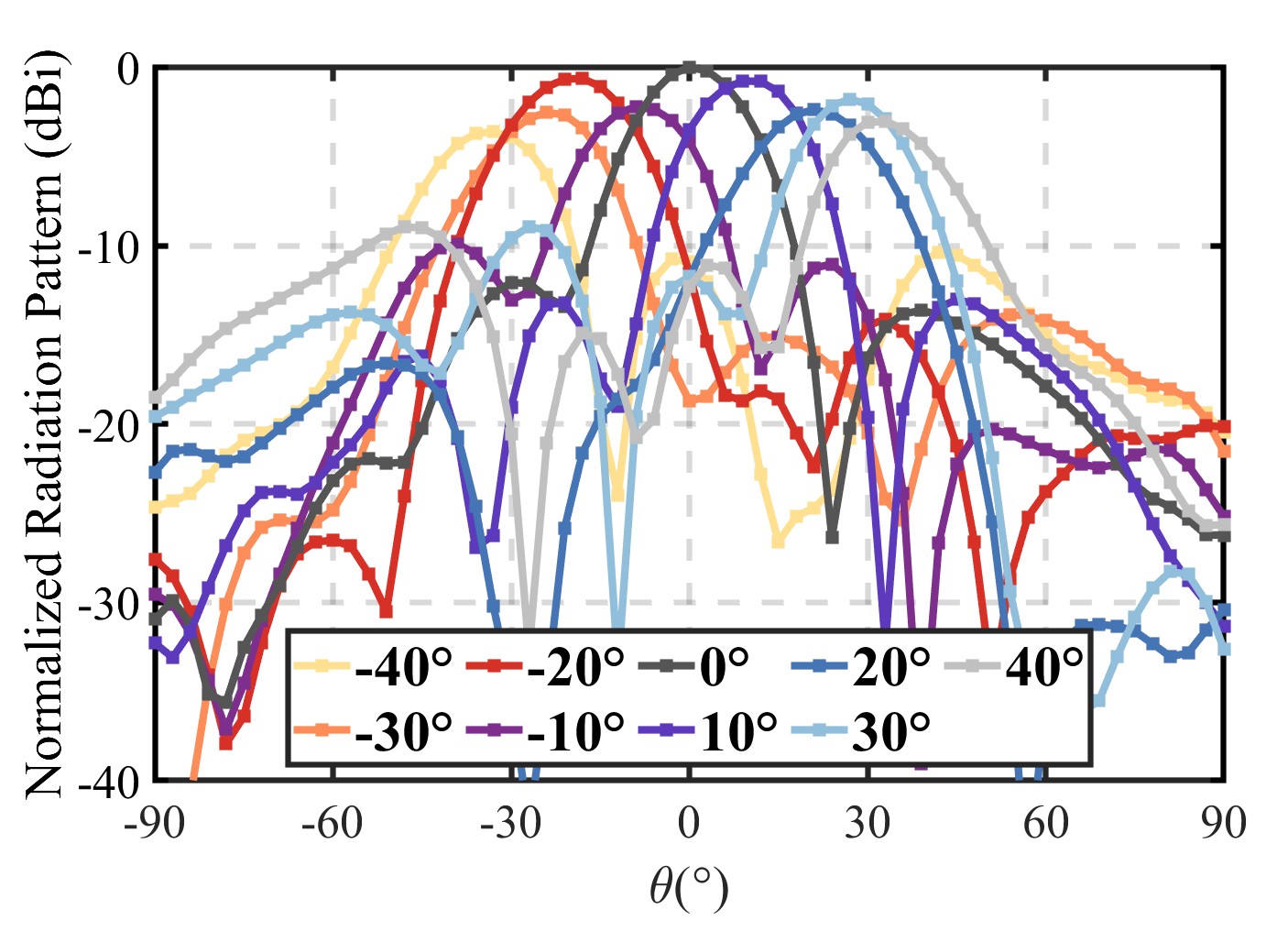}}
		\par\end{centering}
	\begin{raggedright}
		\hspace*{0.45\columnwidth} (d)\hspace*{0.5\columnwidth} (e)\hspace*{0.5\columnwidth}
		(f)
		\par\end{raggedright}
	\caption{Simulated and measured cm-wave radiation patterns for E-plane scanning
		at different frequencies. (a)-(c) Simulated results at 10.2, 10.4, and 10.6 GHz, respectively. (d)-(f) Corresponding measured results.}
	\label{sim and mea Eplane scanning performance}
\end{figure*}

\begin{figure*}[t]
	\begin{centering}
		\textsf{\includegraphics[width=0.55\columnwidth]{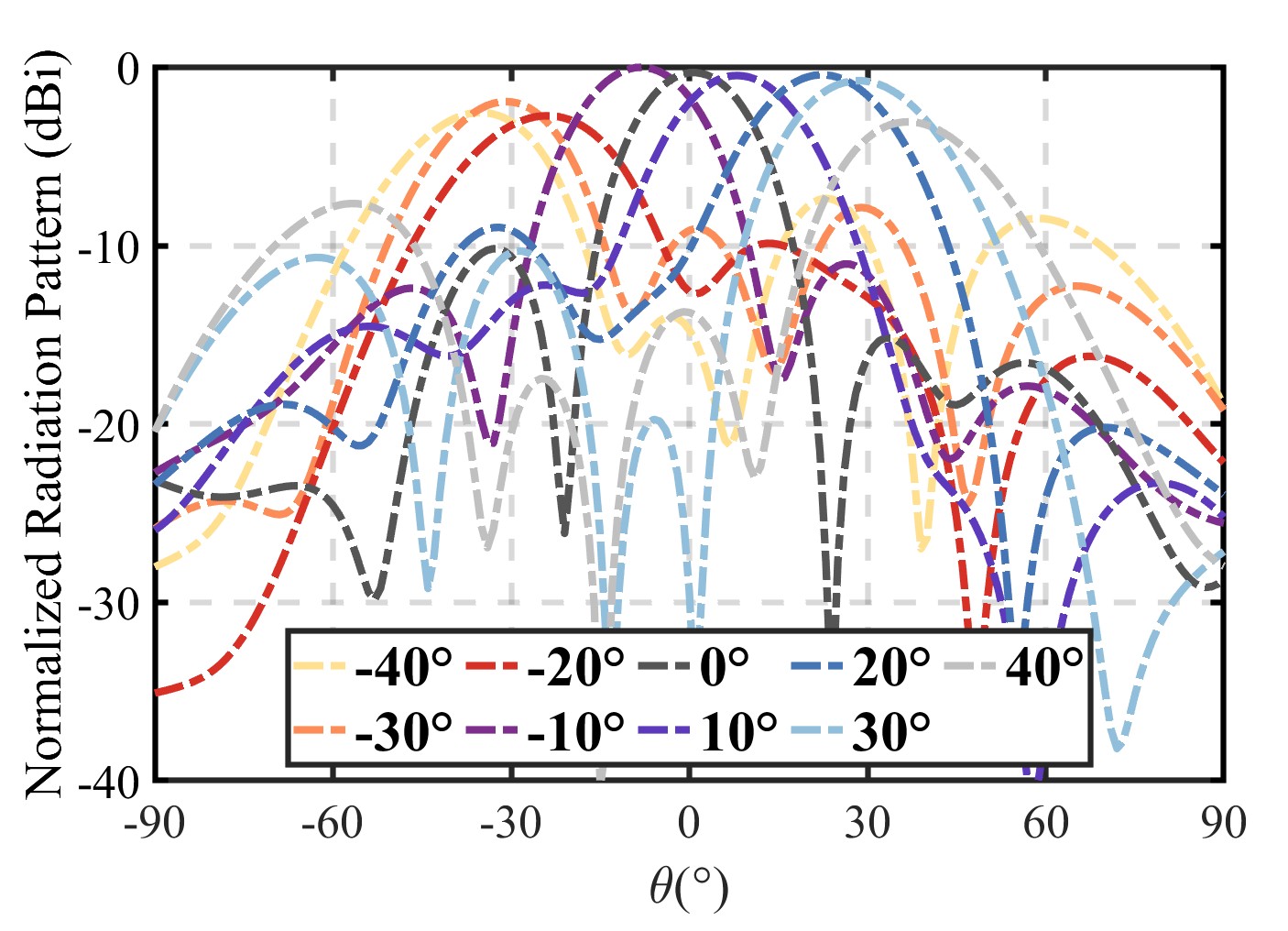}\includegraphics[width=0.55\columnwidth]{\string"Figures/10.4G_Hplane_sim_all_Board3\string".jpg}\includegraphics[width=0.55\columnwidth]{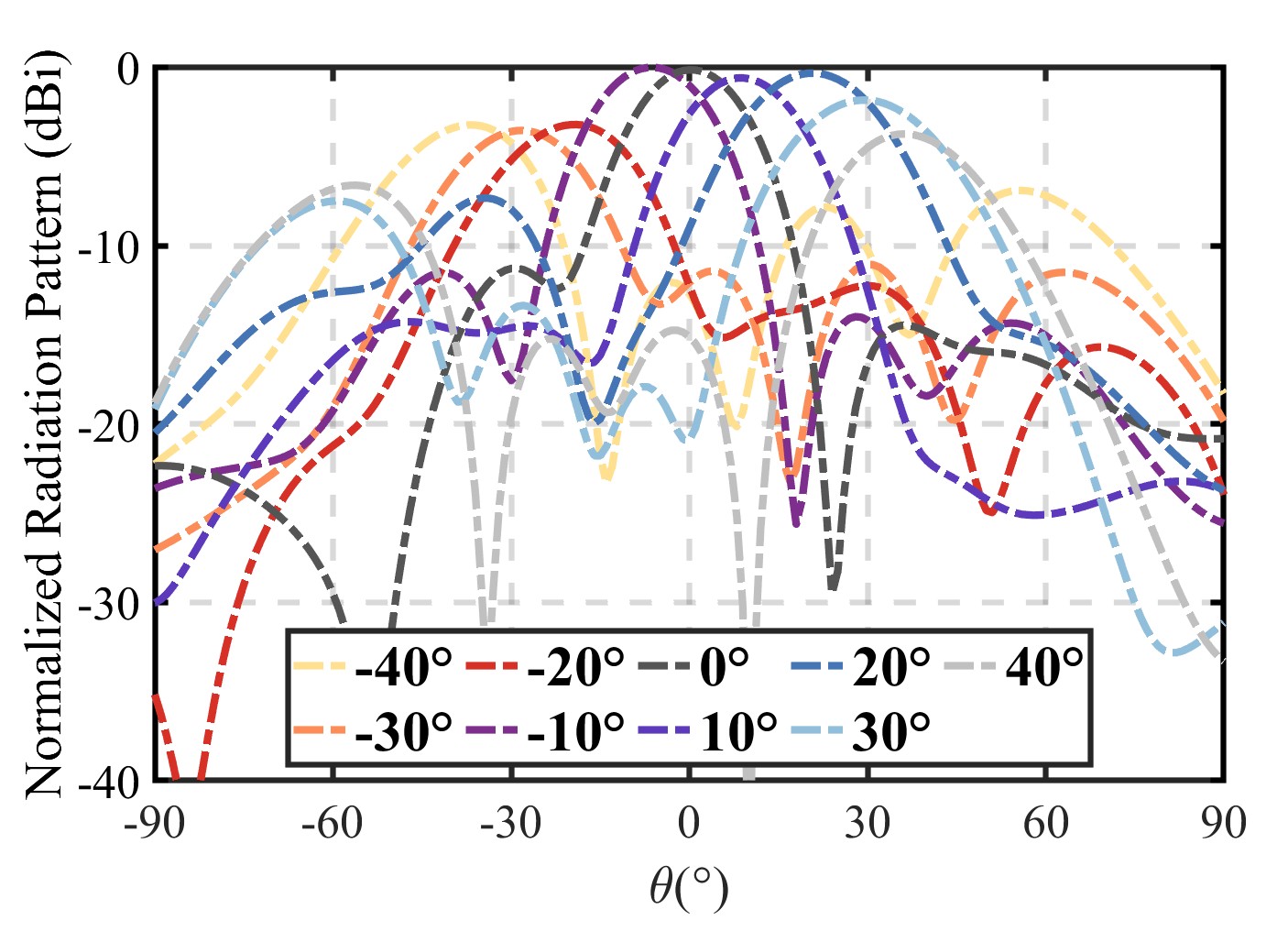}}
		\par\end{centering}
	\begin{raggedright}
		\hspace*{0.45\columnwidth} (a)\hspace*{0.5\columnwidth} (b)\hspace*{0.5\columnwidth}
		(c)
		\par\end{raggedright}
	\begin{centering}
		\textsf{\includegraphics[width=0.55\columnwidth]{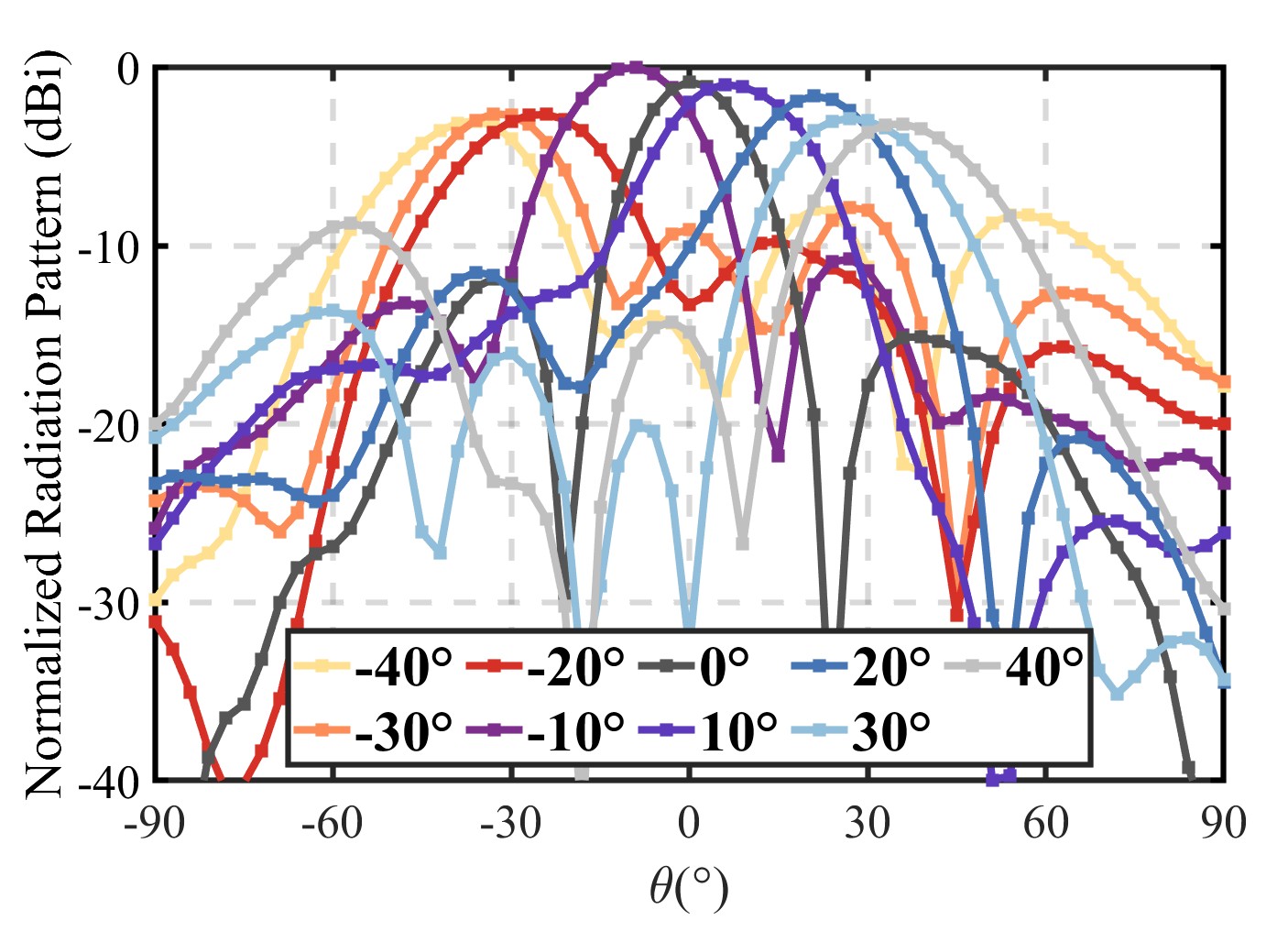}\includegraphics[width=0.55\columnwidth]{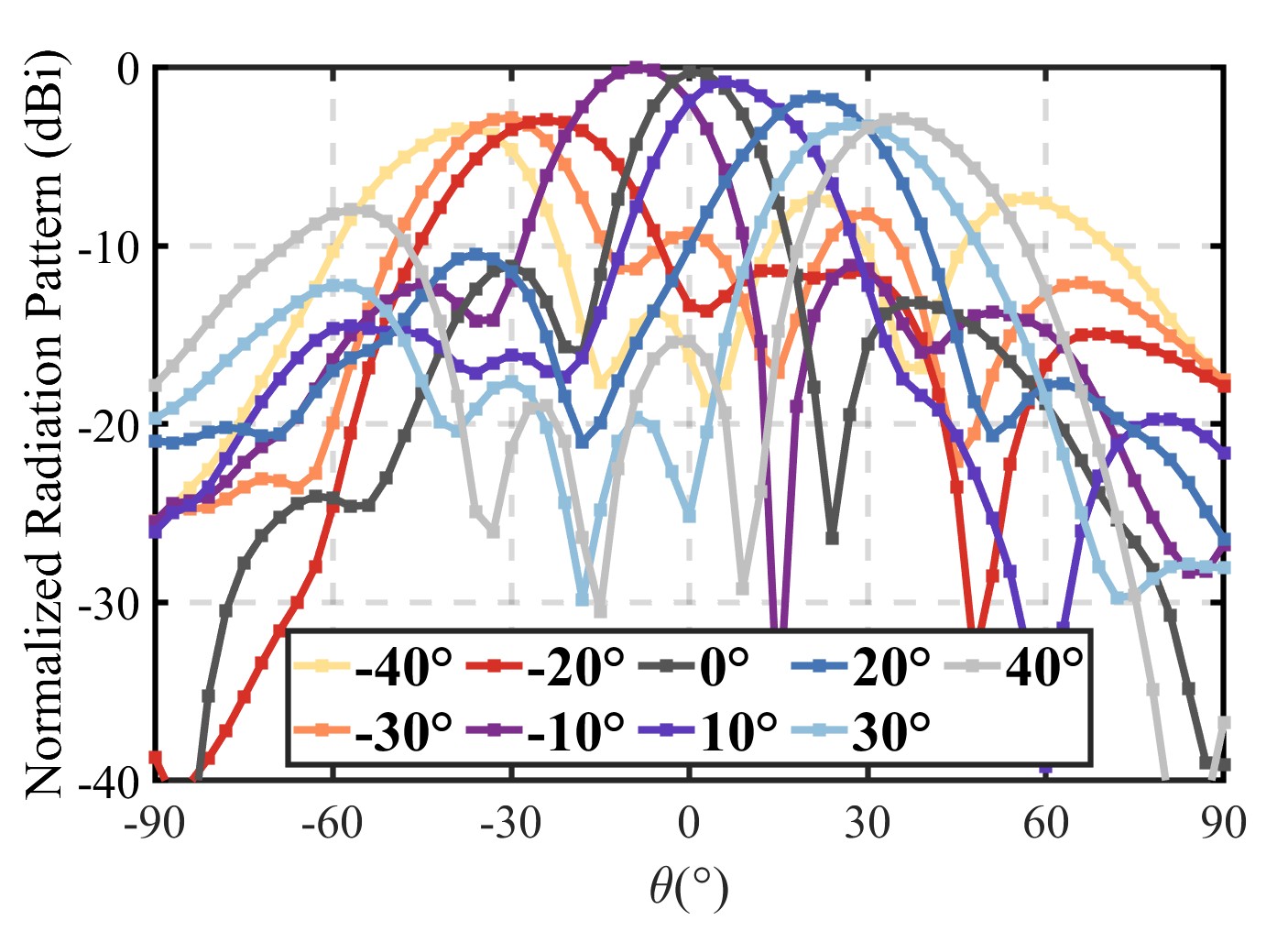}\includegraphics[width=0.55\columnwidth]{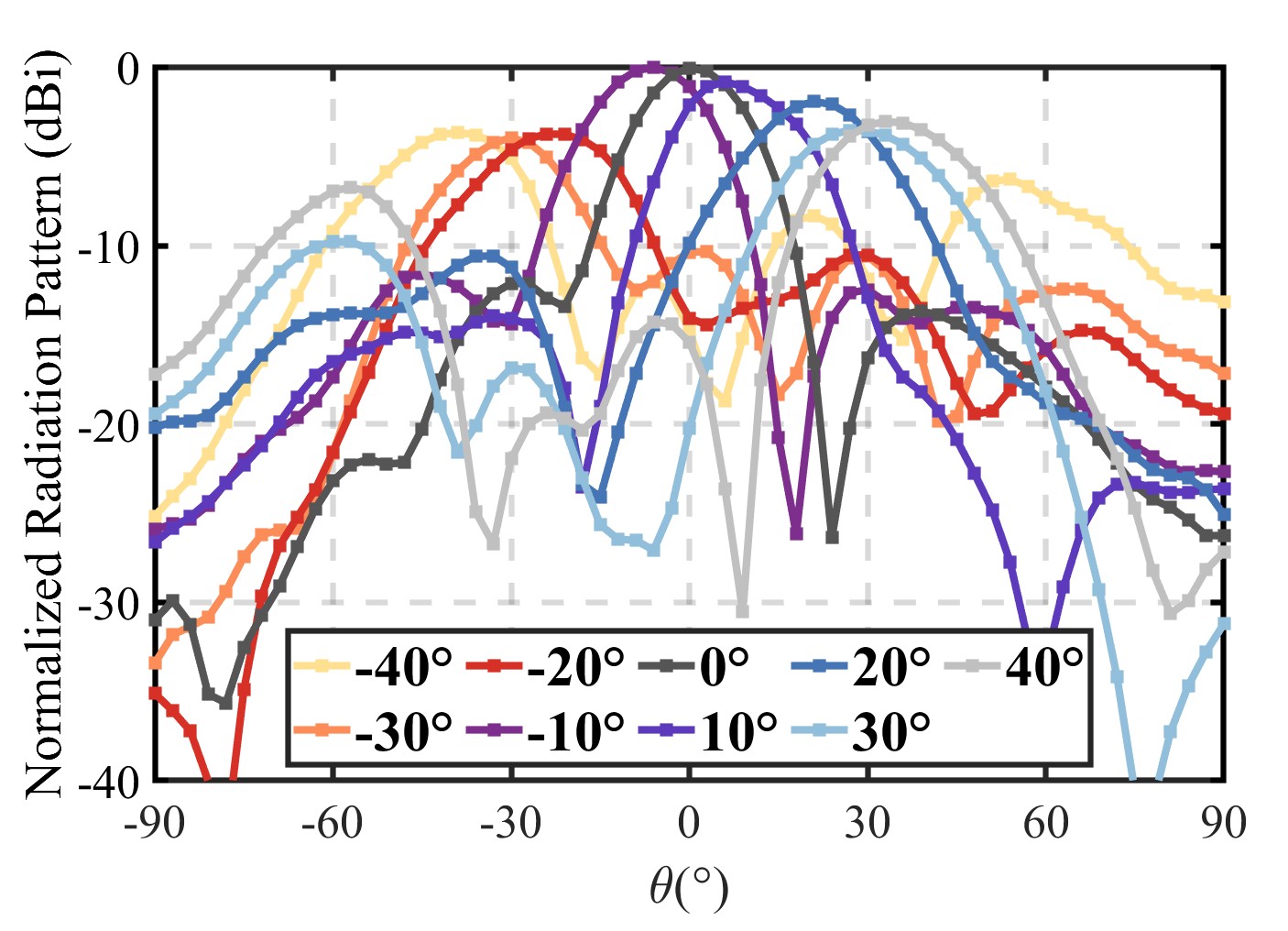}}
		\par\end{centering}
	\begin{raggedright}
		\hspace*{0.45\columnwidth} (d)\hspace*{0.5\columnwidth} (e)\hspace*{0.5\columnwidth}
		(f)
		\par\end{raggedright}
	\caption{Simulated and measured cm-wave radiation patterns for H-plane scanning
		at different frequencies. (a)-(c) Simulated results at 10.2, 10.4, and
		10.6 GHz, respectively. (d)-(f) Corresponding measured results.}
	\label{sim and mea Hplane scanning performance}
\end{figure*}

The simulated and measured radiation patterns for E-plane scanning
at 10.2, 10.4, and 10.6 GHz are shown in Fig. \ref{sim and mea Eplane scanning performance}.
The corresponding H-plane scanning results are shown in Fig. \ref{sim and mea Hplane scanning performance}.
In both planes, the 4$\times$4 quasi-2-bit cm-wave array achieves beam
scanning up to $\pm40^{\circ}$ with a sidelobe level below -5 dB.
The realized peak gain reaches 14.6 dBi at 10.4 GHz in both the E-plane
and H-plane scanning cases.

\begin{figure}[t]
	\begin{centering}
		\textsf{\includegraphics[width=0.5\columnwidth]{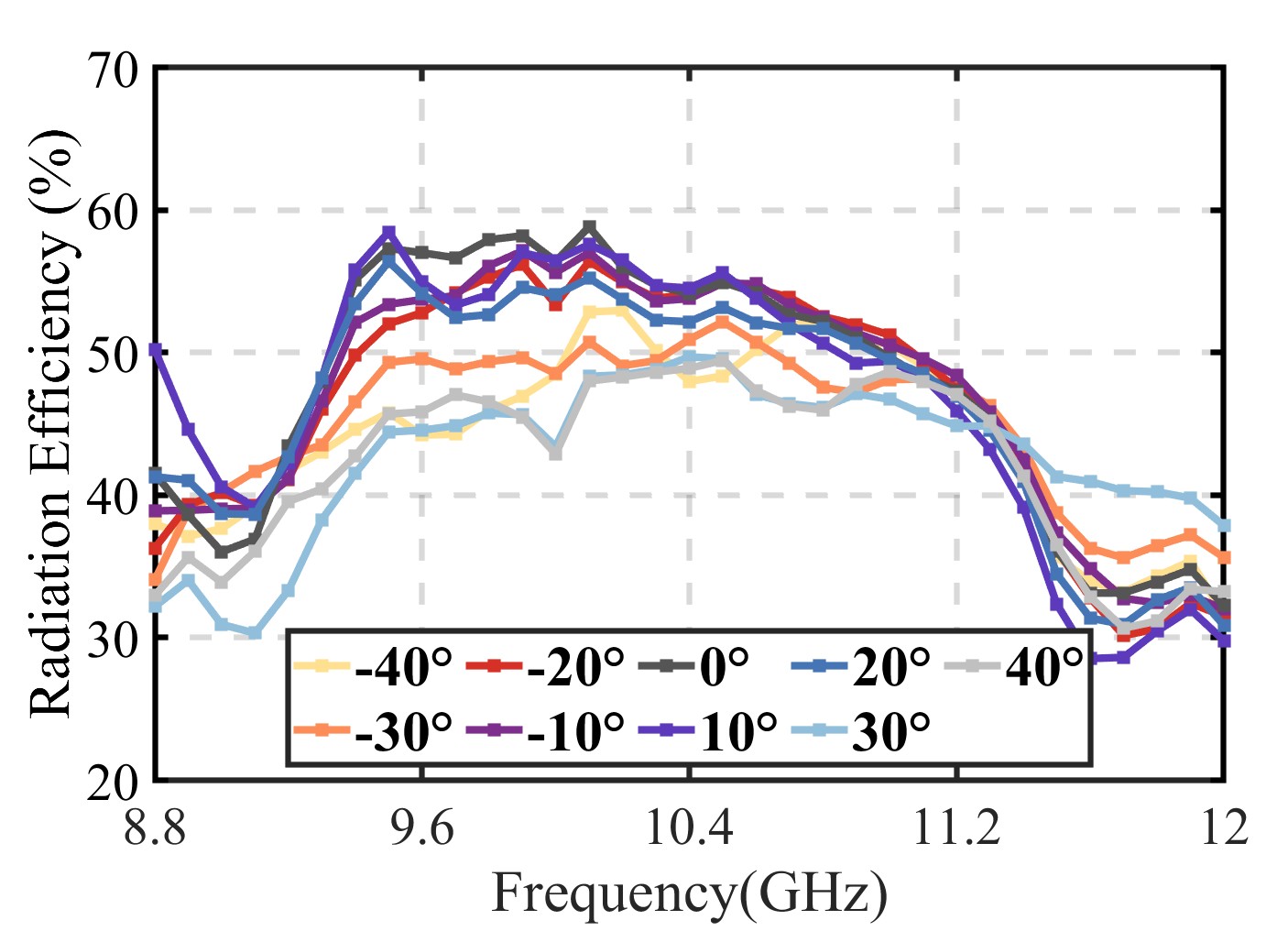}\includegraphics[width=0.5\columnwidth]{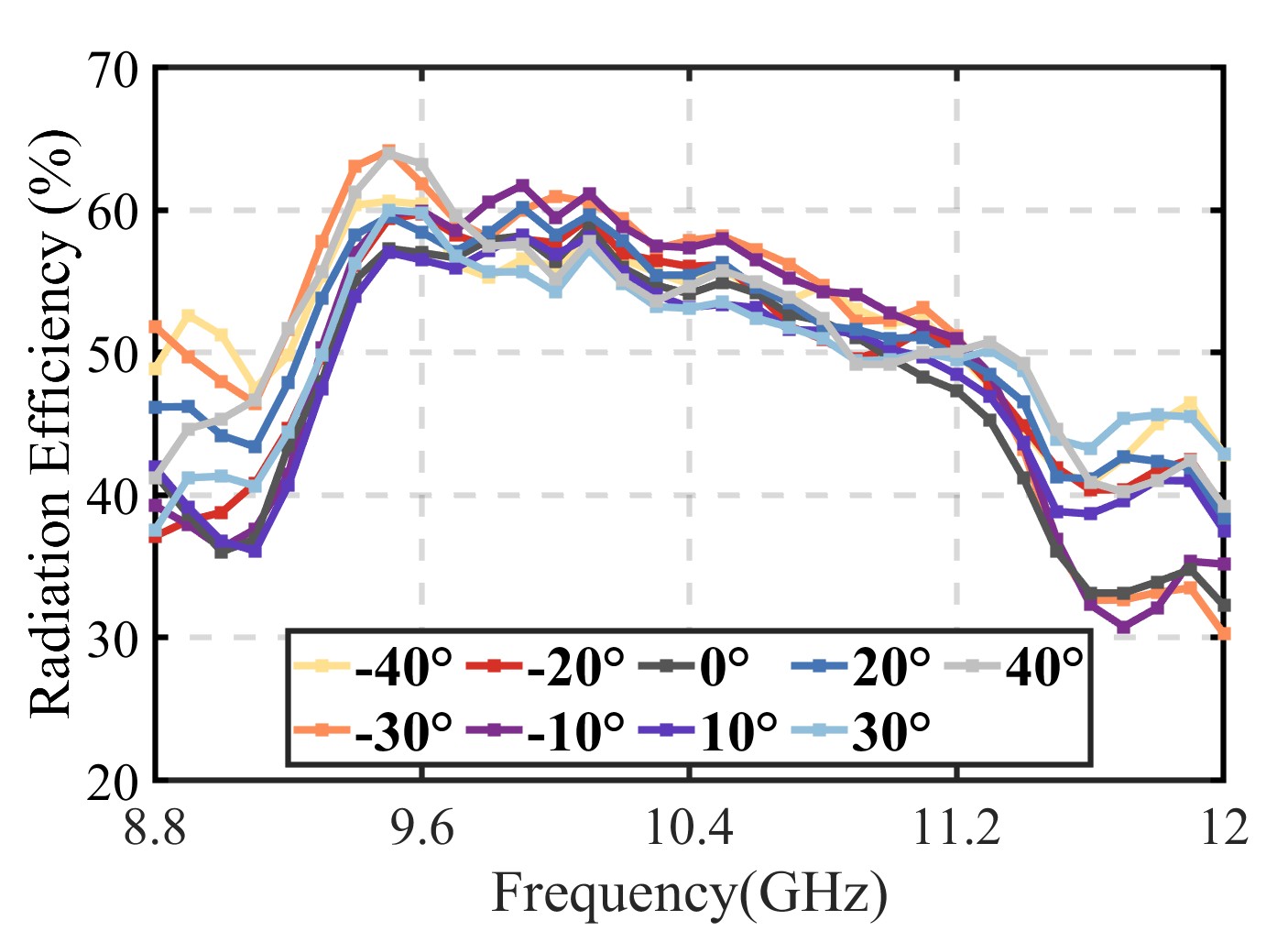}}
		\par\end{centering}
	\begin{raggedright}
		\hspace*{0.24\columnwidth} (a)\hspace*{0.44\columnwidth} (b)
		\par\end{raggedright}
	\caption{Measured radiation efficiencies of the cm-wave array for the nine scanning
		states. (a) E-plane scanning. (b) H-plane scanning.}
	\label{Mea cm efficiency}
\end{figure}

The measured radiation efficiencies of the cm-wave scanning states
are shown in Fig. \ref{Mea cm efficiency}. Most scanning states exhibit
radiation efficiencies higher than 53\% within the operating band,
although a 5\% to 10\% degradation is observed compared with simulation.
In addition to the parasitic effects of the lumped components, PIN
diodes, and cables, this degradation is partly caused by the limited
FPGA drive voltage. In the prototype, the FPGA ground is biased to
$-1.6$ V, so each I/O pin with a 3.3-/0-V output provides either $-1.6$
V or 1.6 V. A higher drive voltage could further reduce the ON-state
resistance of the switches in future implementations. For E-plane scanning
angles larger than $30^{\circ}$, the efficiency is further reduced
by about 5\% to 10\%, which is mainly attributed to stronger coupling
between adjacent sub-6-GHz and cm-wave elements and to stronger perturbation
from the DC control lines arranged along the E-plane direction.

Overall, the measured results verify that the proposed shared-aperture
array provides independent reconfigurable operation in both the sub-6-GHz
and cm-wave bands.

\section{Comparison and Discussion}

\begin{table*}[tp]
\centering{}
\caption{Comparison with related work}
\label{Comparison}%
{\scriptsize
\setlength{\tabcolsep}{5.1pt}
\renewcommand{\arraystretch}{1}
\begin{tabular}{ccccccccccc}
\toprule
Ref. & Freq. & \begin{cellvarwidth}[m]
\centering
Shared-aperture\\phased array
\end{cellvarwidth} & \begin{cellvarwidth}[m]
\centering
Reconfigurable\\antenna array
\end{cellvarwidth} & Reconf. & \begin{cellvarwidth}[m]
\centering
Array\\scale
\end{cellvarwidth} & Profile & \begin{cellvarwidth}[m]
\centering
Scanning\\performance
\end{cellvarwidth} & \begin{cellvarwidth}[m]
\centering
Peak gain (dBi)\\(Apert. eff.)
\end{cellvarwidth} & \begin{cellvarwidth}[m]
\centering
Max SLL\\(dB)
\end{cellvarwidth} & \begin{cellvarwidth}[m]
\centering
Isolation\\(dB)
\end{cellvarwidth}\tabularnewline
\midrule
\cite{yang2021dual} & 2.7/4.9 & \yesmark & \nomark & N.A. & \begin{cellvarwidth}[m]
\centering
1$\times$8;\\1$\times$16
\end{cellvarwidth} & 0.25$\lambda_{0}$ & $\pm60^{\circ}$ (1-D) & \begin{cellvarwidth}[m]
\centering
13.8 (73.3\%);\\15.7 (68.2\%)
\end{cellvarwidth} & -10 & 18/20\tabularnewline
\addlinespace[4pt]
\cite{hao2022k} & 19.7/29 & \yesmark & \nomark & N.A. & \begin{cellvarwidth}[m]
\centering
8$\times$8;\\8$\times$8
\end{cellvarwidth} & N.R. & $\pm60^{\circ}$ (2-D) & $>$20 (N.R.) & -7 & $>$40\tabularnewline
\addlinespace[4pt]
\cite{kim2023dual} & 14/28 & \yesmark & \nomark & N.A. & \begin{cellvarwidth}[m]
\centering
1$\times$3;\\1$\times$5
\end{cellvarwidth} & 0.1$\lambda_{0}$ & \begin{cellvarwidth}[m]
\centering
$\pm41^{\circ}$ (2-D);\\$\pm43^{\circ}$ (2-D)
\end{cellvarwidth} & \begin{cellvarwidth}[m]
\centering
10.3 (N.R.);\\13.1 (N.R.)
\end{cellvarwidth} & -3/-8 & $>$30/20\tabularnewline
\addlinespace[4pt]
\cite{mmwave-wang2022low} & 27 & \nomark & \yesmark & 1-bit reconfig. & 1$\times$4 & N.R. & $\pm37^{\circ}$ (1-D) & 8.9 (N.R.) & -5.4 & N.A.\tabularnewline
\addlinespace[4pt]
\cite{li20221} & 9.5 & \nomark & \yesmark & 1-bit reconfig. & 8$\times$8 & 1.15$\lambda_{0}$ & $\pm60^{\circ}$ (2-D) & 16.1 (N.R.) & -6 & N.A.\tabularnewline
\addlinespace[4pt]
\cite{liu2019circularly} & 3.65 & \nomark & \yesmark & 2-bit reconfig. & 1$\times$8 & N.R. & $\pm49^{\circ}$ (1-D) & 11.8 (60\%) & -7.5 & N.A.\tabularnewline
\addlinespace[4pt]
\cite{yin2023modular} & 18.7 & \nomark & \yesmark & 2-bit reconfig. & 16$\times$7 & N.R. & $\pm60^{\circ}$ (2-D) & 23.3 (24.3\%) & -7 & N.A.\tabularnewline
\addlinespace[4pt]
\cite{wang2022low} & 11 & \nomark & \yesmark & 4-bit reconfig. & 8$\times$8 & 0.09$\lambda_{0}$ & \begin{cellvarwidth}[m]
\centering
$\pm60^{\circ}$ (E);\\$\pm40^{\circ}$ (H)
\end{cellvarwidth} & 16 (21.9\%) & -9 & N.A.\tabularnewline
\addlinespace[4pt]
\textbf{This work} & 5.2/10.4 & \yesmark & \yesmark & \begin{cellvarwidth}[m]
\centering
2-bit /\\quasi-2-bit reconfig.
\end{cellvarwidth} & \begin{cellvarwidth}[m]
\centering
2$\times$2;\\4$\times$4
\end{cellvarwidth} & 0.14$\lambda_{0}$ & \begin{cellvarwidth}[m]
\centering
Sub-6: 11 states;\\cm-wave: $\pm40^{\circ}$ (2-D)
\end{cellvarwidth} & \begin{cellvarwidth}[m]
\centering
10.5 (50.4\%);\\14.6 (32.2\%)
\end{cellvarwidth} & N.A./-5 & 34/36\tabularnewline
\midrule
\multicolumn{11}{l}{N.R.: not reported; N.A.: not applicable; Apert. eff.: aperture efficiency; SLL: sidelobe level.}\tabularnewline
\end{tabular}}
\end{table*}

Table \ref{Comparison} compares the proposed dual-band reconfigurable
shared-aperture antenna array with related work. The first three designs
in the table are dual-band shared-aperture phased arrays \cite{yang2021dual,hao2022k,kim2023dual}.
They provide wide beam-scanning ranges in two bands within a
shared aperture, but they require additional T/R modules or beamforming
networks. These components increase the system complexity, power consumption,
volume, weight, and cost. In contrast, the proposed array realizes
independent beam control in both bands using reconfigurable antenna
elements, compact phase shifters, and fixed phase-delay lines, without
requiring separate T/R modules or beamforming networks.

Compared with reported reconfigurable antenna arrays \cite{mmwave-wang2022low,li20221,liu2019circularly,yin2023modular},
the proposed design extends phase reconfigurability from a single band
to two frequency bands within the same physical aperture. The demonstrated
array also maintains competitive scanning range, gain, sidelobe level,
profile, and port isolation. At the same time, the present prototype
uses relatively small array scales, and the bandwidth, sidelobe level,
and polarization diversity can be further improved. Future work may
therefore scale up the aperture, broaden the operating bandwidth, and
introduce polarization-diverse reconfigurable states.

\section{Conclusion}

A dual-band reconfigurable shared-aperture antenna array with independent
sub-6-GHz and cm-wave beam control has been presented. The array integrates
a 2$\times$2 sub-6-GHz microstrip dipole array and a 4$\times$4 cm-wave
stacked patch array within the same aperture. Slot-coupled feeding
separates the radiators from the RF feeding and DC bias networks, while
PIN-diode switching provides 1-bit phase reconfigurability for both
bands. A compact reconfigurable $90^{\circ}$ phase shifter further
improves the phase resolution, resulting in 2-bit sub-6-GHz elements
and quasi-2-bit cm-wave subarrays. A double-layer EBG structure is
also introduced to suppress cm-wave surface waves and reduce the higher-order
sub-6-GHz modes excited by the cm-wave elements.

A prototype was fabricated and measured to validate the proposed design.
In the sub-6-GHz band, 11 representative reconfigurable patterns are
obtained, including two difference patterns and nine directional beams,
with a measured peak realized gain of 10.5 dBi. In the cm-wave band,
the array achieves two-dimensional beam scanning up to $\pm40^{\circ}$
with a measured peak realized gain of 14.6 dBi. These results show
that the proposed architecture integrates dual-band shared-aperture
operation and reconfigurable beam control without dedicated T/R modules
or conventional beamforming networks, and therefore provides a compact
antenna platform for future multifunctional wireless systems.

\appendices{}
\begin{lyxlist}{00.00.0000}
\item [{}]~
\end{lyxlist}
\bibliographystyle{IEEEtran}
\bibliography{Ref}

\end{document}